\documentclass[a4paper,fleqn,usenatbib]{mnras}

\usepackage{newtxtext,newtxmath}
\usepackage[T1]{fontenc}
\usepackage{ae,aecompl}

\usepackage{adjustbox}
\usepackage{multirow}

\usepackage{graphicx}	% Including figure files
\usepackage{amsmath}	% Advanced maths commands
\usepackage{amssymb}	% Extra maths symbols
\usepackage{epsfig,color,pifont,array}
\usepackage[outdir=./]{epstopdf}
\usepackage{bm}
\usepackage{etoolbox}

\makeatletter

\patchcmd{\NAT@citex}
  {\@citea\NAT@hyper@{%
     \NAT@nmfmt{\NAT@nm}%
     \hyper@natlinkbreak{\NAT@aysep\NAT@spacechar}{\@citeb\@extra@b@citeb}%
     \NAT@date}}
  {\@citea\NAT@nmfmt{\NAT@nm}%
   \NAT@aysep\NAT@spacechar\NAT@hyper@{\NAT@date}}{}{}

\patchcmd{\NAT@citex}
  {\@citea\NAT@hyper@{%
     \NAT@nmfmt{\NAT@nm}%
     \hyper@natlinkbreak{\NAT@spacechar\NAT@@open\if*#1*\else#1\NAT@spacechar\fi}%
       {\@citeb\@extra@b@citeb}%
     \NAT@date}}
  {\@citea\NAT@nmfmt{\NAT@nm}%
   \NAT@spacechar\NAT@@open\if*#1*\else#1\NAT@spacechar\fi\NAT@hyper@{\NAT@date}}
  {}{}

\makeatother

\newcommand\Tstrut{\rule{0pt}{2.6ex}}         % = `top' strut
\newcommand\Bstrut{\rule[-1.5ex]{0pt}{2.6ex}}   % = `bottom' strut
\defcitealias{2016MNRAS.455.1309B}{Paper I}
\defcitealias{2018MNRAS.473.3454B}{Paper II}

\title[Dynamics and survival of fractal clouds in galactic winds]{On the dynamics and survival of fractal clouds in galactic winds}

\author[W.~E.~Banda-Barrag\'{a}n et al.]{W.~E.~Banda-Barrag\'{a}n,$^{1,2}$\thanks{E-mail: wlady.bsc@gmail.com (WBB)}
F.~J.~Zertuche,$^{3,2}$
C.~Federrath,$^{4}$
J.~Garc\'{i}a Del Valle,$^{2}$
\newauthor % starts a new line in the
M.~Br\"{u}ggen,$^{1}$ and
A.~Y.~Wagner$^{5}$\\
$^{1}$Hamburger Sternwarte, Universit\"{a}t Hamburg, Gojenbergsweg 112, D-21029 Hamburg, Germany\\
$^{2}$Facultad de Ingenier\'{i}a Civil y Mec\'{a}nica, Universidad T\'{e}cnica de Ambato, Av. Los Chasquis y R\'{i}o Payamino S/N, Ambato 180206, Ecuador\\
$^{3}$Centro de Investigación Biom\'{e}dica, Universidad Tecnol\'{o}gica Equinoccial, Av. Mariscal Sucre y Mariana de Jes\'{u}s, Quito 170105, Ecuador\\
$^{4}$Research School of Astronomy and Astrophysics, Australian National University, Canberra, ACT 2611, Australia\\
$^{5}$Center for Computational Sciences, University of Tsukuba, 1-1-1 Tennodai, Tsukuba, Ibaraki 305-8577, Japan\\
}

\date{Accepted XXX. Received YYY; in original form ZZZ}

\pubyear{2019}

\begin{document}
\label{firstpage}
\pagerange{\pageref{firstpage}--\pageref{lastpage}}
\maketitle

\begin{abstract}
Recent observations suggest that dense gas clouds can survive even in hot galactic winds. Here we show that the inclusion of turbulent densities with different statistical properties has significant effects on the evolution of wind-swept clouds. We investigate how the initial standard deviation of the log-normal density field influences the dynamics of quasi-isothermal clouds embedded in supersonic winds. We compare uniform, fractal solenoidal, and fractal compressive cloud models in both 3D and 2D hydrodynamical simulations. We find that the processes of cloud disruption and dense gas entrainment are functions of the initial density distribution in the cloud. Fractal clouds accelerate, mix, and are disrupted earlier than uniform clouds. Within the fractal cloud sample, compressive clouds retain high-density nuclei, so they are more confined, less accelerated, and have lower velocity dispersions than their solenoidal counterparts. Compressive clouds are also less prone to Kelvin-Helmholtz and Rayleigh-Taylor instabilities, so they survive longer than solenoidal clouds. By comparing the cloud properties at the destruction time, we find that dense gas entrainment is more effective in uniform clouds than in either of the fractal clouds, and it is more effective in solenoidal than in compressive models. In contrast, mass loading into the wind is more efficient in compressive cloud models than in uniform or solenoidal models. Overall, wide density distributions lead to inefficient entrainment, but they facilitate mass loading and favour the survival of very dense gas in hot galactic winds.
\end{abstract}


%%%%%%%%%%%%%%%%%%%%%%%%%%%%%%%%%%%%%%%%%%%%%%%%%%

%%%%%%%%%%%%%%%%% BODY OF PAPER %%%%%%%%%%%%%%%%%%

\section{Introduction}
\label{sec:Intro}
Multi-wavelength observations of star-forming galaxies reveal that galactic winds, driven by stellar feedback, are large-scale, multi-phase outflows comprised of several gas, dust, and cosmic-ray components (e.g., see \citealt*{1997A&A...320..378S,2002ApJ...576..745C,2005ApJ...621..227M}; \citealt{2015ApJ...804...46M,2017MNRAS.467.4951L,2018MNRAS.476.1756H,2018NatAs...2..901M}). Within the gas component, galactic winds have a hot ($\sim 10^7\,\rm K$), ionised phase that typically moves at speeds of $500-1500\,\rm km\,s^{-1}$, plus a cold ($\sim 10^2-10^4\,\rm K$), atomic/molecular phase that typically moves at speeds of $50-300\,\rm km\,s^{-1}$ (e.g., see \citealt{1998ApJ...493..129S,2005ApJS..160..115R,2009ApJ...697.2030S,2015ApJ...814...83L}). A current, open problem in the theory of galactic winds is understanding how dense gas in the cold phase survives in the hot outflow and how it reaches distances $\sim100-1500\,\rm pc$ above and below the galactic planes (e.g., see \citealt*{2005ARA&A..43..769V}; \citealt{2012ApJS..199...12M,2013ApJ...770L...4M,2016ApJ...826..215L}). Several theories have been proposed to explain the presence of cold, dense clouds and filaments in galactic winds. We mention two of them here: the first one relies on momentum-driven acceleration as the mechanism to transport clouds from low to high latitudes (e.g., see \citealt{2005ApJ...618..569M}), whilst the second one relies on thermal instabilities as the trigger for the in-situ formation of clouds at high latitudes (e.g., see \citealt{2018ApJ...862...56S}).\par

In the first scenario, clouds near the galactic plane are advected from low to high latitudes by either the thermal-gas ram pressure (e.g., see \citealt{2000MNRAS.314..511S,2008ApJ...674..157C,2009ApJ...703..330C}; \citealt*{2012MNRAS.421.3522H}), the radiation pressure (e.g., see \citealt*{2011ApJ...735...66M,2012MNRAS.424.1170Z}), or the cosmic-ray pressure (e.g., see \citealt*{1991A&A...245...79B}; \citealt{2008ApJ...674..258E}) of the outflowing material. A challenge of this scenario is explaining how dense clouds survive the disruptive effects of pressure gradients and dynamical instabilities to become entrained in the wind. In the second scenario, the hot gas in the outflowing wind cools down as it moves outwards and becomes thermally unstable in the process, thus triggering the (re)formation of dense clouds at high latitudes via clumping and warm gas condensation (\citealt{1995ApJ...444..590W,2016MNRAS.455.1830T,2017MNRAS.468.4801Z,2018MNRAS.480L.111G}). For this scenario to work, however, the wind needs to be sufficiently mass loaded at the launching site. Thus, the question remains open and the parameter space to explore in all models is still too broad to draw strong conclusions. The reader is referred to \cite{Heckman2016,2018Galax...6..114Z,2018Galax...6..138R} for recent reviews of galactic wind models and observations.\par

In this paper we concentrate on the first of the above-mentioned scenarios and study clouds that are being ram-pressure accelerated by supersonic winds. We use hydrodynamical, numerical simulations to study wind-cloud models in a previously-unexplored parameter space: one in which turbulent fractal density profiles are considered for the initial cloud setups. Wind-cloud and shock-cloud problems have been widely studied in recent years (see a full list of the parameters explored by previous authors in Appendix \ref{sec:AppendixA}). Still, in most previous models clouds have been idealised as spherical clumps of gas with either uniform or smoothed density profiles (e.g., see \citealt*{1994ApJ...420..213K}; \citealt{2006ApJS..164..477N,2016MNRAS.455.1309B,2016MNRAS.457.4470P}). In this idealised scenario, purely hydrodynamical (hereafter HD) models show that adiabatic clouds are disrupted by instabilities before they travel large distances, while radiative and thermally-conducting clouds survive longer but are not effectively accelerated (e.g., see \citealt{2015ApJ...805..158S,2016ApJ...822...31B}). These results suggest that entrainment is not effective in such scenarios, but other models that incorporate magnetised multi-cloud media (e.g., \citealt{2014MNRAS.444..971A}) and k-{$\epsilon$} turbulence models (e.g., \citealt{2009MNRAS.394.1351P}) show that shielding of clouds and strong turbulence in the flow itself can affect wind mass loading and dense-gas entrainment.\par

In this context, the parameter space in wind-cloud models with self-consistent magnetic fields and/or turbulent clouds is less explored. When included, however, both magnetic fields and turbulence have been shown to produce significant effects on the morphology, dynamics, and survival of wind-swept clouds. On the one hand, magnetohydrodynamical (hereafter MHD) models show that clouds threaded by either uniform, tangled, or turbulent magnetic fields are further clumped and more protected against shear instabilities than their uniform counterparts (e.g., see \citealt{2015MNRAS.449....2M,2018MNRAS.473.3454B}). The extra magnetic pressure in and around shearing layers reduces vorticity generation, but the clouds are not effectively accelerated unless the wind is also strongly magnetised (\citealt{2015MNRAS.449....2M}; \citealt*{2017ApJ...840...25A}). In such case, the effective drag force acting upon the cloud is enhanced, aiding cloud acceleration. On the other hand, cloud models with supersonic velocity fields and/or strong, turbulent magnetic fields also favour scenarios in which clouds undergo a period of fast acceleration (\citealt{2018MNRAS.473.3454B}). In these models the initial turbulent kinetic and magnetic energy thermalises, thus allowing the cloud to expand, accelerate, and reach high velocities over short time-scales, without being significantly disrupted by dynamical instabilities, e.g., Kelvin-Helmholtz (hereafter KH) and Rayleigh-Taylor (hereafter RT) instabilities.\par

The aforementioned results suggest that cloud entrainment could be effective if magnetic fields and turbulence were taken into account in wind-cloud models in a self-consistent manner (i.e., in models with turbulent density, velocity, and magnetic fields that are correlated and coupled by the MHD laws). Previous turbulent fractal cloud models by \cite{2009ApJ...703..330C,2017ApJ...834..144S} and ourselves (see \citealt{2018MNRAS.473.3454B}) contrasted uniform and turbulent fractal clouds in different flow regimes, but in all cases a single probability density function (hereafter PDF) for the cloud density field was assumed. \cite{2009ApJ...703..330C} assumed a Kolmogorov-like log-normal distribution for the cloud, motivated by studies on incompressible turbulence. \cite{2017ApJ...834..144S} included turbulent, solenoidal cloud models from \cite{2012ApJ...750L..31R}, and in \cite{2018MNRAS.473.3454B} we only probed clouds characterised by a single PDF taken from a simulation of isothermal turbulence with mixed forcing (by \citealt{2012ApJ...761..156F}). Nevertheless, changes in the standard deviation of the PDFs and in the power-law index of the spectra of densities are expected for different regimes of turbulence (see \citealt*{2008ApJ...688L..79F,2009ApJ...692..364F,2012ApJ...761..156F}).\par

Thus, in this paper we explore the effects of varying the initial density PDF of turbulent fractal clouds (hereafter fractal clouds) between two extreme regimes of turbulence, namely solenoidal (divergence-free) and compressive (curl-free). We do not include supersonic velocity fields or tangled magnetic fields in this paper as we are interested in isolating the effects of changing the initial statistical parameters of the density field upon the morphology, dynamics, and survival of clouds embedded in hot, supersonic winds.\par

The remainder of this paper is organised as follows: In Section \ref{sec:Method} we describe the HD conservation laws, the initial and boundary conditions, the diagnostics, and the reference time-scales we use for our simulations. In Section \ref{sec:Results} we compare uniform versus fractal cloud models and solenoidal versus compressive simulations, we analyse the cloud dynamics and the processes of gas mixing and dispersion that lead to mass loss and cloud destruction, and we discuss dense gas entrainment and wind mass loading. In Section \ref{sec:Future} we discuss the limitations of this work and the main motivations for future studies. In Section \ref{sec:Conclusions} we summarise our findings. In addition, we include three Appendices with extra details on the numerics of wind-cloud interactions.

\section{Method}
\label{sec:Method}

\subsection{Simulation code}
\label{subsec:SimulationCode}
In order to carry out the simulations presented in this paper, we use the {\sevensize PLUTO v4.0} code (see \citealt{2007ApJS..170..228M}) in 3D $(X_1,X_2,X_3)$ and 2D $(X_1,X_2)$ Cartesian coordinate systems. We solve the system of mass, momentum, and energy conservation laws of ideal hydrodynamics using the \verb#HLLC# approximate Riemann solver of \cite*{Toro:1994} jointly with a Courant-Friedrichs-Lewy (CFL) number of $C_{\rm a}=0.3$. The conservation laws read:

\begin{equation}
\frac{\partial \rho}{\partial t}+\bm{\nabla\cdot}\left[{\rho \bm{v}}\right]=0,
\label{eq:MassConservation}
\end{equation}
\begin{equation}
\frac{\partial \left[\rho \bm{v}\right]}{\partial t}+\bm{\nabla\cdot}\left[{\rho\bm{v}\bm{v}}+{\bm{I}}P\right]=0,
\label{eq:MomentumConservation}
\end{equation}

\begin{equation}
\frac{\partial E}{\partial t}+\bm{\nabla\cdot}\left[\left(E+P\right)\bm{v}\right]=0,
\label{eq:EnergyConservation}
\end{equation}

\begin{equation}
\frac{\partial\left[\rho C\right]}{\partial t}+\bm{\nabla\cdot}\left[{\rho C \bm{v}}\right]=0,
\label{eq:tracer}
\end{equation}

\noindent where $\rho$ is the mass density, $\bm{v}$ is the velocity, $P=\left(\gamma-1\right)\rho\epsilon$ is the gas thermal pressure, $E=\rho\epsilon+\frac{1}{2}\rho\bm{v^2}$ is the total energy density, $\epsilon$ is the specific internal energy, and $C$ is a Lagrangian scalar used to track the evolution of gas initially contained in the cloud.\par

The simulations presented in this paper have been designed as scale-free wind-cloud models. Thus, instead of explicitly including radiative cooling as a source term in the above numerical scheme, we approximate the effects of energy losses in the gas by using a soft adiabatic index of $\gamma=1.1$ for all the models (see also \citealt{1994ApJ...420..213K,2006ApJS..164..477N,2016MNRAS.455.1309B,2018MNRAS.473.3454B}). Our choice of adiabatic index correctly describes cold $\rm H\,I$ gas and molecular clouds (see \citealt{2000ApJ...531..350M,2005MNRAS.359..211L}), and it also allows us to achieve numerical convergence without having to resolve the cooling length (which is a problem-specific quantity).

\subsection{Initial and boundary conditions}
\label{subsec:Initial and Boundary Conditions}

\subsubsection{Dimensionless setup}
\label{subsec:Dimensionlessset-up}
Our simulation sample comprises 23 models in total: one model with a 3D uniform cloud and 22 models with turbulent fractal clouds, split into two sets of runs: a 3D set with two models and a 2D set with 20 models. In both sets, the simulation setup consists of a single, spherically- (in 3D) or cylindrically-outlined (in 2D), uniform or turbulent fractal cloud with radius $r_{\rm cloud}$ and average density $\bar{\rho}_{\rm cloud}$, embedded in a supersonic wind with density, $\rho_{\rm wind}$, and Mach number:

\begin{equation}
{\cal M_{\rm wind}}=\frac{|\bm{v_{\rm wind}}|}{c_{\rm wind}}=4.9,
\label{eq:MachNumber}
\end{equation}

\noindent where $|{\bm{v_{\rm wind}}}|\equiv v_{\rm wind}$ and $c_{\rm wind}=\sqrt{\gamma \frac{P}{\rho_{\rm wind}}}$ are the speed and sound speed of the wind, respectively. In all models the density contrast between cloud and wind material is:

\begin{equation}
\chi=\frac{\bar{\rho}_{\rm cloud}}{\rho_{\rm wind}}=10^3.
\label{eq:DensityContrast}
\end{equation}

\noindent Our choices of wind Mach number and cloud-to-wind density contrast in Equations (\ref{eq:MachNumber}) and (\ref{eq:DensityContrast}), respectively, reflect the physical conditions expected in the inner free-wind region of supernova-driven galactic winds (see \citealt{2008ApJ...674..157C,2009ApJ...703..330C,2017ApJ...834..144S}). In addition, they allow us to directly compare the results from this study with our previous wind-cloud models (see \citealt{2016MNRAS.455.1309B,2018MNRAS.473.3454B}), with similar recent studies on galactic winds (in particular, \citealt{2015ApJ...805..158S,2016ApJ...822...31B,2017ApJ...834..144S}), and with the extensive literature on wind/shock-cloud models (see Appendix \ref{sec:AppendixA}).\par

In addition, the initial density distribution in turbulent fractal clouds is described by a log-normal function:

\begin{equation}
{\cal P}(\rho_{\rm cloud})=\frac{1}{s\sqrt{2\pi}}{\rm e}^{-\frac{[\ln(\rho_{\rm cloud})-m]^2}{2s^2}},
\label{eq:PDF}
\end{equation}

\noindent where $\rho_{\rm cloud}$ is the cloud gas density, $m$ and $s$ are the mean and the standard deviation of the natural logarithm of the density field (see \citealt{2007ApJS..173...37S}). The mean and variance of the density field are $\bar{\rho}_{\rm cloud}={\rm e}^{(m+\frac{s^2}{2})}$ and $\sigma_{\rho_{\rm cloud}}^2=\bar{\rho}_{\rm cloud}^2({\rm e}^{s^2}-1)$, respectively. Thus, the normalised standard deviation of the initial log-normal PDF is:

\begin{equation}
\sigma_{\rm cloud}=\frac{\sigma_{\rho_{\rm cloud}}}{\bar{\rho}_{\rm cloud}}.
\label{eq:PDFsigma}
\end{equation}

\begin{figure}
\begin{center}
  \begin{tabular}{c c c}
    1a) 3Dsol  & 1b) 3Dcomp & \hspace{-5.5mm}$\frac{\rho C_{\rm cloud}}{\rho_{\rm wind}}$\\
  \hspace{-0cm}\resizebox{35mm}{!}{\includegraphics{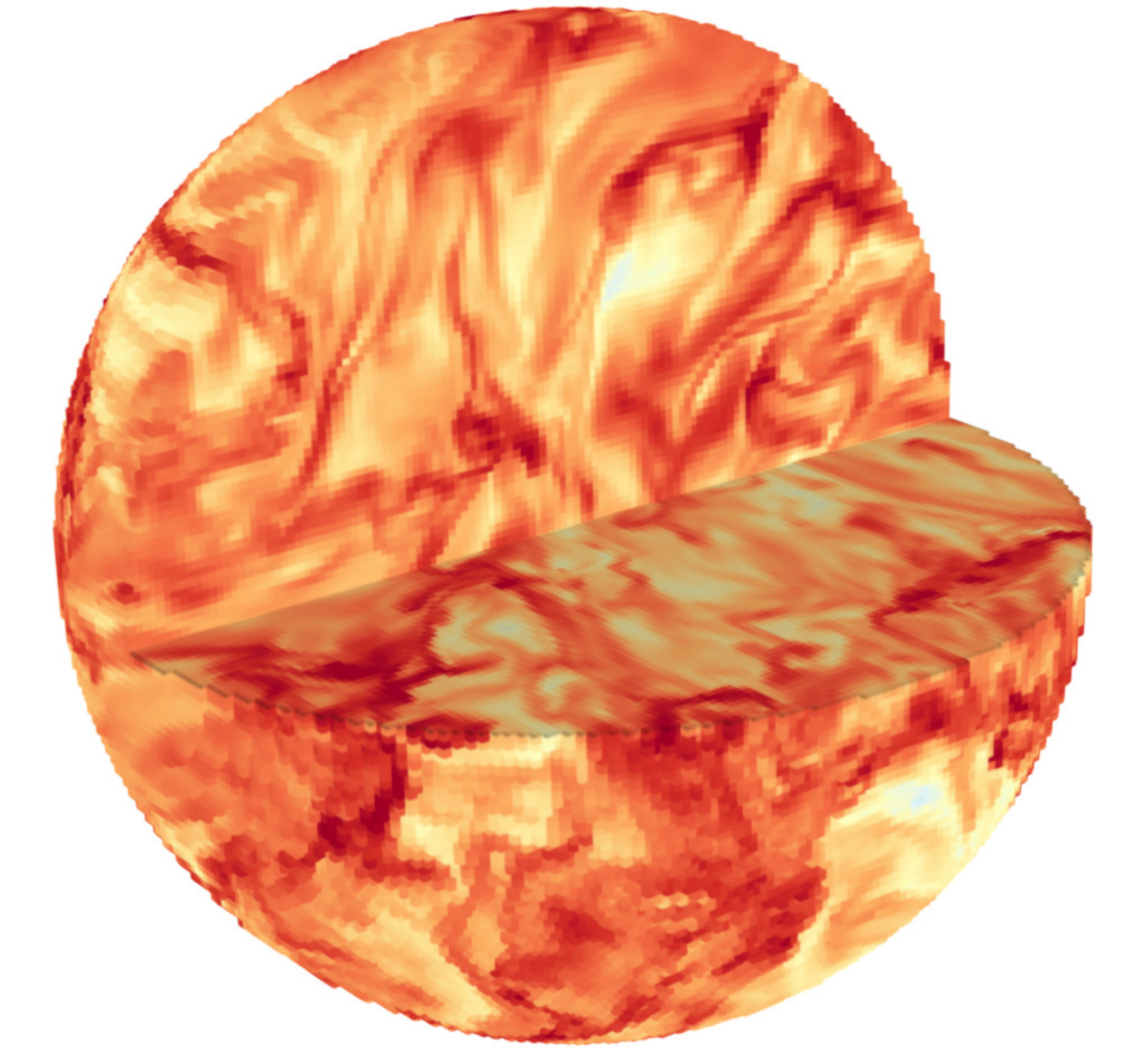}} & \hspace{-0.1cm}\resizebox{35mm}{!}{\includegraphics{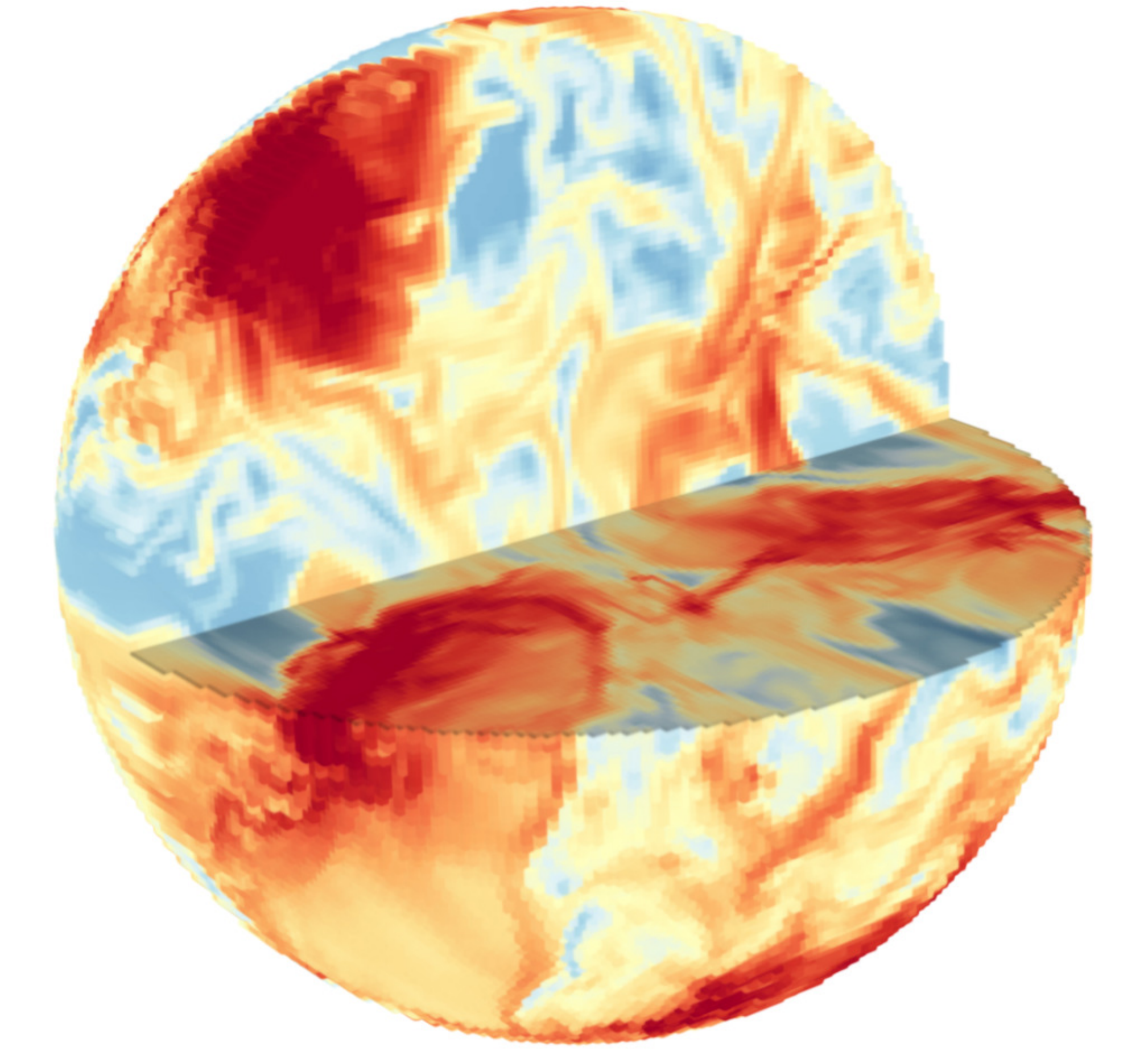}} \\
 1c) 2Dsol & 1d) 2Dcomp\\
   \hspace{-0.45cm}\resizebox{40mm}{!}{\includegraphics{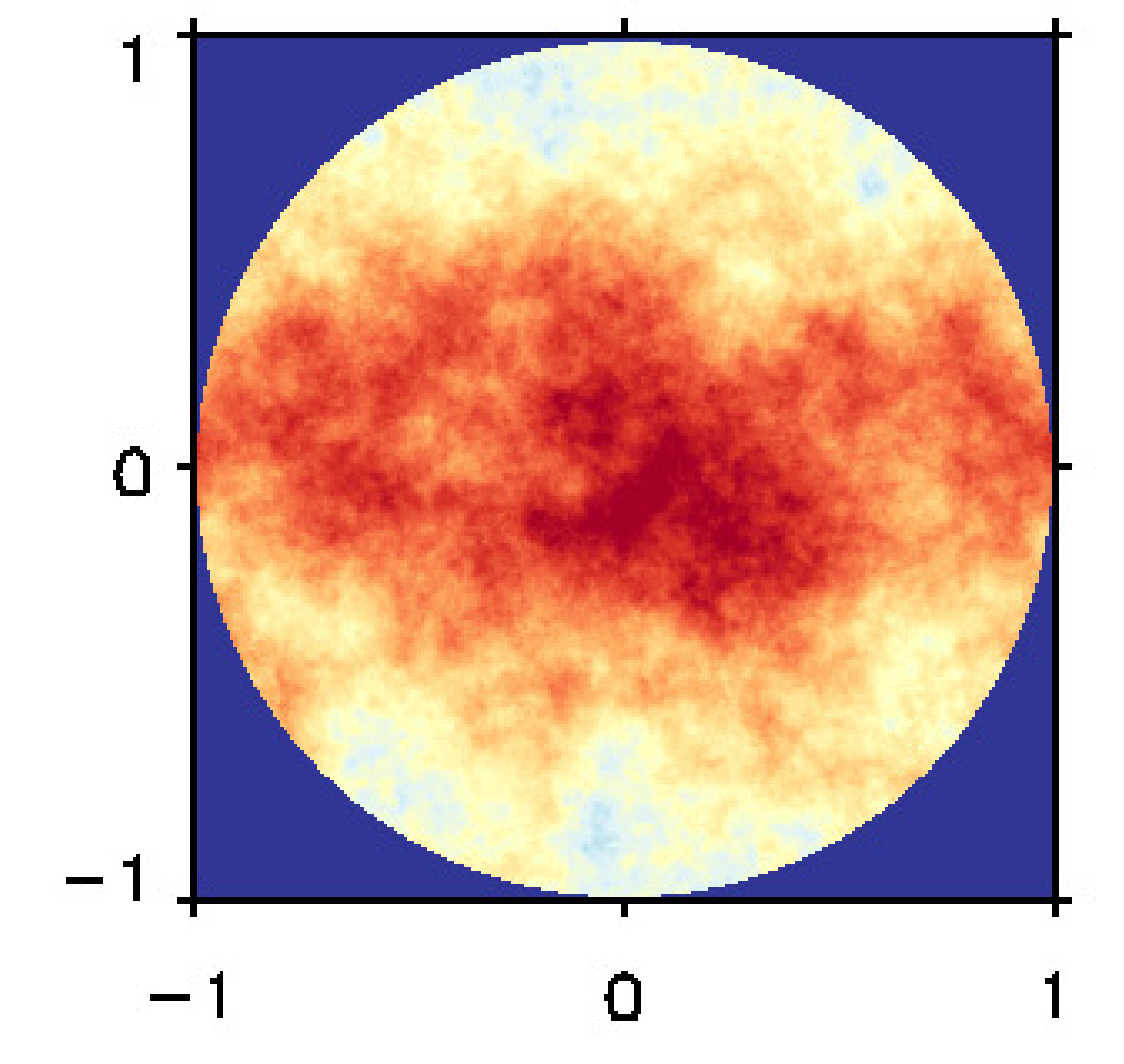}} & \hspace{-0.57cm}\resizebox{40mm}{!}{\includegraphics{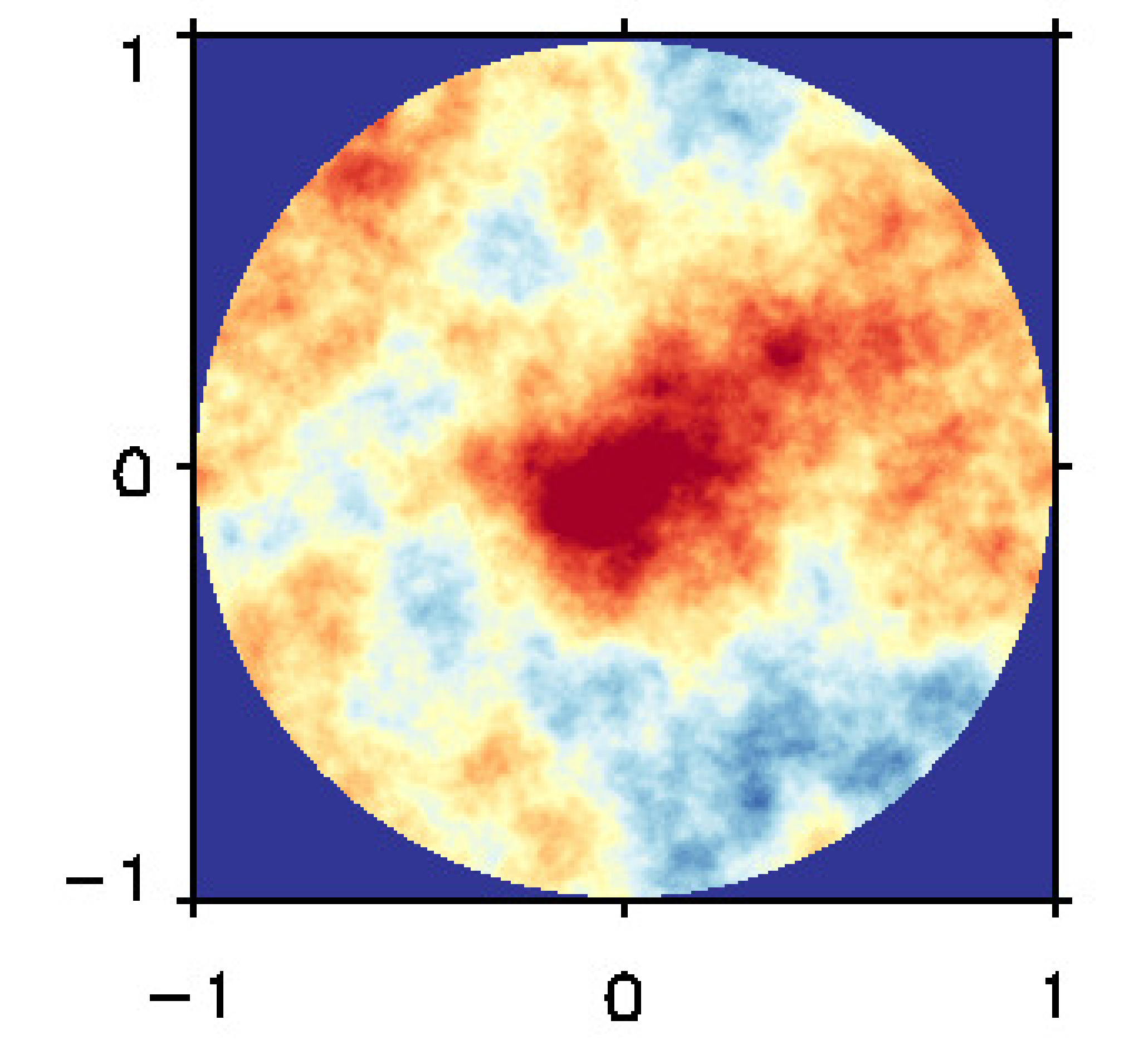}} & \multirow[t]{2}{*}{\hspace{-0.5cm}{\includegraphics[width=9.2mm]{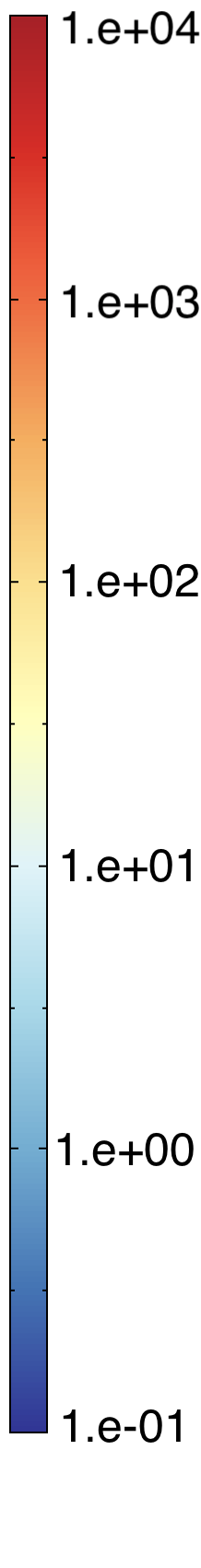}}}\\
   \multicolumn{3}{c}{\hspace{+2mm}1e) Initial density PDFs}\\
   \multicolumn{3}{c}{\resizebox{65mm}{!}{\includegraphics{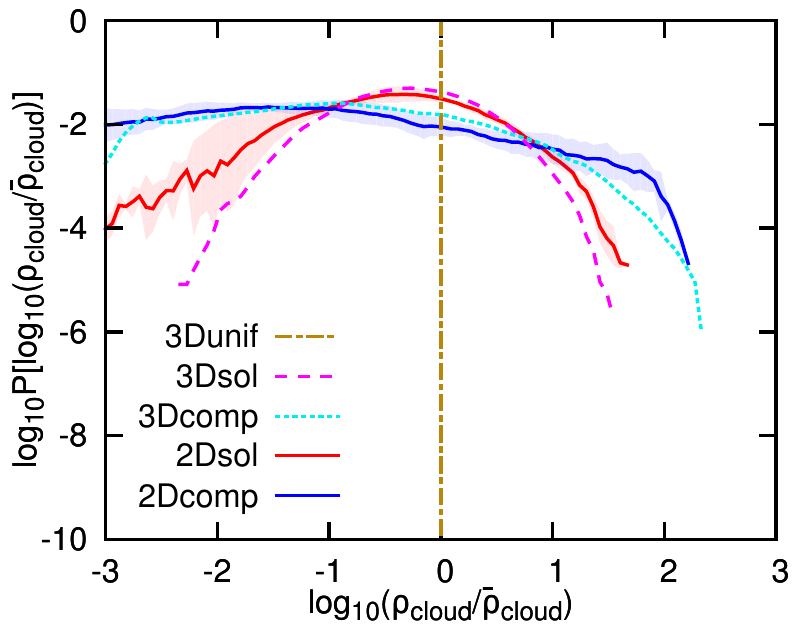}}}\vspace{-2mm}\\
  \end{tabular}
  \caption{Initial density structure of the 3D and 2D solenoidal (panels 1a and 1c) and compressive (panels 1b and 1d) fractal clouds presented in this paper. In the 3D models we clipped a quarter of the volume, so that the renderings show the internal structure of the clouds. Panel 1e shows the initial density distributions of uniform, solenoidal, and compressive cloud models, in both 3D and 2D.} 
  \label{Figure1}
\end{center}
\end{figure}

Note that the above set-up for the cloud density field implies that some regions inside the 3D and 2D fractal clouds are $\sim 10^5$ times denser than the wind (see Figure \ref{Figure1}).

\subsubsection{3D fractal cloud models}
\label{subsec:3Dmodels}
In the 3D set, we initialise the clouds with log-normal density fields taken from snapshots of simulations of solenoidal (divergence-free, $\sigma_{\rm cloud}=1.3$) and compressive (curl-free, $\sigma_{\rm cloud}=4.1$), supersonic turbulence (with an rms Mach number, ${\cal M}_{\rm turb}\sim4.5\pm0.5$) reported by \cite{2017MNRAS.466.2272H}. We set up two models, 3Dsol and 3Dcomp, with solenoidal and compressive clouds, respectively (see panels 1a and 1b of Figure \ref{Figure1}), by following a four-step process: 1) we mask regions in the clouds' domain outside a radius $r_{\rm cloud}$; 2) we scale the average density to $\bar{\rho}_{\rm cloud}$ in both clouds, 3) we interpolate the resulting density data cube into the 3D simulation domain described below, and 4) we initialise the simulations with the clouds in thermal pressure equilibrium ($P$) with the ambient medium. This process allows us to compare the evolution of both models and ensure that all simulations start with clouds of the same initial mass and average density.

\subsubsection{3D domain and resolution}
\label{subsec:3Ddomainandresolution}
The 3D fractal clouds are centred on the origin $(0,0,0)$ of the computational domain, which consists of a prism with a spatial range $-5\,r_{\rm cloud}\leq X_1\leq5\,r_{\rm cloud}$, $-2\,r_{\rm cloud}\leq X_2\leq28\,r_{\rm cloud}$, and $-5\,r_{\rm cloud}\leq X_3\leq5\,r_{\rm cloud}$. The numerical resolution is $R_{\rm 64}$ (i.e., $64$ cells cover the cloud radius), which corresponds to a uniform grid resolution of $(N_{\rm X_{1}}\times N_{\rm X_{2}}\times N_{\rm X_{3}})=(640\times1920\times640)$. This resolution is adequate to describe the overall evolution of 3D turbulent cloud models as shown in \cite{2018MNRAS.473.3454B} for a similar configuration. Note also that uniform-grid simulations, albeit more expensive than moving-mesh simulations, have the advantage of capturing the high-density gas in the cloud, the wind-cloud interface (where instabilities grow), and the low-density mixed gas at identical resolution in all models.\par

\subsubsection{2D fractal cloud models}
\label{subsec:2Dmodels}
The above-mentioned 3D models are computationally expensive, so we can only investigate wind-cloud models in small simulation domains at the resolution required for convergence. This means that we can only follow the evolution of 3D wind-swept clouds for $2-3$ dynamical time-scales before a significant amount of cloud mass leaves the computational domain. Therefore, in order to follow the evolution of solenoidal and compressive cloud models for longer time-scales and larger spatial scales, we also investigate 2D wind-cloud models in both regimes of turbulence (see panels 1c and 1d of Figure \ref{Figure1}).\par

In the 2D set, we also initialise the clouds with density profiles described by log-normal distributions, but in this set we generate these scalar fields using the pyFC library\footnote{Available at: \url{https://bitbucket.org/pandante/pyfc}} instead of taking snapshots from turbulence simulations. The pyFC library uses a Fourier method developed by \cite{Lewis:2002ug} to generate random log-normal density fields with user-defined power-law spectra. For this set, we configure two sets of models, 2Dsol and 2Dcomp, with 10 solenoidal (divergence-free, $\sigma_{\rm cloud}=1.9$) and 10 compressive clouds (curl-free, $\sigma_{\rm cloud}=5.9$), respectively. We follow a four-step process to set up these clouds: 1) we use the pyFC library to iteratively produce $512^2$-sized data cubes containing clouds with log-normal density distributions with standard deviations and fractal dimensions characteristic of solenoidal and compressive clouds; 2) we mask regions in the fractal cloud domain outside a radius $r_{\rm cloud}$; 3) we scale the average density of each cloud to $\bar{\rho}_{\rm cloud}$ so that all models start with the same mass and initial mean density; and 4) we interpolate the clouds into the simulation domain described below and initialise the simulations with the wind and cloud in thermal pressure equilibrium.\par

Note that $7$ out of the $10$ clouds in each 2D set are generated with different seeds while the remaining three have the same seeds as others in the sample, but are rotated by $90^{\circ}$ (counterclockwise) with respect to the $X_2$ axis. The chosen standard deviations for the density PDFs of the 2D models correspond to supersonic (${\cal M}_{\rm turb}\sim5.5\pm0.6$) turbulence (see \citealt{2008ApJ...688L..79F,2010A&A...512A..81F,2011PhRvL.107k4504F}).\par

\subsubsection{2D domain and resolution}
\label{subsec:2Ddomainandresolution}
In the 2D set, the clouds are also centred on the origin $(0,0)$ of the computational domain, which consists of a rectangular area with a spatial range $-40\,r_{\rm cloud}\leq X_1\leq40\,r_{\rm cloud}$, $-2\,r_{\rm cloud}\leq X_2\leq158\,r_{\rm cloud}$. The numerical resolution is $R_{\rm 128}$ ($128$ cells cover the cloud radius), which corresponds to a uniform grid resolution of $(N_{\rm X_{1}}\times N_{\rm X_{2}})=(10240\times20480)$. This resolution is adequate to describe the overall evolution of 2D fractal cloud models as shown in Appendix \ref{sec:AppendixB}. Table \ref{Table1} presents a summary of the models and initial conditions described above.\par

\subsubsection{Boundary conditions}
\label{subsec:Boundaries}
In both sets (3D and 2D), we prescribe diode boundary conditions on the lateral and back sides of the simulation domains and an inflow boundary condition on the front side. The inflow zone is located at the ghost zone that faces the leading edge of the cloud and injects a constant supply of wind gas into the computational domain.\par

\begin{table*}\centering
\caption{Initial conditions for the 3D and 2D models. Column 1 provides the name of the model set and column 2 indicates the number of runs in each set. Columns 3, 4, and 5 show the normalised domain size, the number of grid cells along each axis, and the effective numerical resolution in terms of number of cells per cloud radius, respectively. Columns 6 and 7 show the polytropic index of the gas and wind Mach number, respectively. Column 8 shows the type of cloud density field, and columns 9 and 10 list the cloud-to-wind density contrast and the normalised standard deviation of the initial cloud densities, $\sigma_{\rm cloud}=\sigma_{\rho_{\rm cloud}}/\bar{\rho}_{\rm cloud}$, respectively.}
\begin{adjustbox}{max width=\textwidth}
\begin{tabular}{c c c c c c c c c c}
\hline
\textbf{(1)} & \textbf{(2)} & \textbf{(3)} & \textbf{(4)} & \textbf{(5)} & \textbf{(6)} & \textbf{(7)} & \textbf{(8)} & \textbf{(9)} & \textbf{(10)}\Tstrut\\
\textbf{Model} & \textbf{Runs}  & \textbf{Domain size} & \textbf{Grid cells} &\textbf{Resolution} & $\gamma$ & $\cal M_{\rm wind}$ & \textbf{Density Field} & $\chi$ & $\sigma_{\rm cloud}$\Tstrut\Bstrut \\ \hline
3Dunif & 1 & $(10\times30\times10)\,r_{\rm cloud}$ & $(640\times1920\times640)$ & $R_{64}$ &  $1.1$ & 4.9 & Uniform & $10^3$ & $0$ \Tstrut \\
3Dsol & 1 & $(10\times30\times10)\,r_{\rm cloud}$ & $(640\times1920\times640)$ & $R_{64}$ &  $1.1$ & 4.9 &  Fractal (turbulence) & $10^3$ & $1.3$\\
3Dcomp & 1 & $(10\times30\times10)\,r_{\rm cloud}$ & $(640\times1920\times640)$ & $R_{64}$ &  $1.1$ & 4.9 & Fractal (turbulence) & $10^3$ & $4.1$\\
2Dsol & 10 & $(80\times160)\,r_{\rm cloud}$ & $(10240\times20480)$ & $R_{128}$ &  $1.1$ & 4.9 & Fractal (generator) & $10^3$ & $1.9$\Tstrut \\
2Dcomp & 10 & $(80\times160)\,r_{\rm cloud}$ & $(10240\times20480)$ & $R_{128}$ &  $1.1$ & 4.9 & Fractal (generator) & $10^3$ & $5.9$\Bstrut\\\hline
\end{tabular}
\end{adjustbox}
\label{Table1}
\end{table*} 

\subsection{Diagnostics}
\label{subsec:Diagnostics}
To compare the results from different simulations we use mass-weighted quantities:

\begin{equation}
\langle~{\cal G}~\rangle=\frac{\int {\cal G}\rho\, C\, dV}{M_{\rm cloud}}=\frac{\int {\cal G}\rho\, C\, dV}{\int \rho\, C\, dV},
\label{eq:IntegratedG}
\end{equation}

\noindent where $G$ is any scalar from the simulation, $V$ is the volume, $C$ is the cloud tracer defined in Section \ref{subsec:SimulationCode}, and $M_{\rm cloud}$ is the cloud mass.

\subsubsection{Cloud dynamics}
\label{subsec:Morphological}
Using Equation (\ref{eq:IntegratedG}), we define the displacement of the centre-of-mass and the bulk speed (along $X_2$) of cloud gas above predefined density thresholds as $\langle~X_{{\rm 2},{\rm cloud_{threshold}}}~\rangle$ and $\langle~v_{2,{\rm cloud_{threshold}}}~\rangle$, respectively. The total cloud displacement and speed are $\langle~X_{{\rm 2},{\rm cloud}}~\rangle$ and $\langle~v_{2,{\rm cloud}}~\rangle$. Henceforth, the former of these quantities is normalised with respect to the initial cloud radius, $r_{\rm cloud}$; while the latter is normalised with respect to the wind speed, $v_{\rm wind}$.\par

In addition, we calculate the effective cloud acceleration, ${\rm a}_{\rm eff}$, by computing the time-derivatives of the above cloud bulk speeds as follows:
 
\begin{equation}
{\rm a}_{\rm eff}(t)=\frac{d}{dt}v(t)=\sum_{i=1}^{\infty} b_{i}\,\frac{d}{dt}g_{i}(t)
\label{eq:Acceleration}
\end{equation}

\noindent where $v(t)\equiv\langle~v_{2,{\rm cloud}}(t)~\rangle=\sum_{i=1}^{\infty} b_{i}\,g_{i}(t)$, $g_{i}$ is a set of basis functions called B-splines (see \citealt*{DBLP:books/lib/HastieTF09}), and $b_{i}$ are the decomposition coefficients, which are estimated from the set of simulated data $\left[t_i,v(t_i)\right]$ by solving a linear regression problem\footnote{For further details, see: \url{https://github.com/notblank/fda/blob/master/fractal\%20clouds.ipynb}}. The accelerations are normalised with respect to the drag acceleration, $a_{\rm drag}=v_{\rm wind}/t_{\rm drag}$, where $t_{\rm drag}$ is the drag time-scale (see \citealt{1994ApJ...420..213K}).

\subsubsection{Gas dispersion}
\label{subsec:Dispersion}
We define the dispersion of the $\rm j$-component (i.e., along each axis, $\rm j=1,2,3$) of the velocity, $\delta_{{\rm v}_{{\rm j},{\rm cloud}}}$, as:

\begin{equation}
\delta_{{\rm v}_{{\rm j},{\rm cloud}}}=\left(\langle~v^2_{{\rm j},{\rm cloud}}~\rangle-\langle~v_{{\rm j},{\rm cloud}}~\rangle^2\right)^{\frac{1}{2}},
\label{eq:rmsVelocityComponent}
\end{equation}

\noindent where $\langle~v_{{\rm j},{\rm cloud}}~\rangle$ and $\langle~v^2_{{\rm j},{\rm cloud}}~\rangle^{1/2}$ are the average cloud velocity and its rms, respectively\footnote{Note that in 2D simulations we only have two axes, so $\rm j=1,2$.}. Based on these quantities, we define the transverse velocity dispersion as $\delta_{{\rm v}_{\rm cloud}}\equiv|\bm{\delta_{{\rm v}_{\rm cloud}}}|=(\sum_{\rm j}\delta_{{\rm v}_{{\rm j},{\rm cloud}}}^2)^{1/2}$ for $\rm j=1,3$, which we also normalise with respect to the wind speed, $v_{\rm wind}$.

\subsubsection{Mixing and mass-loss}
\label{subsec:Mixing}
In order to understand cloud disruption and how gas with different densities evolves, we measure the fraction of gas mixing occurring between cloud and wind material using the following definition:

\begin{equation}
f_{{\rm mix}_{\rm cloud}}=\frac{\int \rho\,C_{\rm mix}\,dV}{M_{{\rm cloud},0}},
\label{eq:MixingFraction}
\end{equation}

\noindent where the numerator represents the mass of cloud gas mixed with the wind, $0.1\leq C_{\rm mix}\leq 0.9$ is the tracer tracking mixed gas, and $M_{{\rm cloud},0}$ is the initial mass of the cloud.\par

In addition, we define cloud mass fractions at or above a density threshold, $\rho_{\rm threshold}$, as:

\begin{equation}
F_{1/{\rm threshold}}=\frac{M_{1/{\rm threshold}}}{M_{{\rm cloud},0}}=\frac{\int \left[\rho\,C\right]_{\,\rho\geq\rho_{\rm threshold}}\,dV}{M_{{\rm cloud},0}},
\label{eq:MassFraction}
\end{equation}

\noindent where $M_{1/{\rm threshold}}$ is the total mass of cloud gas with densities at or above a density threshold. Using Equation (\ref{eq:MassFraction}) we define $F_{1/500}$,  $F_{1/100}$, $F_{1/3}$, and $F_{1}$, as the fractions of cloud mass with densities at or above $\bar{\rho}_{\rm cloud}/500$, $\bar{\rho}_{\rm cloud}/100$, $\bar{\rho}_{\rm cloud}/3$, and $\bar{\rho}_{\rm cloud}$, respectively.

\subsection{Reference time-scales}
\label{subsec:DynamicalTime-Scales}
The dynamical time-scales relevant for the simulations presented here are:

a) The cloud-crushing time (see \citealt{1994ApJ...432..194J,1996ApJ...473..365J}),

\begin{equation}
t_{\rm cc}=\frac{2r_{\rm cloud}}{v_{\rm shock}}=\left(\frac{\bar{\rho}_{\rm cloud}}{\rho_{\rm wind}}\right)^{\frac{1}{2}}\frac{2r_{\rm cloud}}{{\cal M_{\rm wind}} c_{\rm wind}}=\chi^{\frac{1}{2}}\frac{2r_{\rm cloud}}{{\cal M_{\rm wind}} c_{\rm wind}},
\label{eq:CloudCrushing}
\end{equation}

\noindent where $v_{\rm shock}={\cal M_{\rm wind}}\,c_{\rm wind}\,\chi^{-\frac{1}{2}}$ is the approximate speed of the shock refracted into the cloud after the initial wind-cloud collision. Note, however, that the actual speed of the transmitted shock inside fractal clouds varies strongly with position, owing to their intrinsic density variations. Henceforth, we use the cloud-crushing time to normalise all time-scales\footnote{Note that in some previous studies authors used the cloud radius rather than its diameter to definite the cloud-crushing time. The cloud diameter is the appropriate quantity for wind-cloud models as the refracted shock is predominantly transmitted into the cloud from its front surface, while the cloud radius is the appropriate quantity for shock-cloud models as the transmitted shock moves into the cloud from all sides and converges approximately at the centre. Thus, if the reader wishes to compare our time-scales to such studies, the definition needs to be contrasted with ours and, if needed, all the times reported in this paper should be multiplied by a factor of $2$.}.

b) The KH instability growth time (see \citealt{1961hhs..book.....C}):

\begin{equation}
t_{\rm KH}\simeq \frac{\chi^{1/2}_{\rm eff}}{k_{\rm KH}\,(v'_{\rm wind}-v'_{\rm cloud})},
\label{KHtime}
\end{equation}

\noindent where $\chi_{\rm eff}$ is the effective density contrast between cloud gas denser than $\bar{\rho}_{\rm cloud}/3$ and the wind, $k_{\rm KH}=\frac{2\pi}{\lambda_{\rm KH}}$ is the wavenumber of the KH perturbations, and the primed velocities correspond to their values at the location of shear layers.

c) The RT instability growth time (see \citealt{1961hhs..book.....C}):

\begin{equation}
t_{\rm RT}\simeq\frac{1}{\left[k_{\rm RT}\,(a_{\rm eff})\right]^{1/2}},
\label{RTtime}
\end{equation}

\noindent where $k_{\rm RT}=\frac{2\pi}{\lambda_{\rm RT}}$ is the wavenumber of the RT perturbations, and $a_{\rm eff}$ is the effective cloud acceleration (see Equation \ref{eq:Acceleration}). Note that both Equations (\ref{KHtime}) and (\ref{RTtime}) were originally derived for incompressible fluids, so any KH and RT time-scales mentioned hereafter should be considered solely as indicative values for the wind-cloud models presented in this paper. Note also that, in general, $t_{\rm KH}\lesssim t_{\rm cc}$ and $t_{\rm RT}\lesssim t_{\rm cc}$ for $\lambda_{\rm KH}\lesssim r_{\rm cloud}$ and $\lambda_{\rm RT}\lesssim r_{\rm cloud}$, respectively, so both instabilities are dynamically important in these models.\par

d) In 3D, the simulation time is $t_{\rm sim}/t_{\rm cc}=3.3$, $2.5$, and $2.9$ in the uniform, solenoidal, and compressive models, respectively. In 2D, the simulation time is $t_{\rm sim}/t_{\rm cc}=8.0$ in all cases. In the diagnostic plots presented in Section \ref{sec:Results}, we only display the curves up to the time when we can ensure that at least $75$ per cent of the original cloud material is still in the computational domain.

\section{Results}
\label{sec:Results}

\subsection{Uniform vs. turbulent fractal clouds}
\label{subsec:Evolution}
The disruption process of quasi-isothermal clouds immersed in supersonic winds occurs in four stages in both uniform and fractal cloud models, but the resulting morphology, dynamics, and destruction time-scales of clouds differ depending on their initial density distributions.\par

Figure \ref{Figure2} shows 2D slices at $X_3=0$ of the cloud density, $\rho C_{\rm cloud}$, normalised with respect to the wind density, $\rho_{\rm wind}$, in three 3D models, at six different times in the range $0\leq t/t_{\rm cc}\leq2.5$. Panel 2a of this figure shows the evolution of the uniform cloud model (3Dunif), panel 2b shows the solenoidal cloud model (3Dsol), and panel 2c shows the evolution of the compressive cloud model (3Dcomp). Below, we highlight the main qualitative similarities and differences in the evolution of uniform and fractal cloud models:

\begin{figure*}
\begin{center}
  \begin{tabular}{c c c c c c c}
      \multicolumn{1}{l}{\hspace{-2mm}2a) 3Dunif \hspace{+2mm}$t/t_{\rm cc}=0$} & \multicolumn{1}{c}{\hspace{+7mm}$t/t_{\rm cc}=0.5$} & \multicolumn{1}{c}{\hspace{+7mm}$t/t_{\rm cc}=1.0$} & \multicolumn{1}{c}{\hspace{+7mm}$t/t_{\rm cc}=1.5$} & \multicolumn{1}{c}{\hspace{+7mm}$t/t_{\rm cc}=2.0$} & \multicolumn{1}{c}{\hspace{+7mm}$t/t_{\rm cc}=2.5$} & $\frac{\rho C_{\rm cloud}}{\rho_{\rm wind}}$\\    
       \hspace{-0.37cm}\resizebox{28mm}{!}{\includegraphics{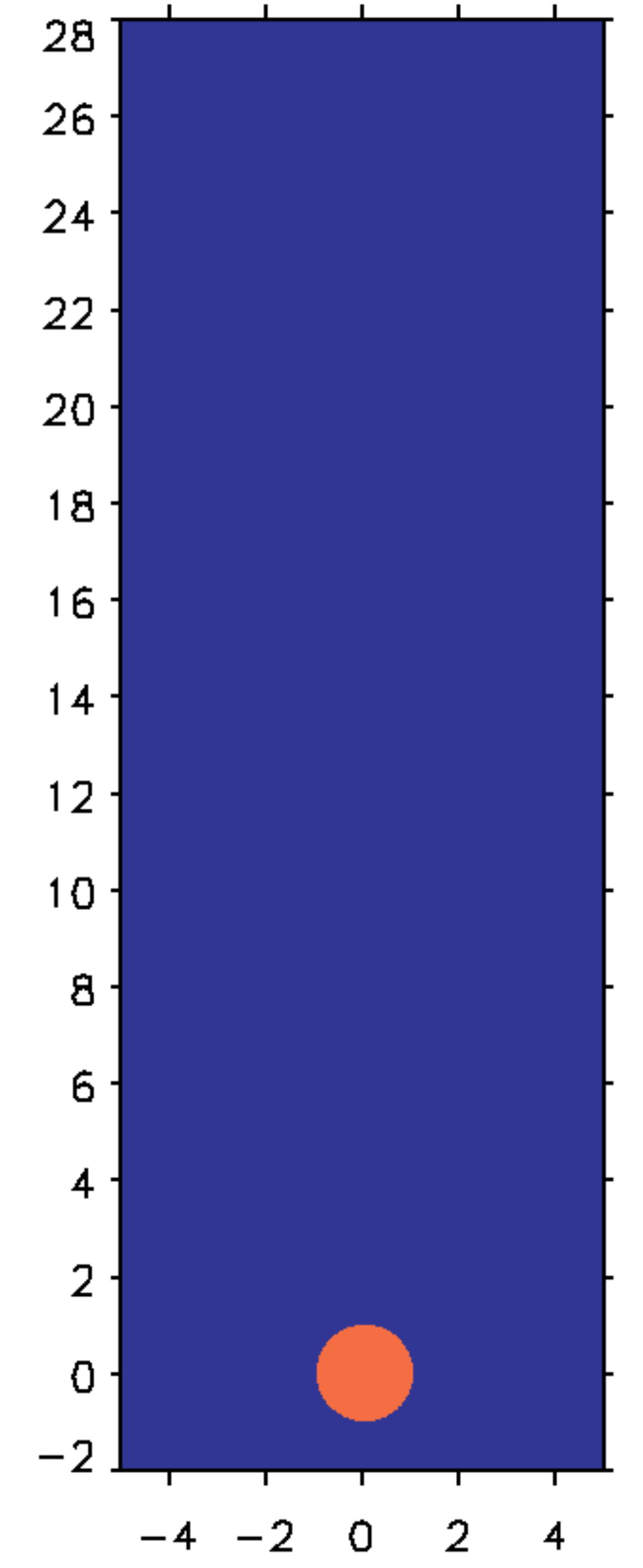}} & \hspace{-0.48cm}\resizebox{28mm}{!}{\includegraphics{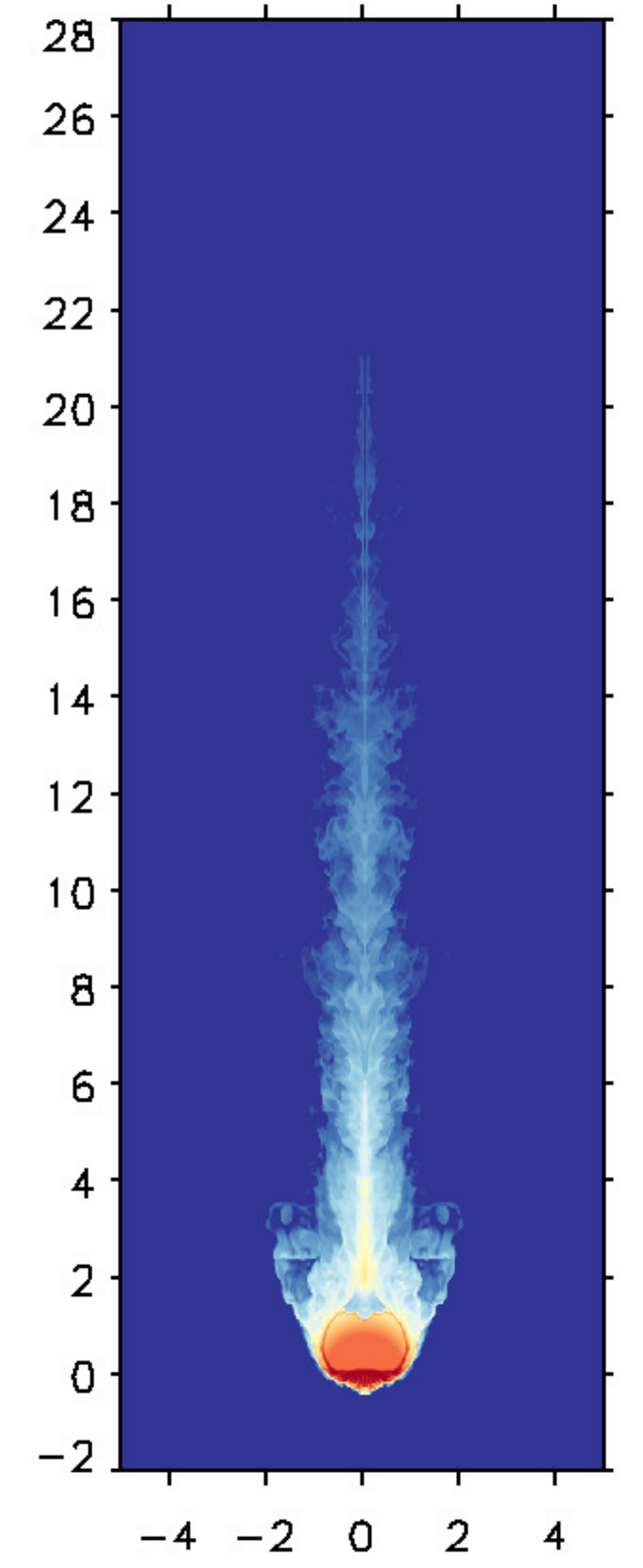}} & \hspace{-0.48cm}\resizebox{28mm}{!}{\includegraphics{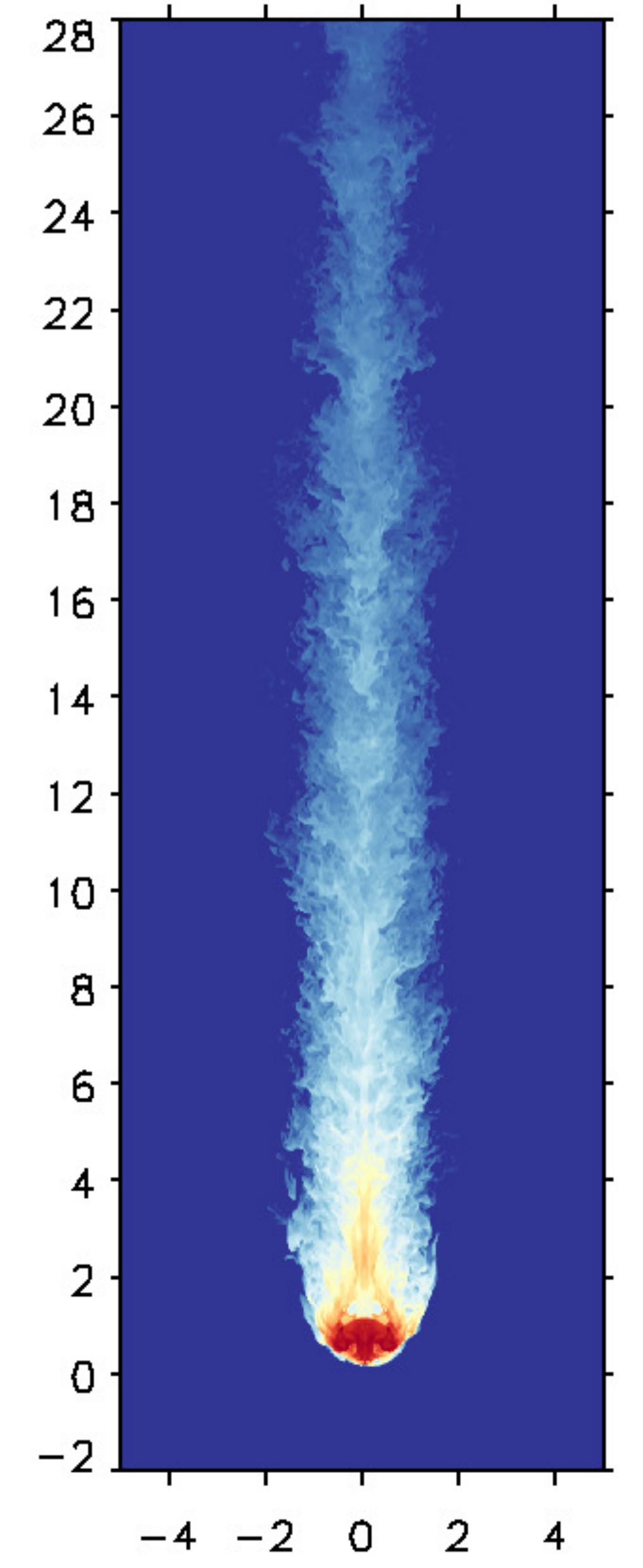}} & \hspace{-0.48cm}\resizebox{28mm}{!}{\includegraphics{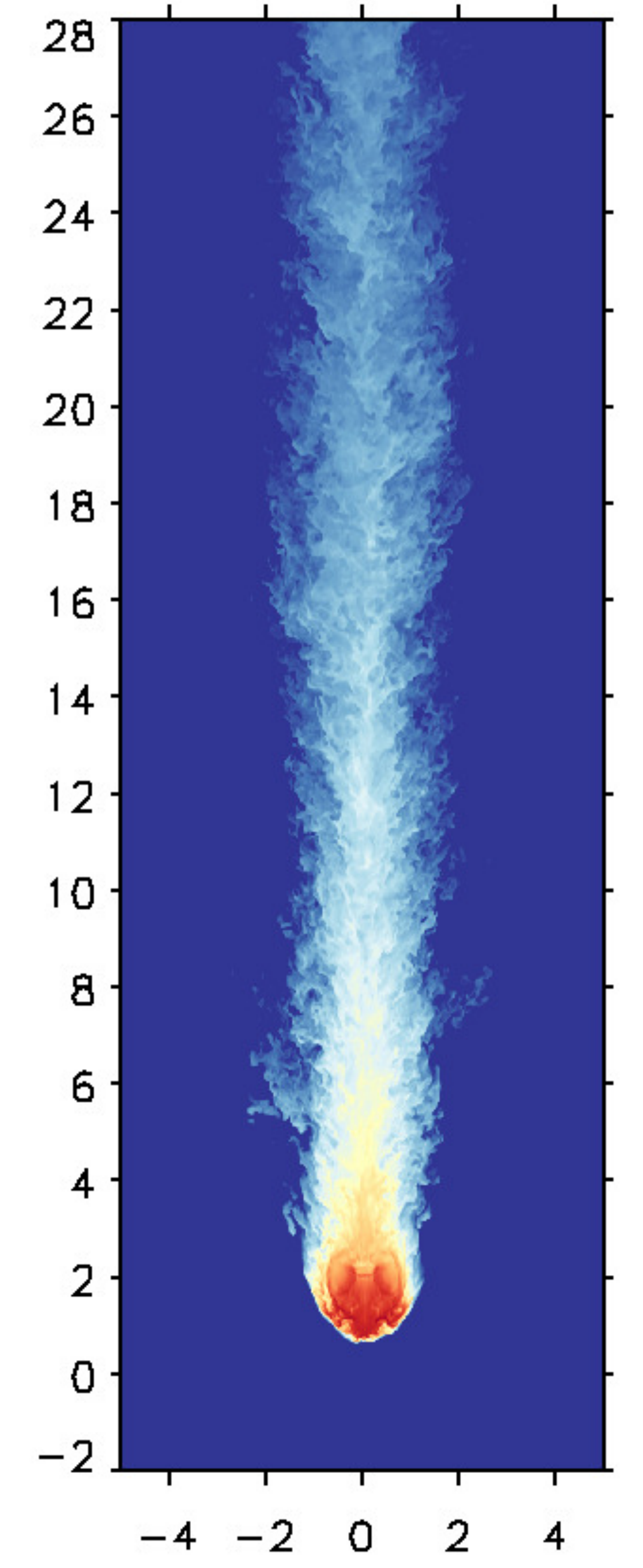}} & \hspace{-0.48cm}\resizebox{28mm}{!}{\includegraphics{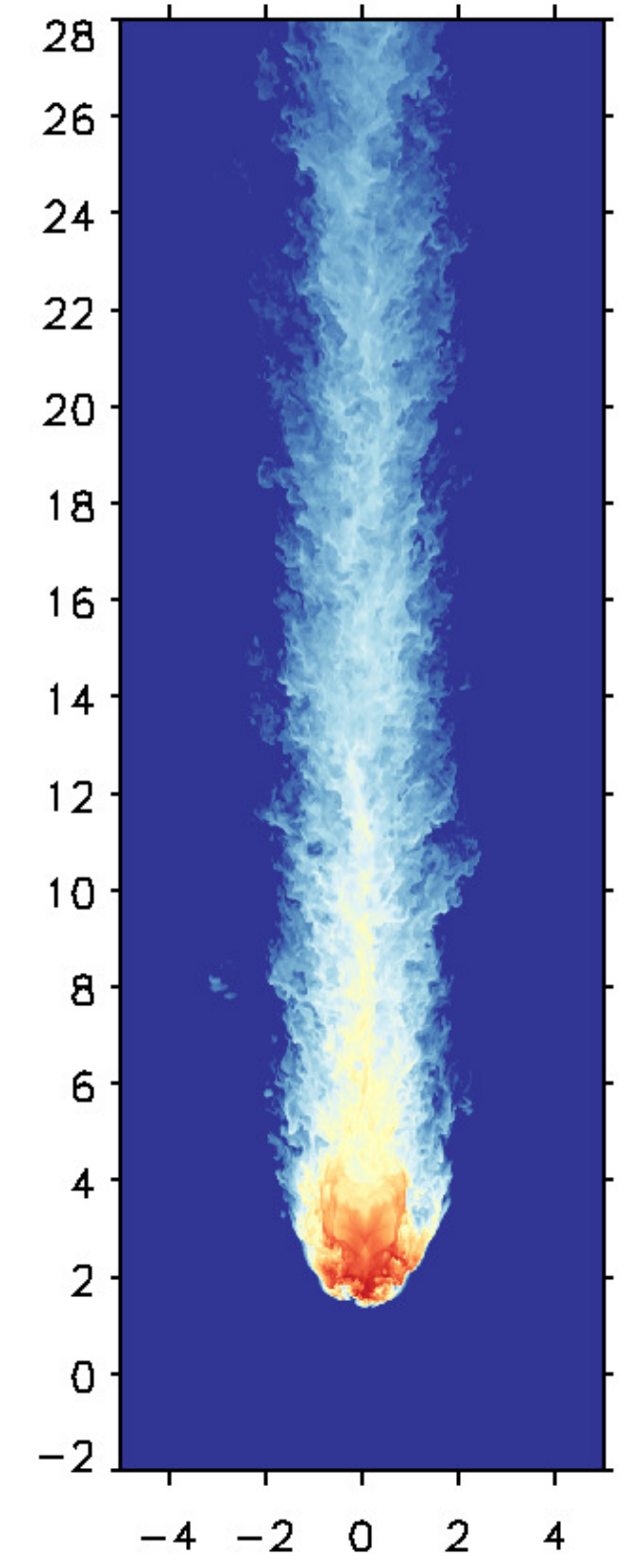}} & \hspace{-0.48cm}\resizebox{28mm}{!}{\includegraphics{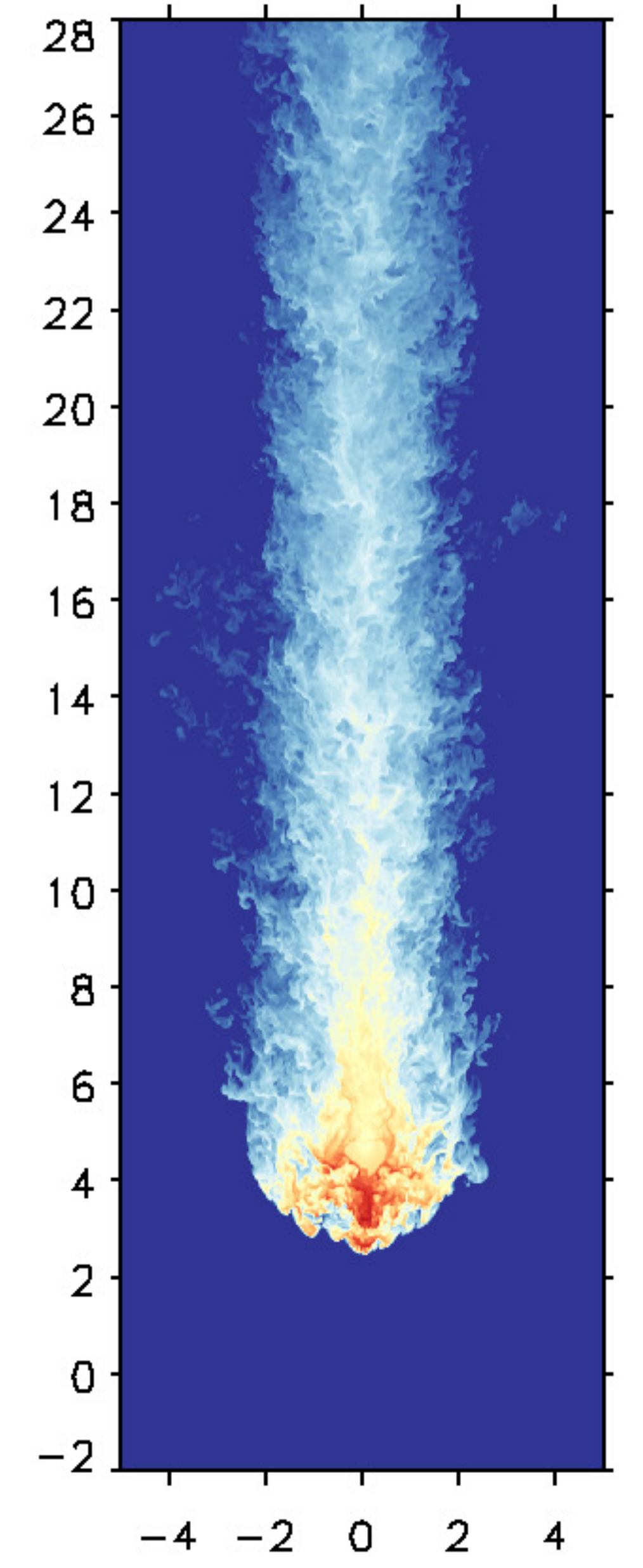}} & \hspace{-0.2cm}{\includegraphics[width=8.5mm]{pbarc.png}}\\
       \multicolumn{1}{l}{\hspace{-2mm}2b) 3Dsol \hspace{+3.5mm}$t/t_{\rm cc}=0$} & \multicolumn{1}{c}{\hspace{+7mm}$t/t_{\rm cc}=0.5$} & \multicolumn{1}{c}{\hspace{+7mm}$t/t_{\rm cc}=1.0$} & \multicolumn{1}{c}{\hspace{+7mm}$t/t_{\rm cc}=1.5$} & \multicolumn{1}{c}{\hspace{+7mm}$t/t_{\rm cc}=2.0$} & \multicolumn{1}{c}{\hspace{+7mm}$t/t_{\rm cc}=2.5$} & $\frac{\rho C_{\rm cloud}}{\rho_{\rm wind}}$\\    
       \hspace{-0.37cm}\resizebox{28mm}{!}{\includegraphics{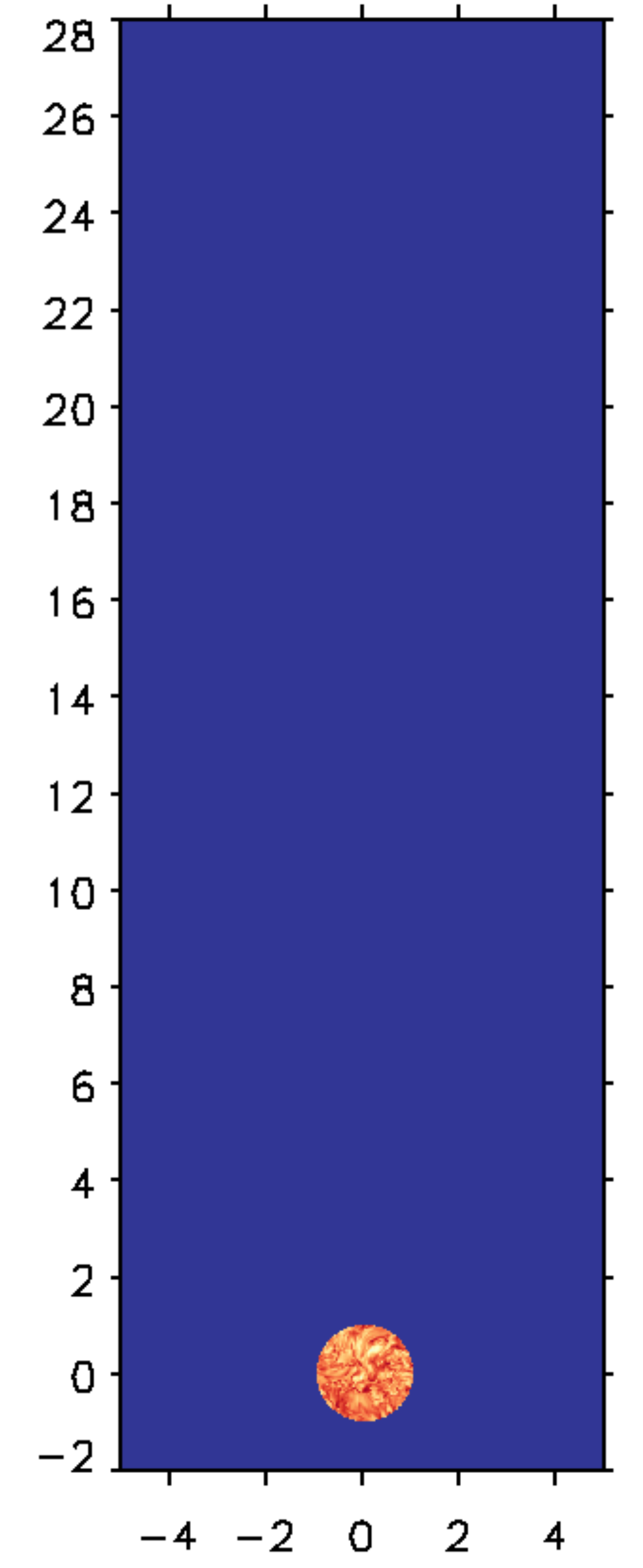}} & \hspace{-0.48cm}\resizebox{28mm}{!}{\includegraphics{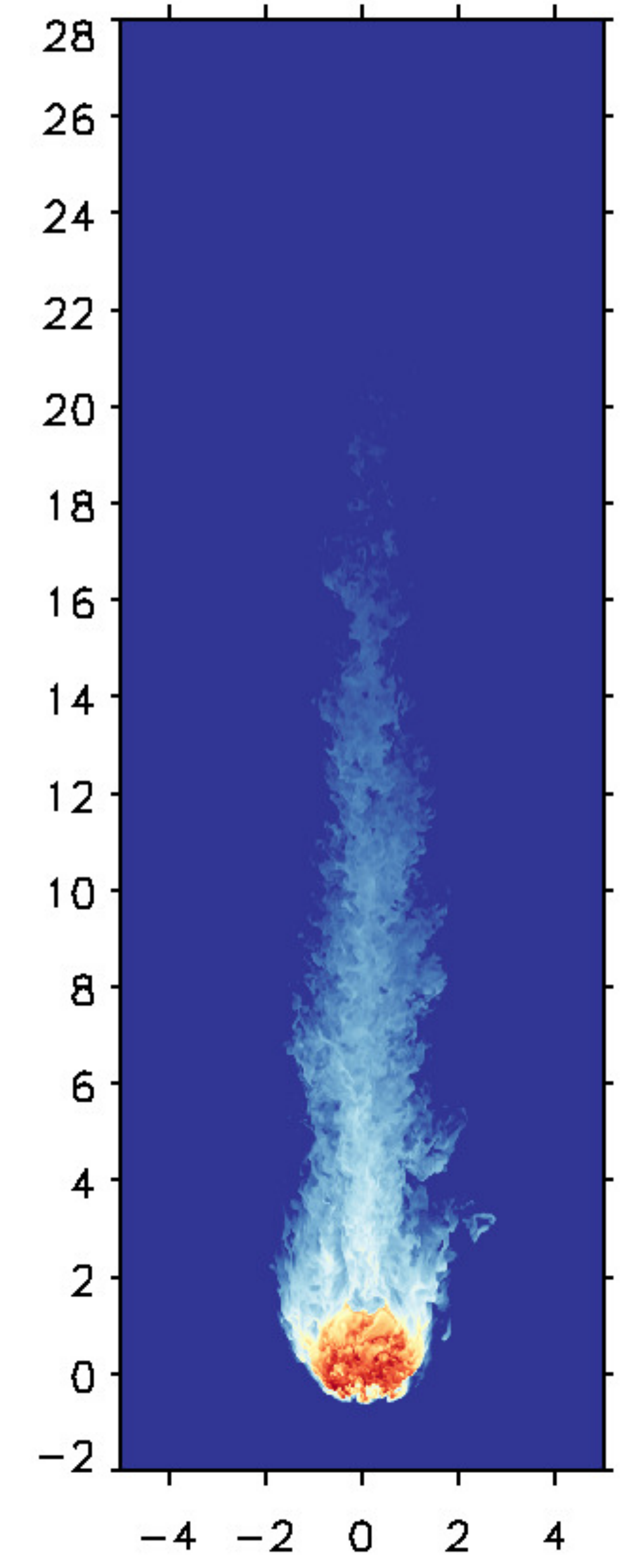}} & \hspace{-0.48cm}\resizebox{28mm}{!}{\includegraphics{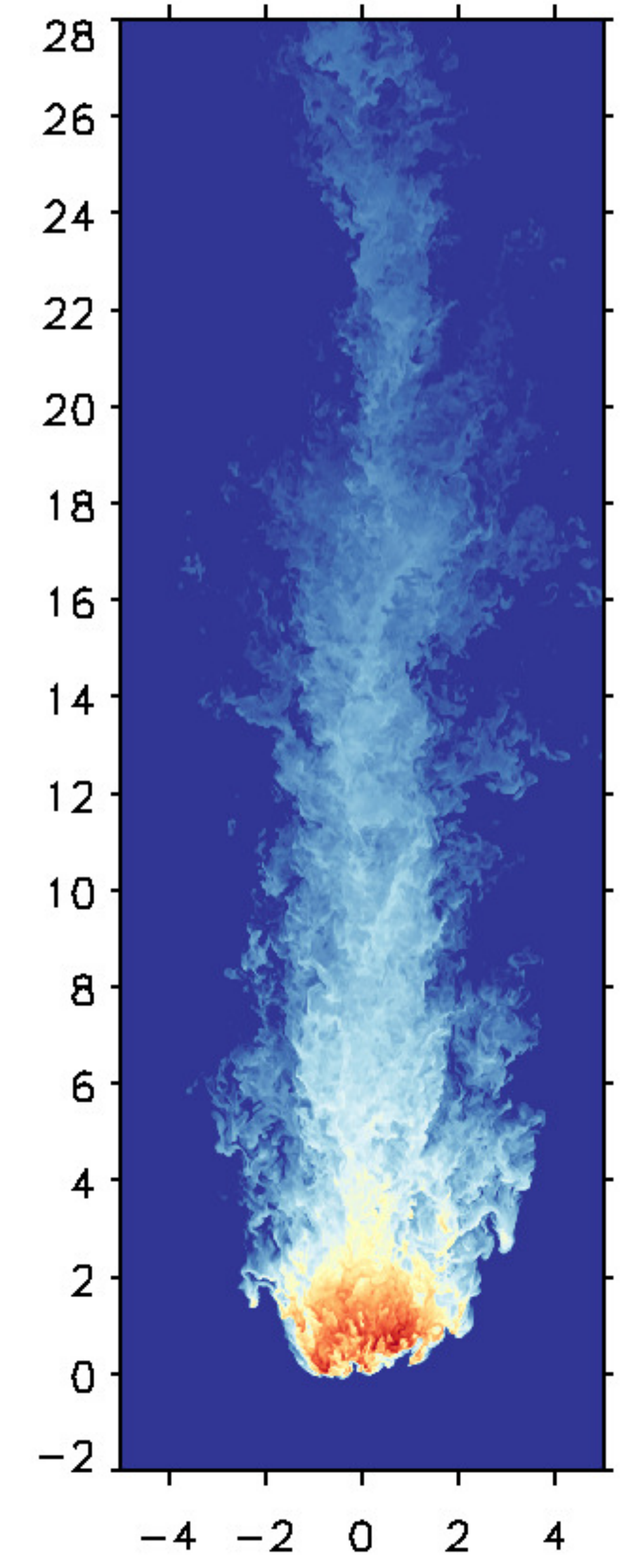}} & \hspace{-0.48cm}\resizebox{28mm}{!}{\includegraphics{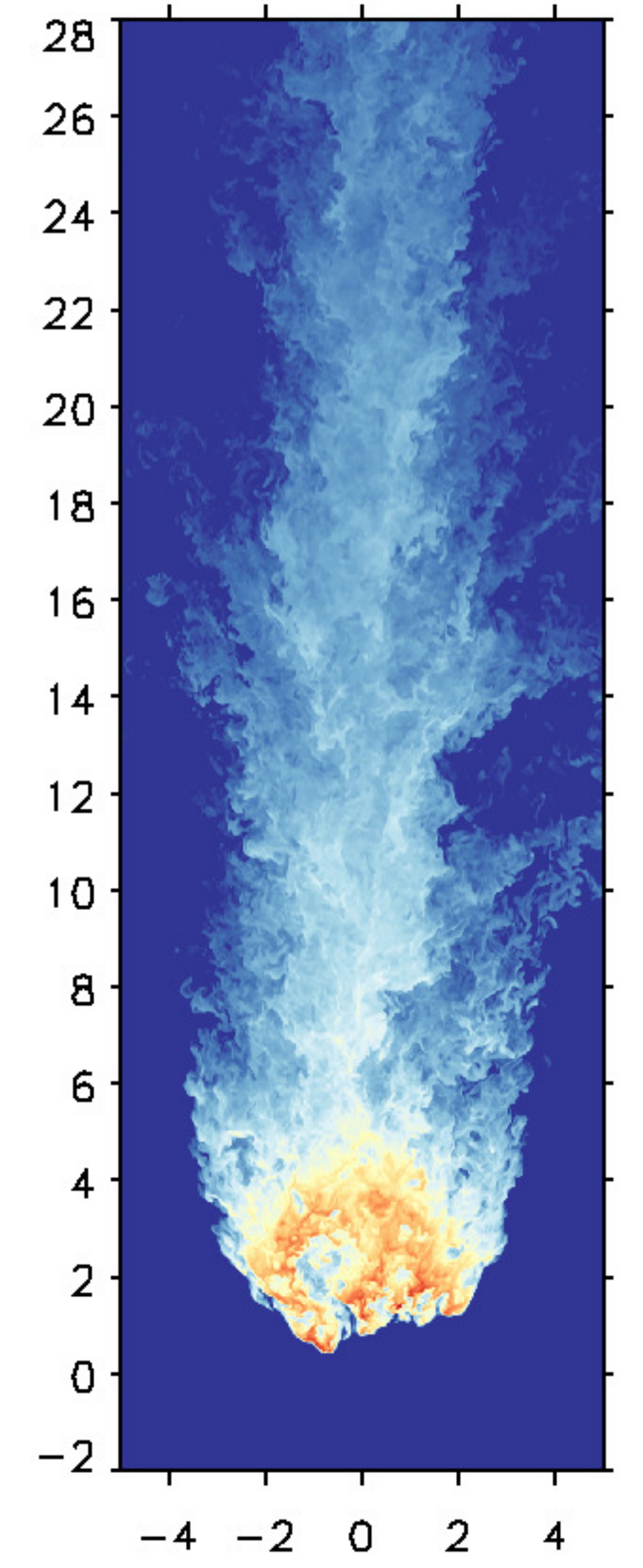}} & \hspace{-0.48cm}\resizebox{28mm}{!}{\includegraphics{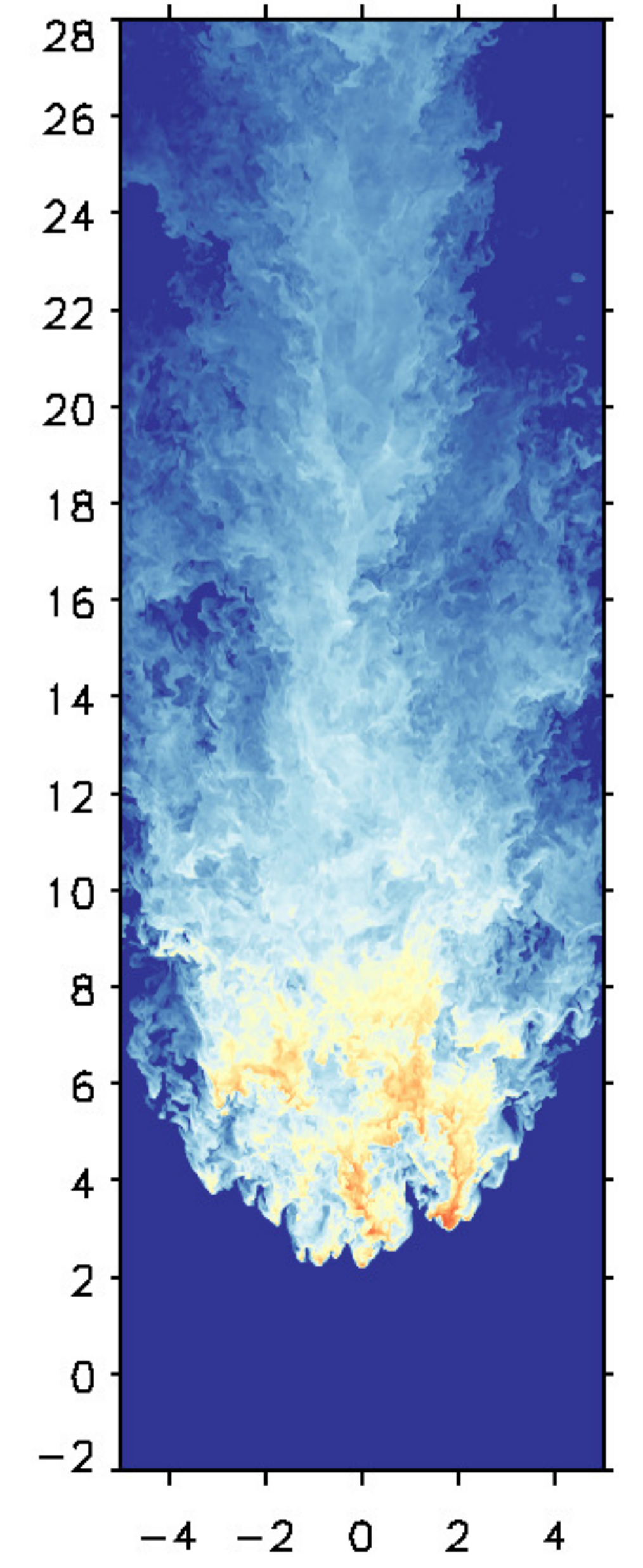}} & \hspace{-0.48cm}\resizebox{28mm}{!}{\includegraphics{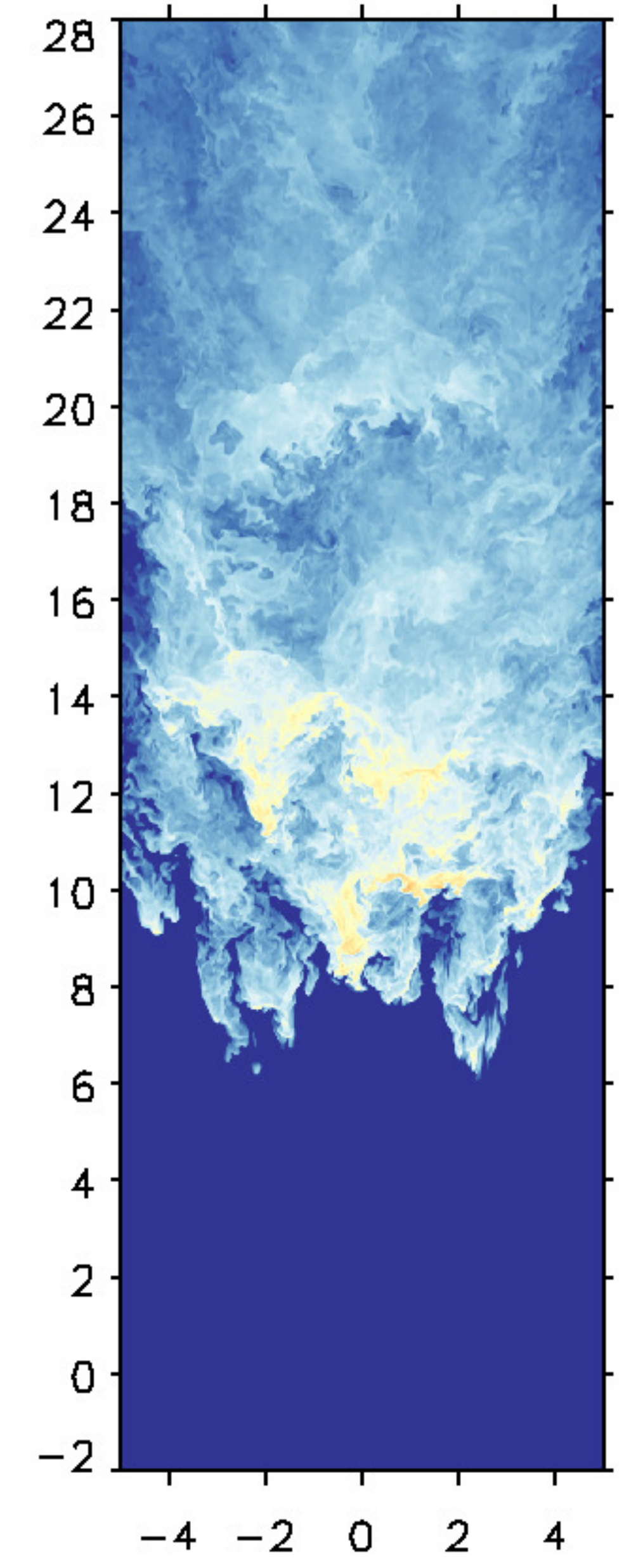}} & \hspace{-0.2cm}{\includegraphics[width=8.5mm]{pbarc.png}}\\
       \multicolumn{1}{l}{\hspace{-2mm}2c) 3Dcomp \hspace{+0.5mm}$t/t_{\rm cc}=0$} & \multicolumn{1}{c}{\hspace{+7mm}$t/t_{\rm cc}=0.5$} & \multicolumn{1}{c}{\hspace{+7mm}$t/t_{\rm cc}=1.0$} & \multicolumn{1}{c}{\hspace{+7mm}$t/t_{\rm cc}=1.5$} & \multicolumn{1}{c}{\hspace{+7mm}$t/t_{\rm cc}=2.0$} & \multicolumn{1}{c}{\hspace{+7mm}$t/t_{\rm cc}=2.5$} & $\frac{\rho C_{\rm cloud}}{\rho_{\rm wind}}$\\    
       \hspace{-0.37cm}\resizebox{28mm}{!}{\includegraphics{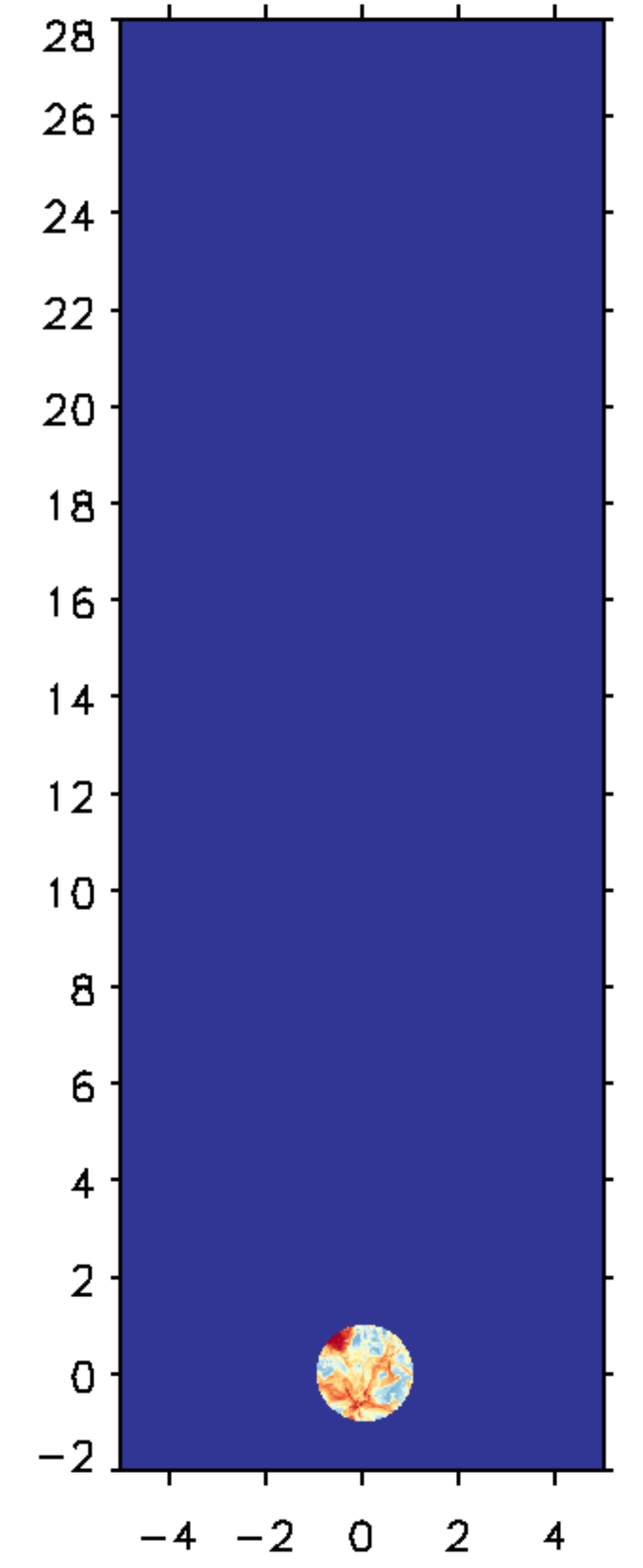}} & \hspace{-0.48cm}\resizebox{28mm}{!}{\includegraphics{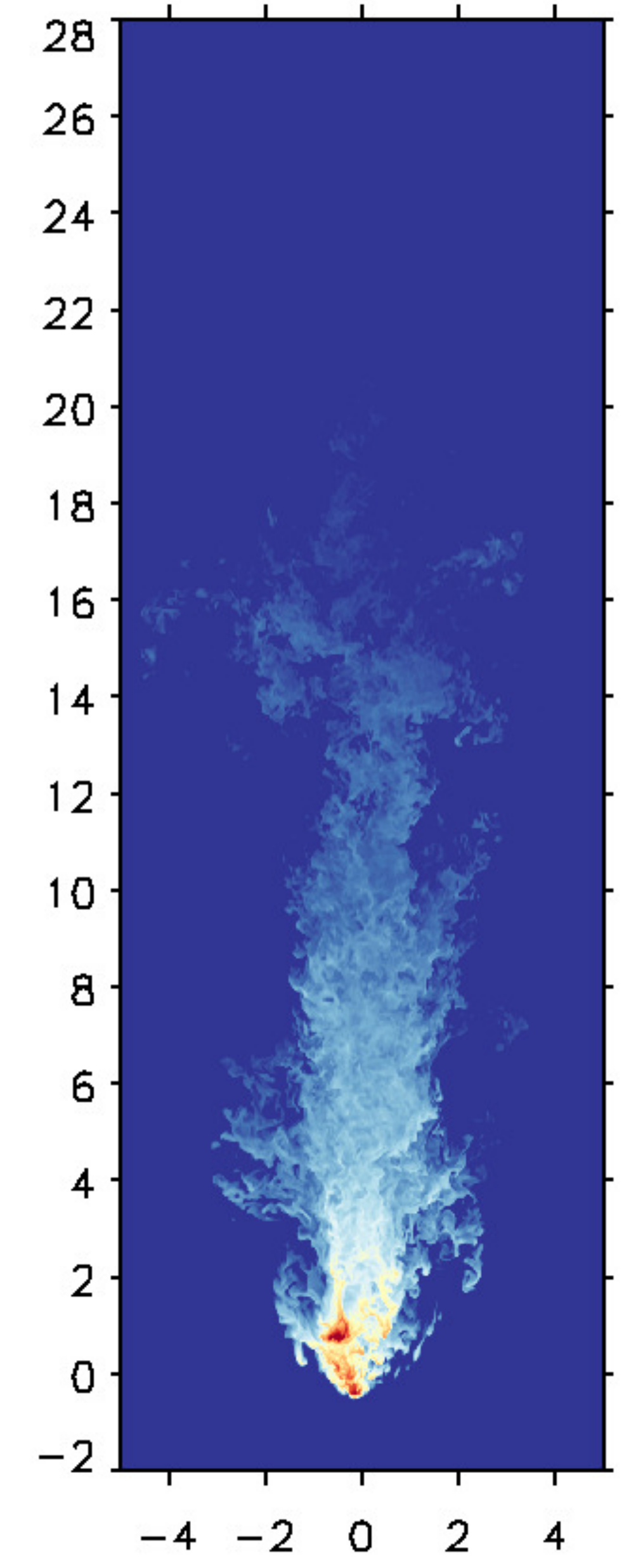}} & \hspace{-0.48cm}\resizebox{28mm}{!}{\includegraphics{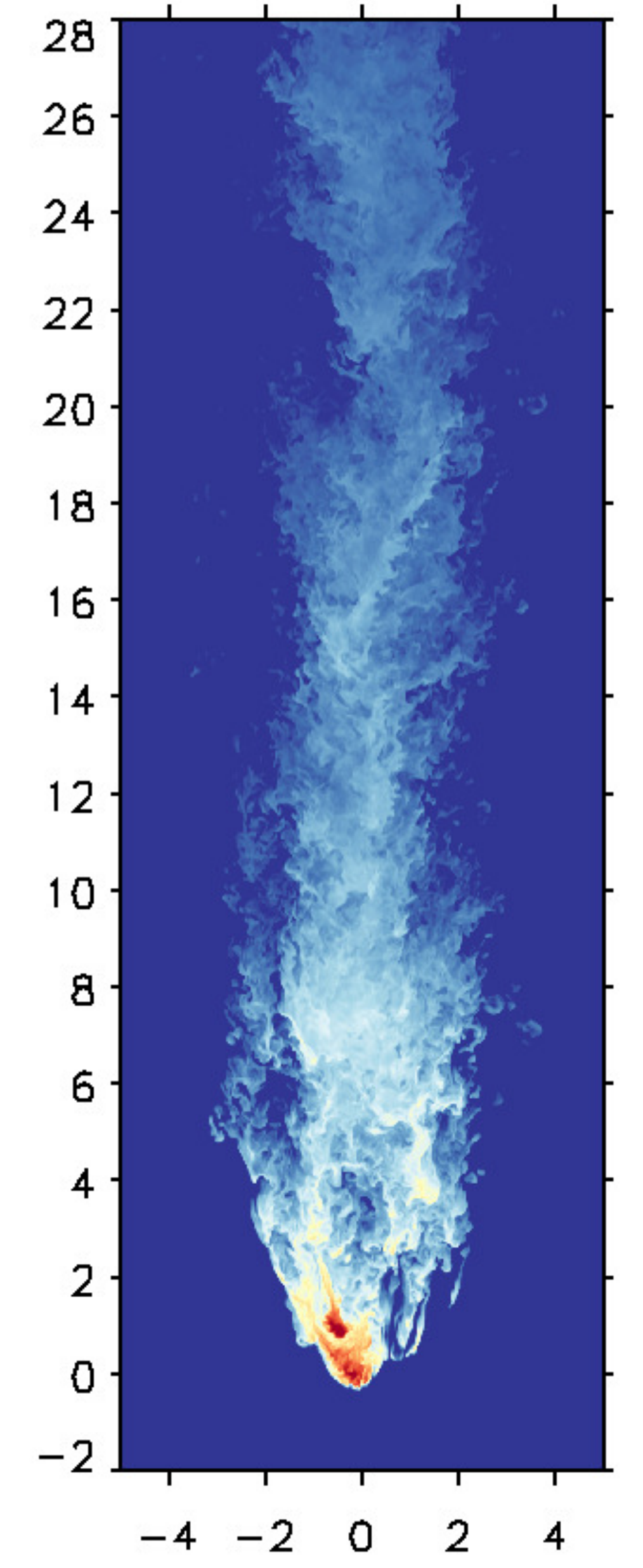}} & \hspace{-0.48cm}\resizebox{28mm}{!}{\includegraphics{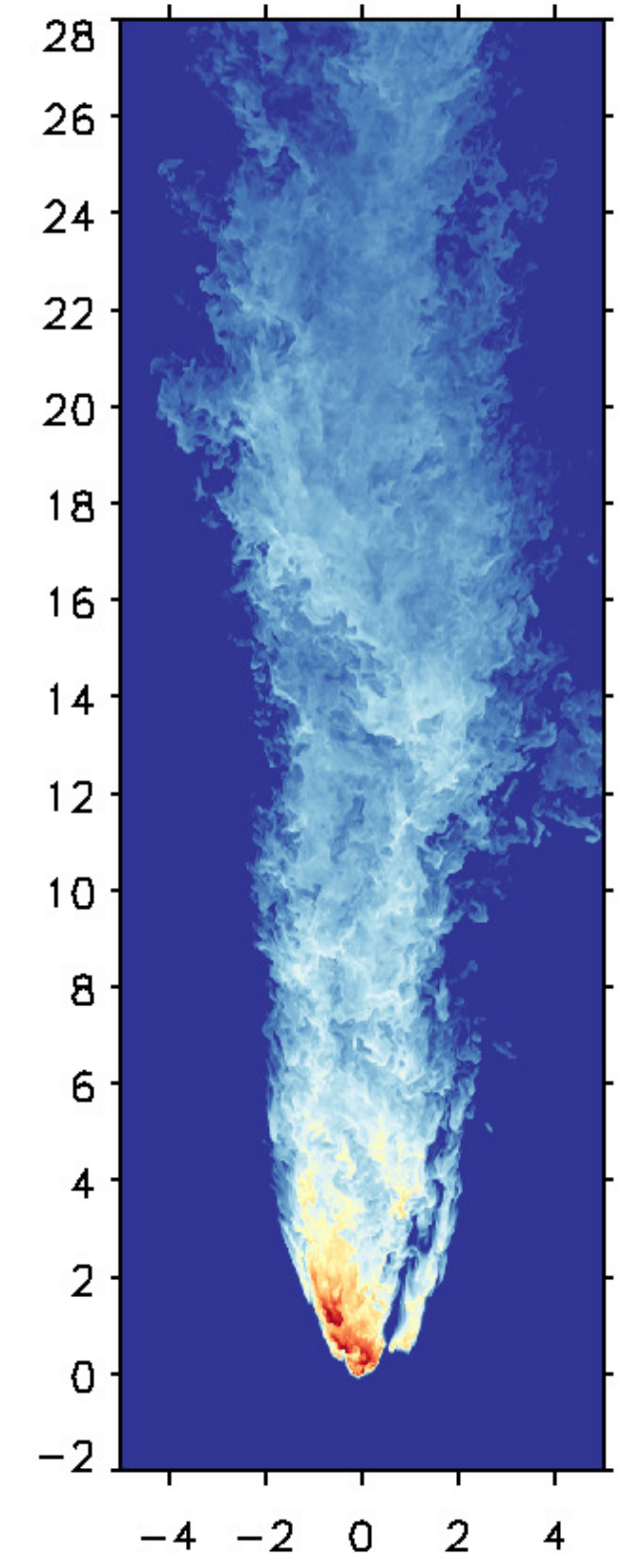}} & \hspace{-0.48cm}\resizebox{28mm}{!}{\includegraphics{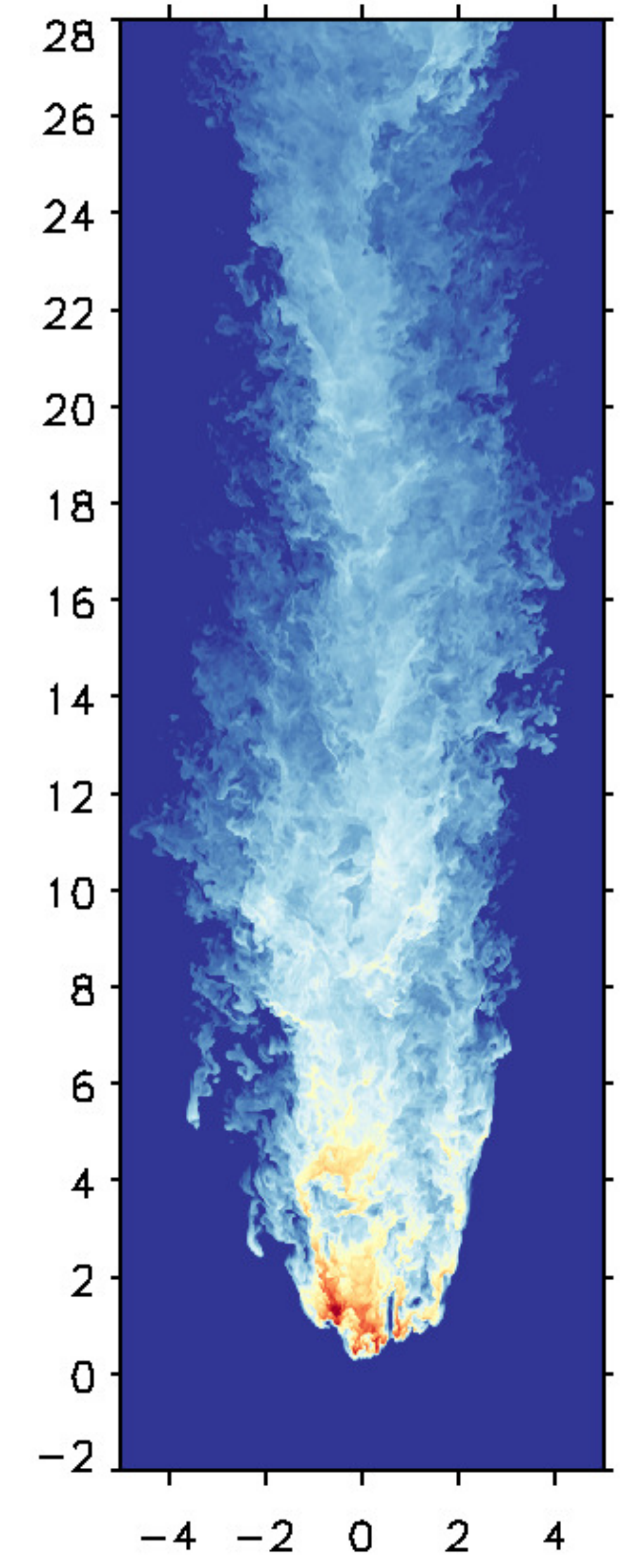}} & \hspace{-0.48cm}\resizebox{28mm}{!}{\includegraphics{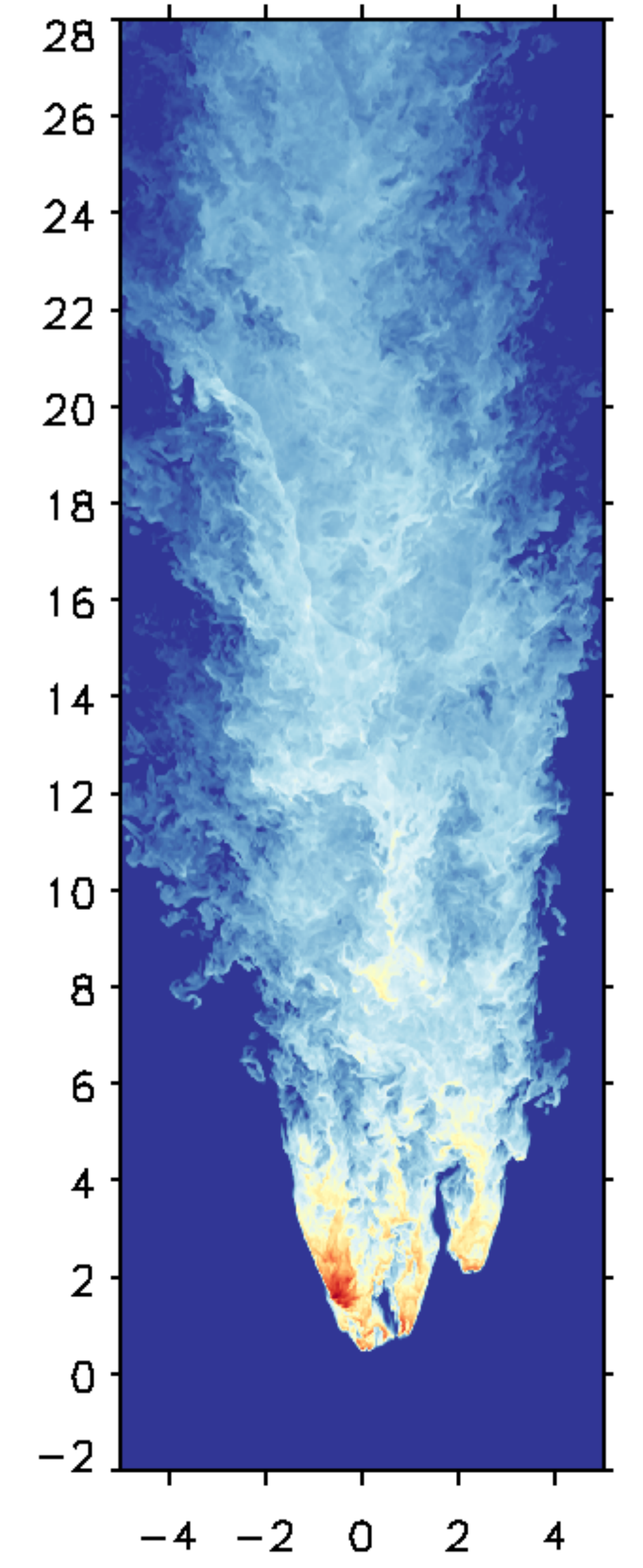}} & \hspace{-0.2cm}{\includegraphics[width=8.5mm]{pbarc.png}}\vspace{-0.2cm}\\
  \end{tabular}
  \caption{2D slices at $X_3=0$ showing the evolution between $0\leq t/t_{\rm cc}\leq2.5$ of the normalised cloud gas density ($\rho C_{\rm cloud}$) for our three 3D models, 3Dunif, 3Dsol, and 3Dcomp, which are representative of the uniform (panel 2a), turbulent solenoidal (panel 2b), and turbulent compressive (panel 2c) regimes, respectively. Fractal clouds are more expanded and turbulent than the uniform cloud. The fractal compressive cloud, which starts off with a higher standard deviation, is slower and more confined than its low-standard-deviation solenoidal counterpart, as it is supported by a higher-density core. Movies of the full-time evolution of these wind-cloud interactions are available online at \url{https://gwcsim.page.link/fractal}.} 
  \label{Figure2}
\end{center}
\end{figure*}

\begin{enumerate}
\item In the first stage, the initial impact of the wind on the clouds triggers both reflected and refracted shocks. The reflected shock creates a bow shock at the leading edge of the clouds while the refracted shock travels through the clouds at speed, $v_{\rm shock}$ (see all panels of Figure \ref{FigureC1} in Appendix \ref{sec:AppendixC}). In fractal cloud models the bow shock is anisotropic and several refracted shocks (instead of a single shock) are transmitted into the cloud (compare panels C1b and C1c of Figure \ref{FigureC1}). This is because the turbulent density fields in fractal clouds have a more intricate substructure than uniform density fields, which favours shock splitting (see also \citealt{2005ApJ...633..240P}).

\begin{figure*}
\begin{center}
  \begin{tabular}{c c c}
  3a) 3Dunif & 3b) 3Dsol & 3c) 3Dcomp \\
  \hspace{-0.2cm}\resizebox{58mm}{!}{\includegraphics{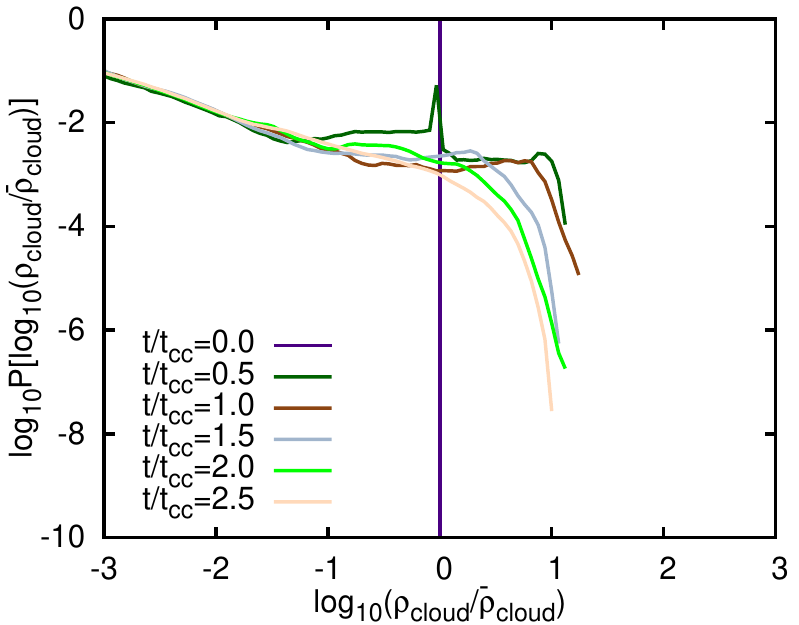}} & \hspace{-0.4cm}\resizebox{58mm}{!}{\includegraphics{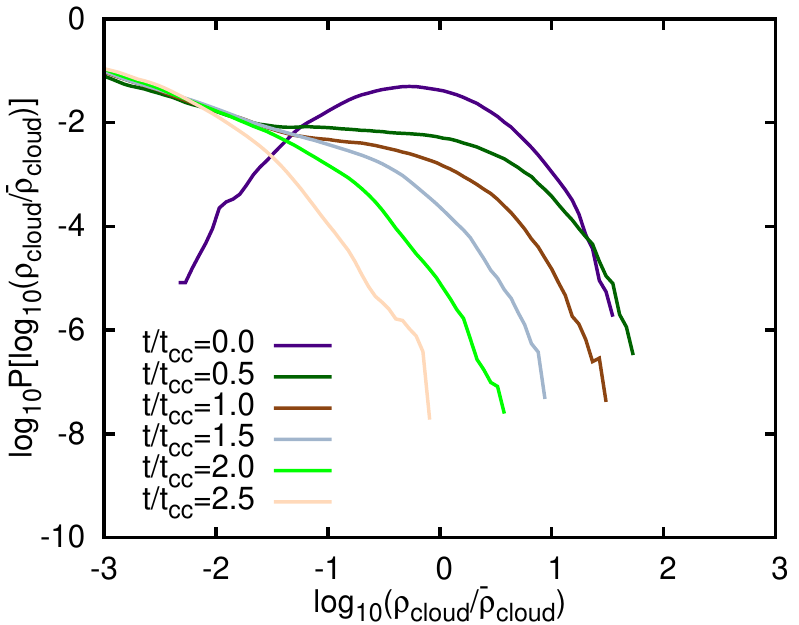}} & \hspace{-0.4cm}\resizebox{58mm}{!}{\includegraphics{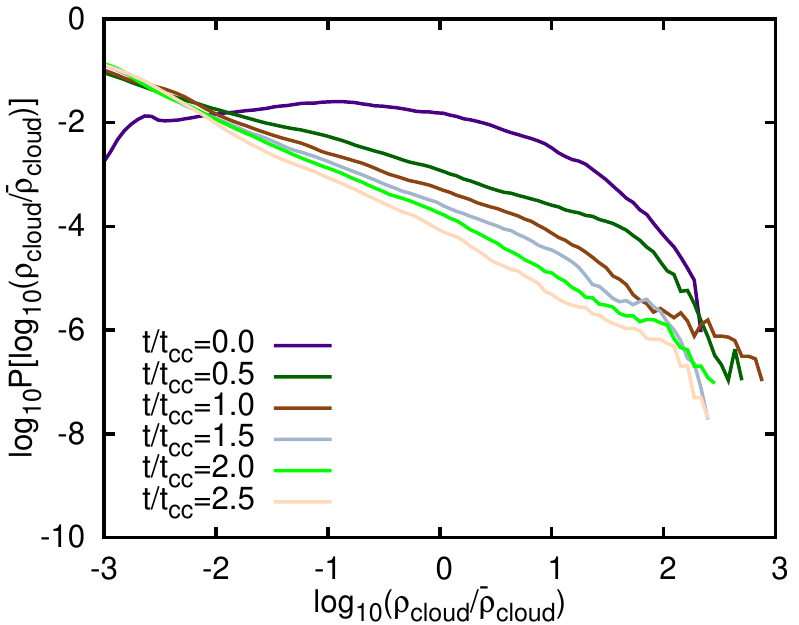}}\vspace{-0.2cm}\\
  \end{tabular}
  \caption{Time evolution of the density PDFs in the 3D cloud models: uniform 3Dunif (panel 3a), solenoidal 3Dsol (panel 3b), and compressive 3Dcomp (panel 3c). Gas mixing creates a flat, low-density tail in all cases and dismantles the log-normal nature of the PDFs in fractal models. The high-density tail of the PDFs evolves differently in solenoidal and compressive cloud models: while it rapidly recedes in the former, it remains coherent in the latter.}
  \label{Figure3}
\end{center}
\end{figure*}

\item In the second stage, the cloud expands as a result of internal shock heating, and pressure-gradient forces start to accelerate it and stretch it downstream. In general, shock-driven expansion\footnote{Note that shock-driven expansion does occur in quasi-isothermal ($\gamma=1.1$) models, but it is less pronounced than in adiabatic ($\gamma=1.67$) models as quasi-isothermal clouds can more efficiently convert the shock-driven heat into kinetic energy. Since a smaller cross section translates into less acceleration, quasi-isothermal clouds are slower and live longer than their adiabatic counterparts (see a discussion in \citealt{2016MNRAS.455.1309B}).} increases the effective cross sectional area of clouds and facilitates the wind-to-cloud momentum transfer. However, this occurs differently for uniform and fractal cloud models. Fractal clouds are porous, so they expand and accelerate faster than uniform clouds because the wind can more easily remove (and move through) the low-density-gas regions in them.

\item In the third stage, acceleration continues and the cloud loses mass via stripping by short-wavelength KH instabilities. KH instabilities mainly grow at the sides of the clouds at locations where velocity shears occur (see all panels of Figure \ref{Figure2}), but in fractal cloud models they also grow in the cloud's interior. Vortical motions remove gas from the cloud and the wind deposits it downstream, thus forming a long-standing, turbulent filamentary tail at the rear side of the cloud. In fractal cloud models wind and cloud gas mix more effectively than in uniform models, so the resulting filament in these models has a more complex structure populated by a collection of knots and sub-filaments (see also \citealt{2009ApJ...703..330C,2018MNRAS.473.3454B}).

\item In the fourth stage, the cloud has accelerated sufficiently for long-wavelength RT instabilities to grow at the leading edge of it, so the cloud/filament breaks up into smaller cloudlets as RT bubbles penetrate the cloud, causing it to further expand. These cloudlets and their tails survive for a few extra dynamical time-scales before dissolving into the ambient medium (see Section \ref{subsec:CloudDestruction}). Uniform and fractal cloud models experience break-up phases differently. While the break-up of uniform clouds is abrupt, fractal clouds undergo a rather steady disruption process as they have several high-density nuclei and each of them undergoes its own break-up phase (at its own time-scale).
\end{enumerate}

\subsection{The role of the initial density PDF}
\label{subsec:DensityPDF}
As discussed in the previous section, the initial density distribution of clouds influences their evolution. Both, the uniform and fractal clouds in our sample are initialised with the same mass and average density. However, they evolve into filaments that are morphologically and dynamically different. These differences translate into uniform and fractal clouds being accelerated and disrupted at different rates. In addition, as we discuss throughout this paper, the same occurs when we consider subsamples of fractal clouds with different statistical properties. For simplicity we study and discuss the disruption of subsamples of fractal clouds in two regimes of turbulence, namely solenoidal (characterised by low PDF standard deviations) and compressive (characterised by high PDF standard deviations).

\subsubsection{Solenoidal versus compressive cloud models}
\label{subsec:3D}

The 2D slices of the normalised cloud density in panels 2b and 2c of Figure \ref{Figure2} show that solenoidal clouds are more expanded, travel faster, lose high-density gas more quickly, and are disrupted earlier than their compressive counterparts. The higher acceleration of the solenoidal cloud makes it more prone to RT instabilities than its compressive counterpart, while the steeper density gradients in the compressive cloud delays the emergence of long-wavelength KH instabilities at shear layers. The differences seen in Figure \ref{Figure2} can also be understood if we compare how the density PDFs of solenoidal and compressive cloud models change over time. Figure \ref{Figure3} presents the evolution of the density PDFs of the 3D models in the uniform case (panel 3a) and in both regimes of turbulence (panels 3b and 3c). The densities in these curves are normalised to the initial average cloud density in all models.\par

Figure \ref{Figure3} shows that in all cases the low-density tails of the PDFs are rapidly flattened as wind and cloud gas mix, implying that low-density gas is the dominant component of ram-pressure accelerated gas in wind-swept clouds. Mixing processes also dismantle the log-normality of the initial PDFs of fractal clouds, although they act differently in solenoidal and compressive models as evidenced by the distinct evolution of the high-density tails of their PDFs. The high-density tail of the PDF in the solenoidal model moves much faster towards lower density values than in the compressive model. By $t/t_{\rm cc}=2.5$ no gas in the solenoidal cloud has densities higher than the initial average cloud density, while the compressive cloud retains some gas with densities $10-10^2$ times higher than the initial average cloud density. This dense gas is contained in several nuclei (see panel 2c of Figure \ref{Figure2}), which act as a long-lived footpoint for the downstream filamentary network.\par

The above analysis shows that the process of cloud disruption is indeed sensitive to the initial cloud substructure, but are the differences seen in the solenoidal and compressive cloud models tied to the initial standard deviations of their density PDFs? Is the evolution of wind-swept fractal clouds rather chaotic, without a clear correlation to the PDF statistical parameters? For instance, an alternative explanation for the observed differences could be that the mass in the compressive model (in this particular sample case) is arranged in such a way that high-density gas is somewhat protected from the wind by upstream gas that prevents it from being ablated, thus delaying its disruption. This would be akin to the shielding processes reported by \citealt{2014MNRAS.444..971A,2018arXiv181012925F} for multi-cloud media and cold gas streams embedded in hot outflows (although the mixing of upstream gas can enhance the turbulence-driven destruction of downstream clouds in some multi-cloud configurations; see e.g., \citealt*{2002ApJ...576..832P}; \citealt{2012MNRAS.425.2212A}). However, as we show below, the differences seen in 3D solenoidal and compressive cloud models can actually be linked to the initial standard deviation of their density PDFs.

\subsubsection{A statistical view of the evolution of solenoidal and compressive cloud models}
\label{subsec:2D}

\begin{figure*}
\begin{center}
  \begin{tabular}{c c c c c c}
\multicolumn{1}{l}{\hspace{-1.5mm}4a) 2Dsol \hspace{+6mm}$t/t_{\rm cc}=1.0$} & \multicolumn{1}{c}{\hspace{+11mm}$t/t_{\rm cc}=2.0$} & \multicolumn{1}{c}{\hspace{+11mm}$t/t_{\rm cc}=4.0$} & \multicolumn{1}{c}{\hspace{+11mm}$t/t_{\rm cc}=6.0$} &  \multicolumn{1}{c}{\hspace{+11mm}$t/t_{\rm cc}=8.0$} & \hspace{-4mm}$\frac{\rho C_{\rm cloud}}{\rho_{\rm wind}}$\\    
       \hspace{-0.3cm}\resizebox{34mm}{!}{\includegraphics{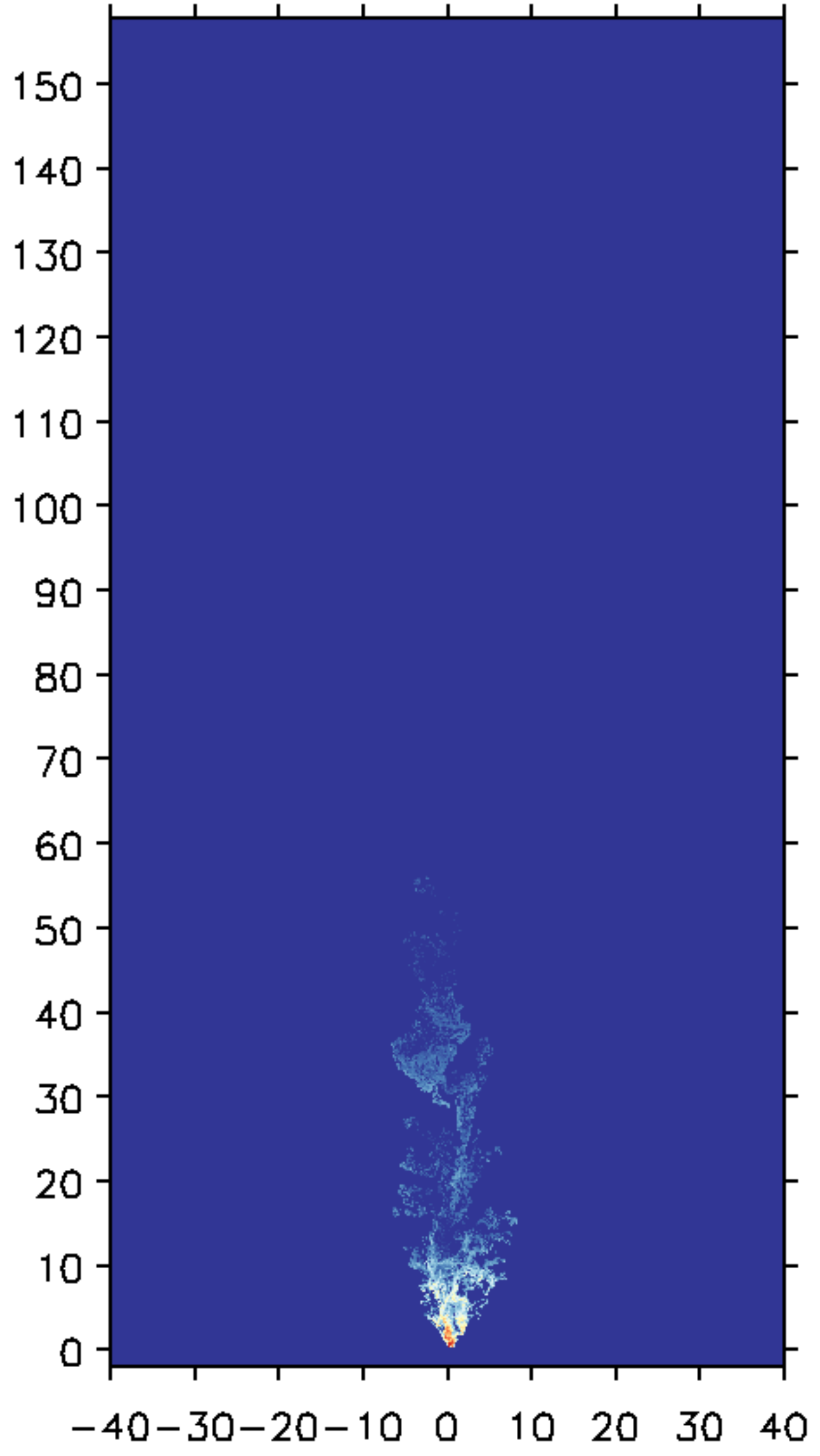}} & \hspace{-0.48cm}\resizebox{34mm}{!}{\includegraphics{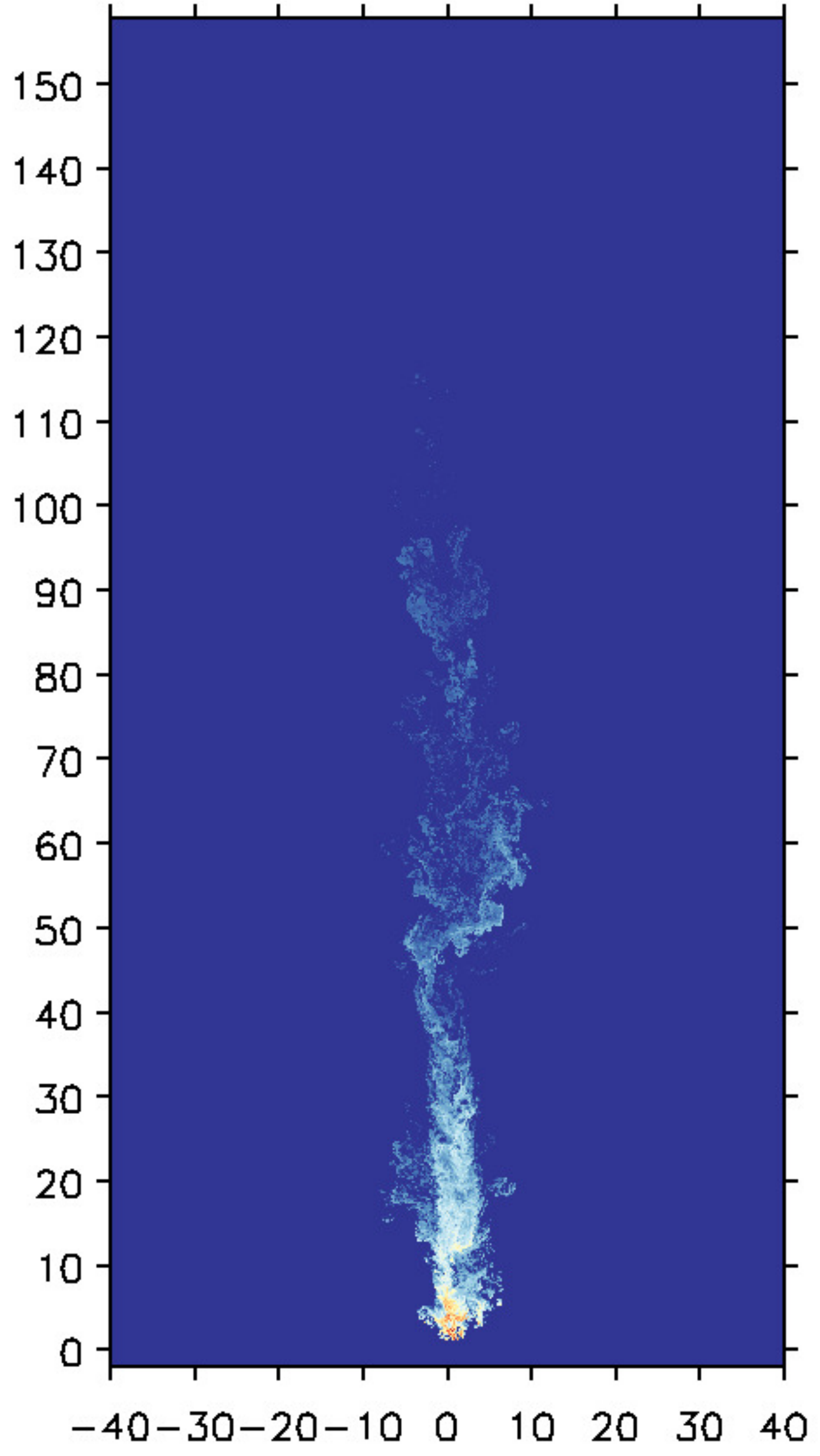}} & \hspace{-0.48cm}\resizebox{34mm}{!}{\includegraphics{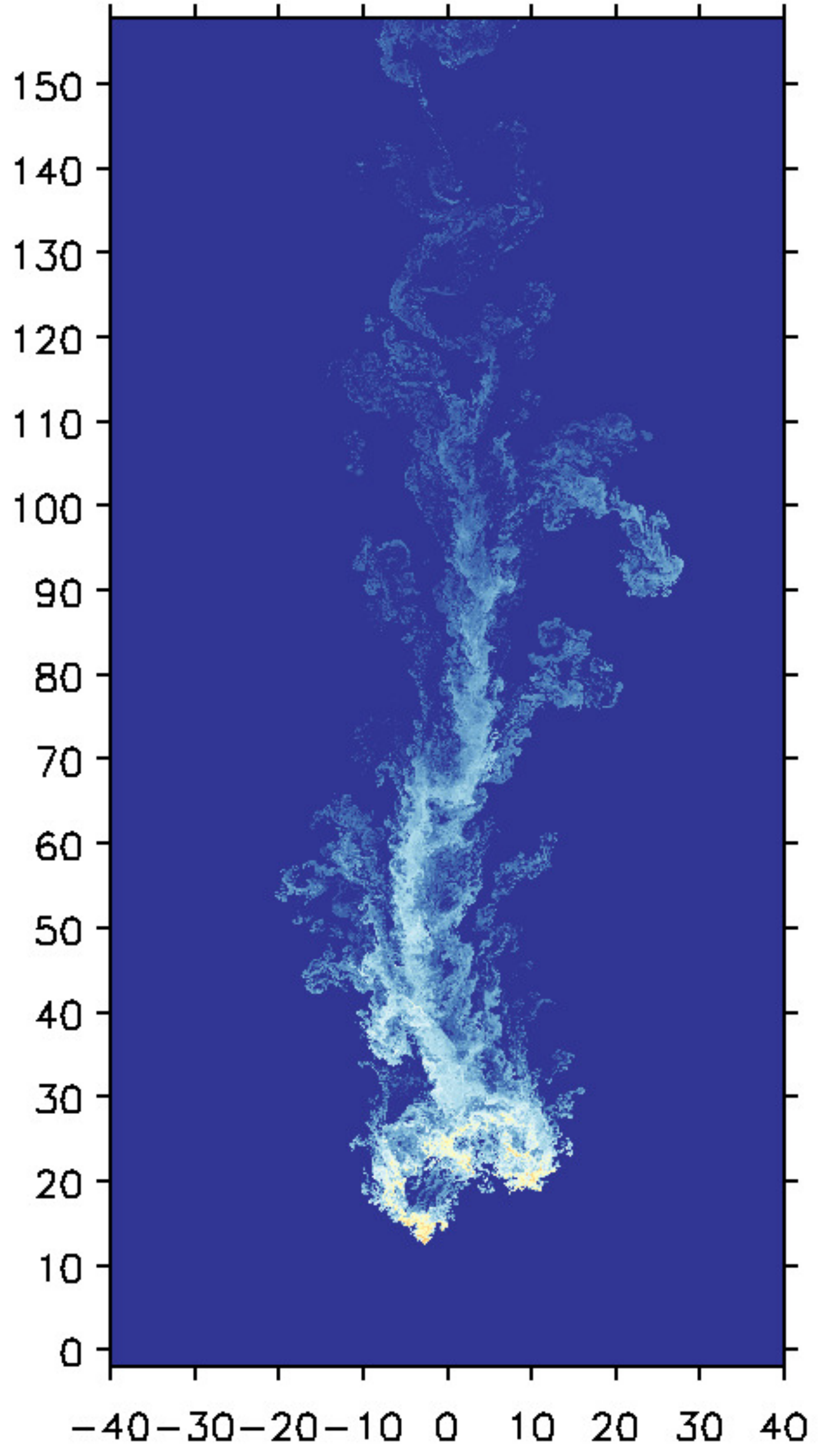}} & \hspace{-0.48cm}\resizebox{34mm}{!}{\includegraphics{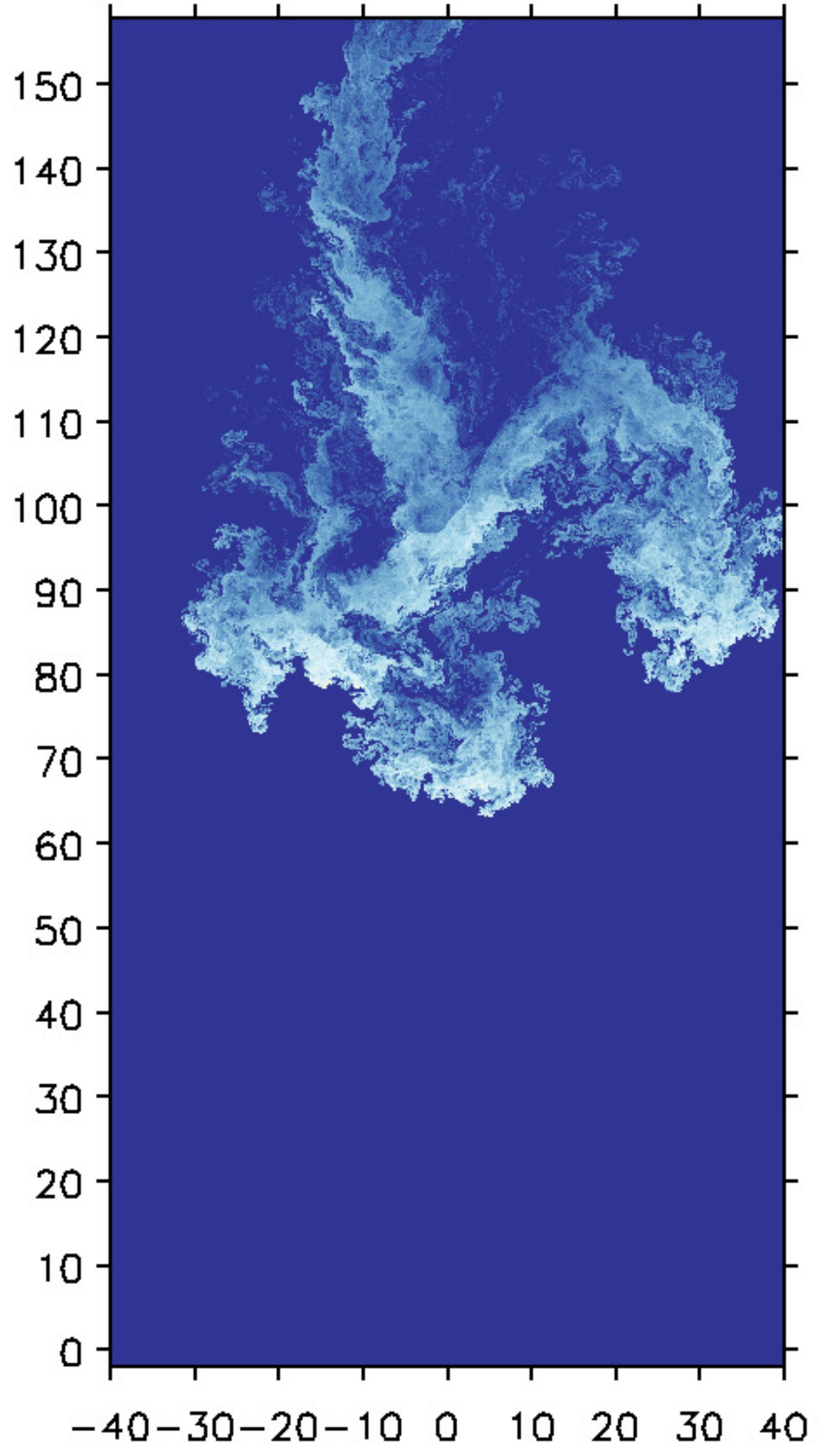}} & \hspace{-0.48cm}\resizebox{34mm}{!}{\includegraphics{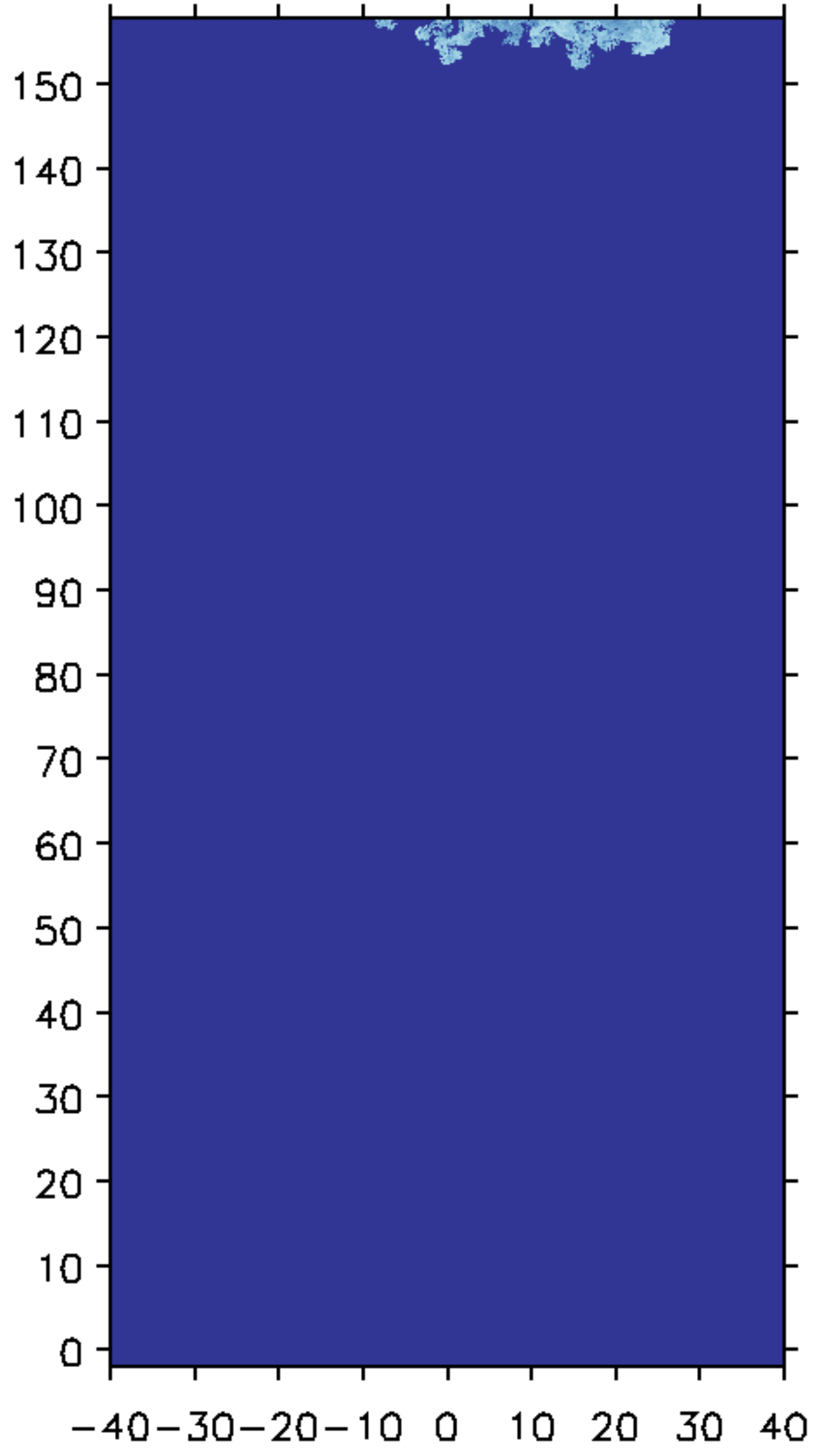}} & \hspace{-0.3cm}{\includegraphics[width=7.4mm]{pbarc.png}}\\
\multicolumn{1}{l}{\hspace{-1.5mm}4b) 2Dcomp \hspace{+6mm}$t/t_{\rm cc}=1.0$} & \multicolumn{1}{c}{\hspace{+11mm}$t/t_{\rm cc}=2.0$} & \multicolumn{1}{c}{\hspace{+11mm}$t/t_{\rm cc}=4.0$} & \multicolumn{1}{c}{\hspace{+11mm}$t/t_{\rm cc}=6.0$} &  \multicolumn{1}{c}{\hspace{+11mm}$t/t_{\rm cc}=8.0$} & \hspace{-4mm}$\frac{\rho C_{\rm cloud}}{\rho_{\rm wind}}$\\    
       \hspace{-0.3cm}\resizebox{34mm}{!}{\includegraphics{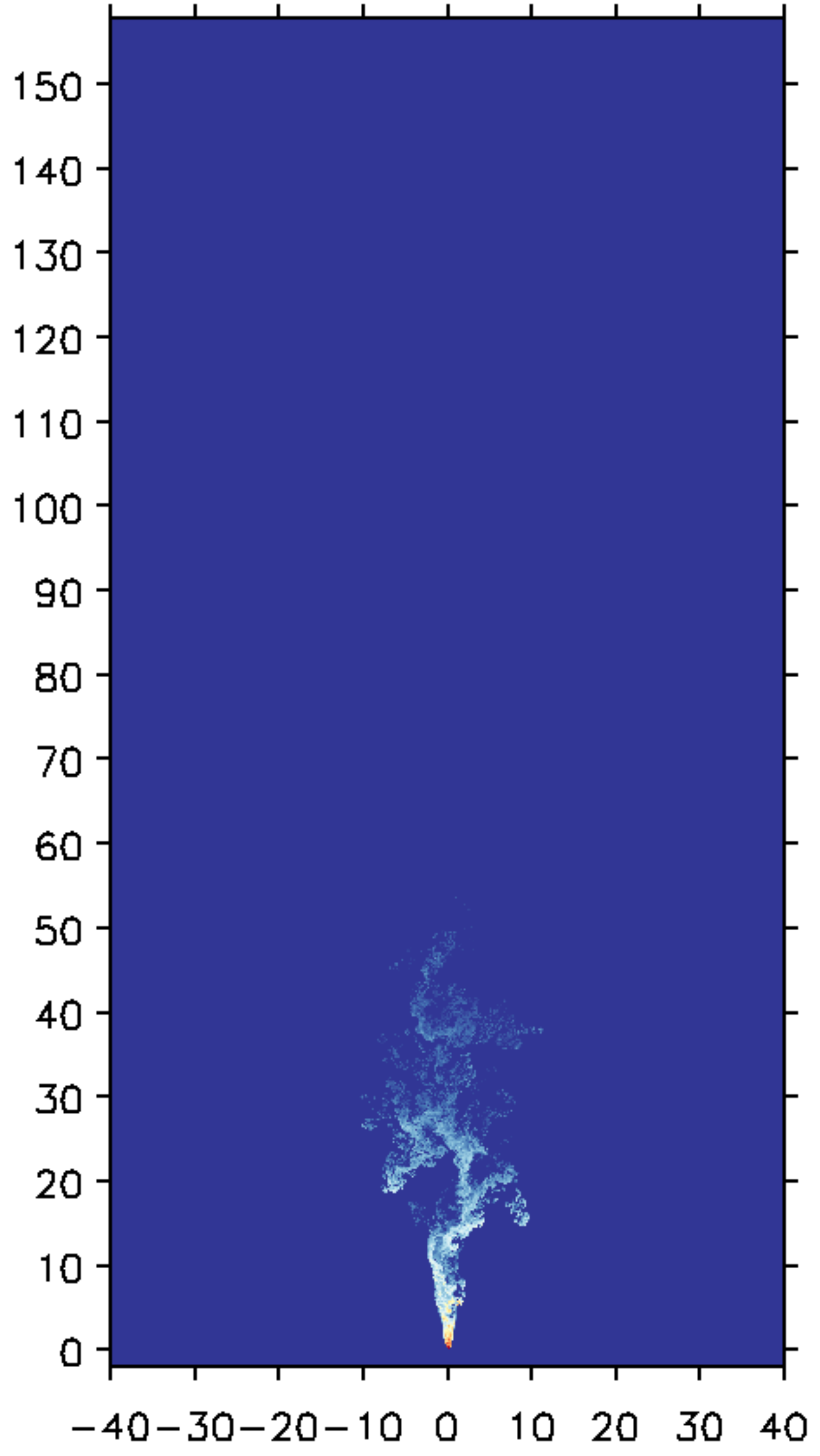}} & \hspace{-0.48cm}\resizebox{34mm}{!}{\includegraphics{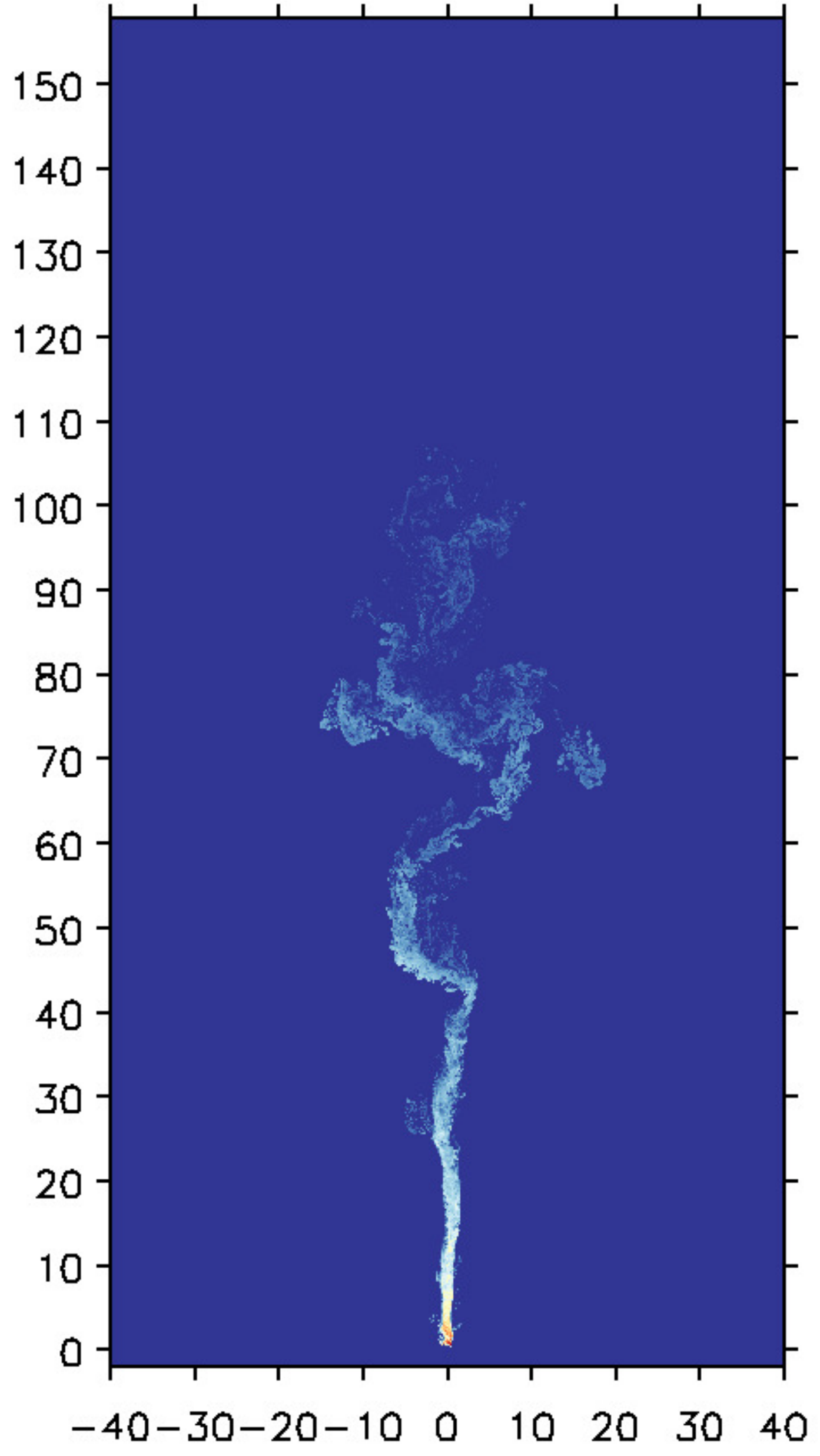}} & \hspace{-0.48cm}\resizebox{34mm}{!}{\includegraphics{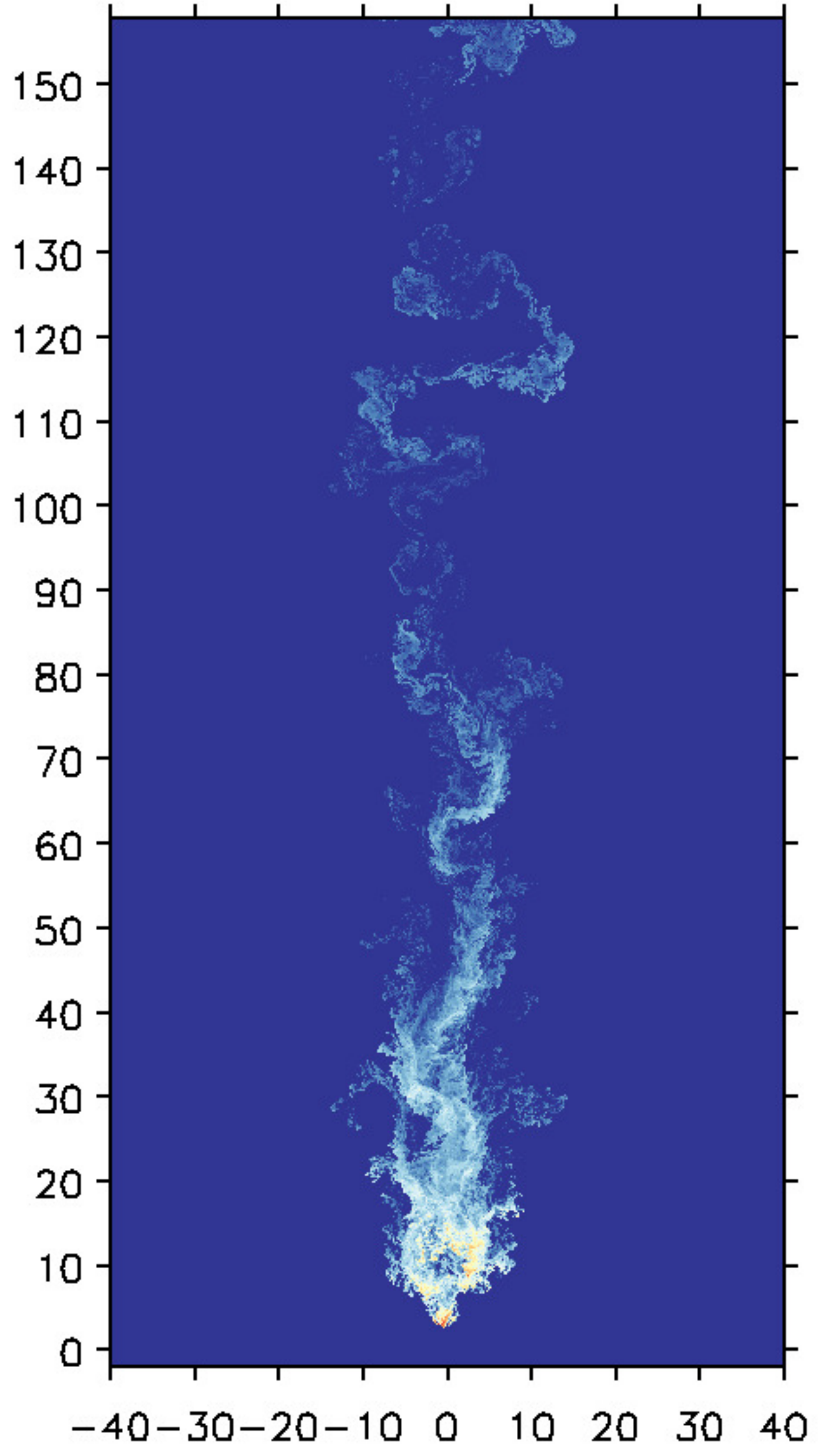}} & \hspace{-0.48cm}\resizebox{34mm}{!}{\includegraphics{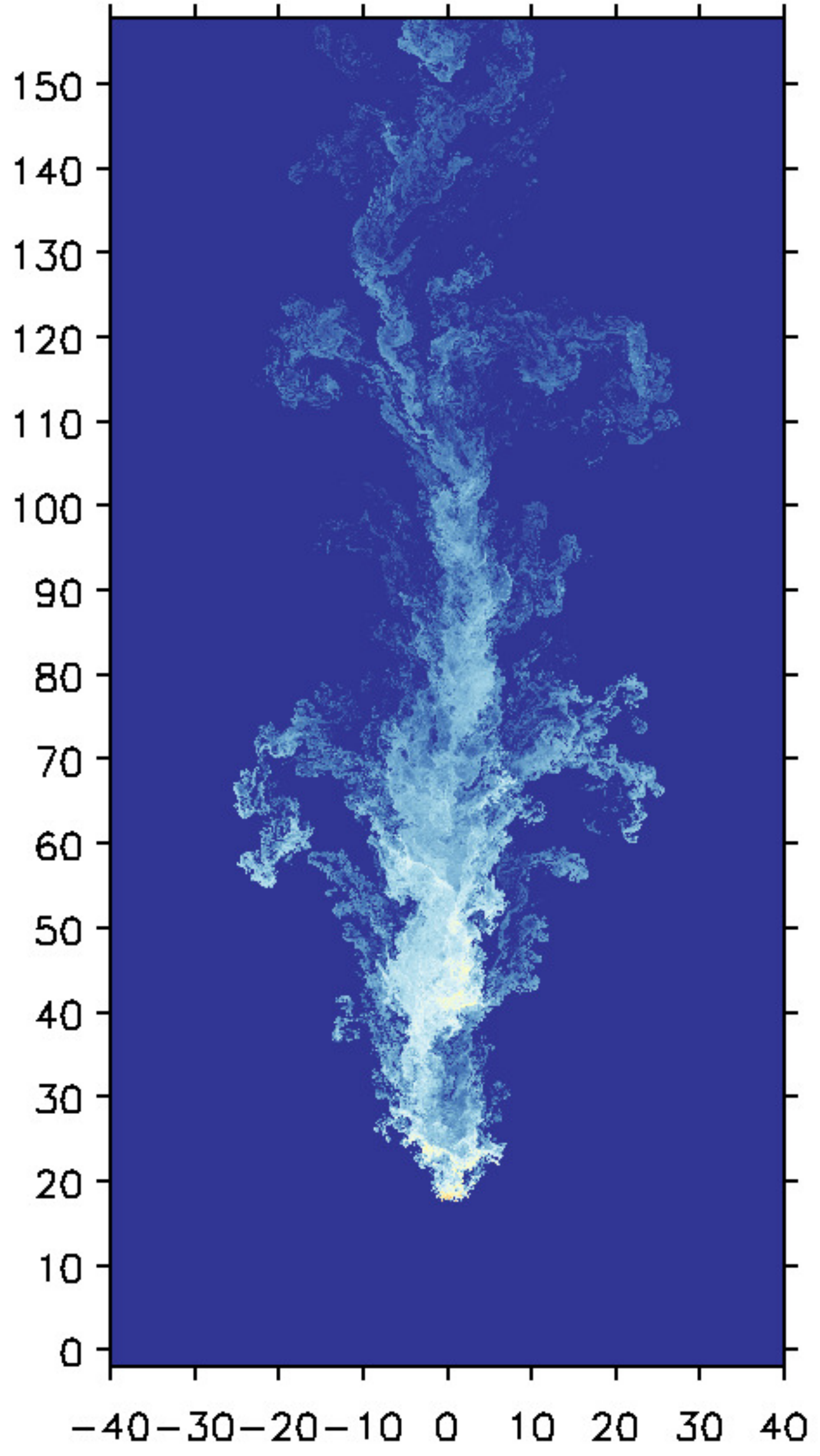}} & \hspace{-0.48cm}\resizebox{34mm}{!}{\includegraphics{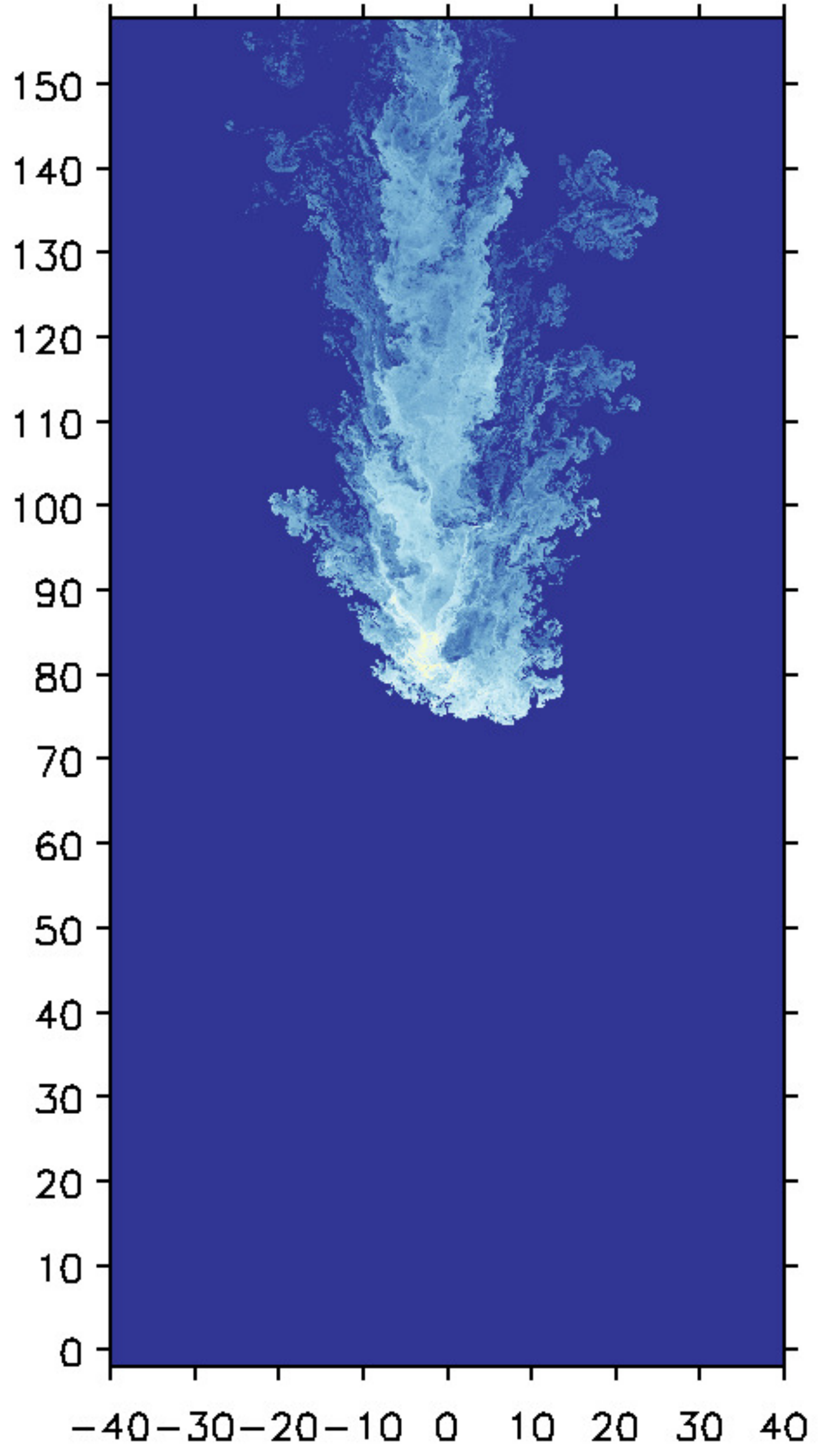}} & \hspace{-0.3cm}{\includegraphics[width=7.4mm]{pbarc.png}}\\  
  \end{tabular}
  \caption{2D plots showing the evolution between $1\leq t/t_{\rm cc}\leq 8$ of the normalised cloud gas density ($\rho C_{\rm cloud}$) in two fiducial 2D models, 2Dsol and 2Dcomp, that are representative of the solenoidal (panel 4a) and compressive (panel 4b) regimes, respectively. Similarly to the 3D models in Figure \ref{Figure2}, these panels show that solenoidal clouds are faster and less confined than their compressive counterparts (which are supported over longer time-scales by high-density nuclei). Runs with differently-seeded and distinctly-rotated clouds display similar behaviours. Movies of the full-time evolution of these wind-cloud interactions are available online at \url{https://gwcsim.page.link/fractal}.} 
  \label{Figure4}
\end{center}
\end{figure*}

Models in 3D are computationally expensive (even in purely hydrodynamical models), so using a 3D geometry constrains both the size of the computational domain and the number of realisations that we can investigate. Thus, we will now look into the 2D set of solenoidal and compressive fractal cloud models. Analysing 2D models allows us to enlarge the computational domain, increase the simulation sample, and build up statistics on the behaviour of solenoidal and compressive cloud models, all at a fraction of the computational resources needed for a similar 3D study.\par

\begin{figure*}
\begin{center}
  \begin{tabular}{c c c}
  \hspace{-2.9cm}5a) $t/t_{\rm cc}=2.0$ & \hspace{-2.9cm}5b) $t/t_{\rm cc}=4.0$ & \hspace{-2.9cm}5c) $t/t_{\rm cc}=6.0$ \\
  \hspace{-0.2cm}\resizebox{58mm}{!}{\includegraphics{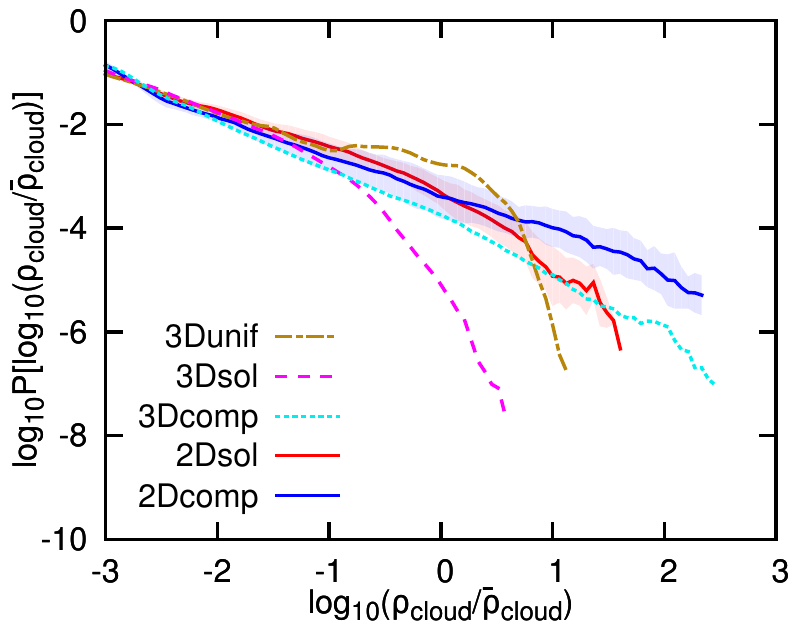}} & \hspace{-0.4cm}\resizebox{58mm}{!}{\includegraphics{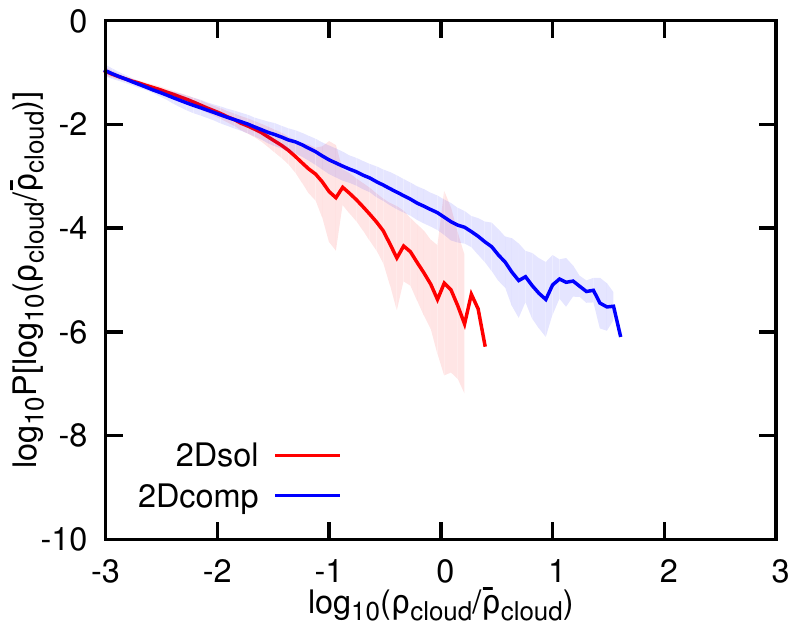}} & \hspace{-0.4cm}\resizebox{58mm}{!}{\includegraphics{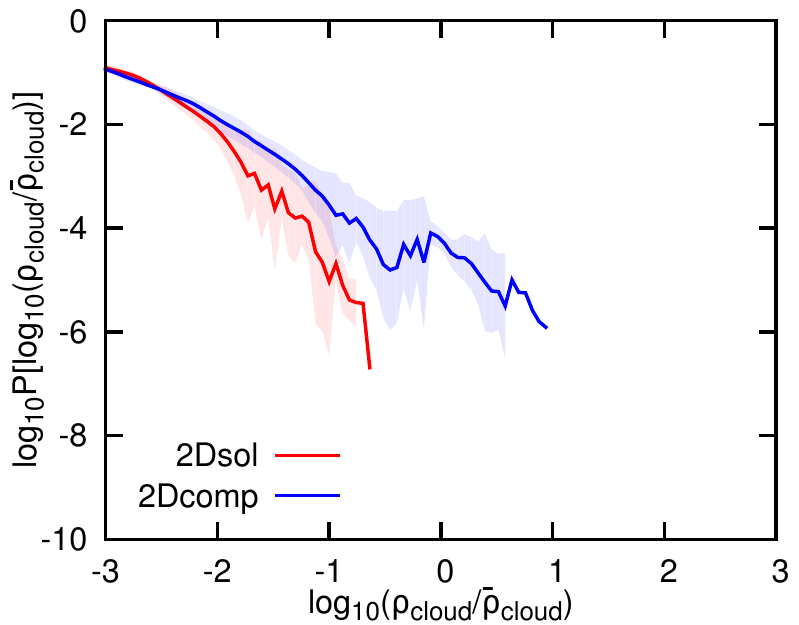}}\vspace{-0.2cm}\\
  \end{tabular}
  \caption{Density PDFs of the 2D fractal cloud models (solenoidal, 2Dsol, and compressive, 2Dcomp) at three different times, namely: $t/t_{\rm cc}=2.0$ (panel 5a, which also shows the 3D models), $t/t_{\rm cc}=4.0$ (panel 5b), and $t/t_{\rm cc}=6.0$ (panel 5c). The gas density in the solenoidal cloud models falls below $10\,\bar{\rho}_{\rm cloud}$ by $t/t_{\rm cc}=4.0$, while in the compressive cloud models this happens later, at $t/t_{\rm cc}=6.0$. Thus, in agreement with 3D fractal cloud models, 2D compressive clouds are disrupted over longer time-scales than 2D solenoidal clouds.}
  \label{Figure5}
\end{center}
\end{figure*}

In this section, we present $20$ fractal cloud models in 2D, $10$ of which are representative of the solenoidal regime and $10$ of the compressive regime. As mentioned in Section \ref{subsec:2Dmodels}, each of the clouds in each regime is either generated with a different seed or with the same seed but it is rotated. Therefore, each cloud in each sample has a different spatial distribution of densities than its pairs, but the same log-normal statistical parameters. Since the cloud evolution depends on the initial density distribution, each cloud also has its own intrinsic evolution. However, we find that wind-swept clouds that are initialised with the same log-normal density parameters, either solenoidal or compressive, do develop common, regime-dependent morphological and dynamical features.\par

In this context, Figure \ref{Figure4} shows the cloud density, $\rho C_{\rm cloud}$, at five different times in the range $1\leq t/t_{\rm cc}\leq8$, normalised with respect to the wind density, $\rho_{\rm wind}$ of two fiducial 2D models. Panel 4a of this figure shows the evolution of a fiducial solenoidal model, while panel 4b shows the evolution of a fiducial compressive model. These panels confirm the conclusions drawn from the analysis of the 3D runs (see Figure \ref{Figure2}), i.e., that solenoidal clouds have larger cross sections, have larger accelerations, and are mixed and disrupted faster than their compressive counterparts.\par

Figure \ref{Figure5} shows the evolution of the density PDFs in 2D solenoidal and compressive cloud models, normalised with respect to the initial average cloud density, at three different times: $t/t_{\rm cc}=2.0$ (panel 5a), $t/t_{\rm cc}=4.0$ (panel 5b), and $t/t_{\rm cc}=6.0$ (panel 5c). The solid lines represent the average values while the shaded areas around them denote the one-standard-deviation limits. Similarly to the 3D models displayed in Figure \ref{Figure3}, the panels of Figure \ref{Figure5} show that the high-density tail in the solenoidal cloud models moves towards low-density values faster than in the compressive clouds (which retain long-lived nuclei). For comparison, the gas density in the solenoidal cloud models has fallen below $10\,\bar{\rho}_{\rm cloud}$ by $t/t_{\rm cc}=4.0$, while in the compressive cloud models this happens at $t/t_{\rm cc}=6.0$. Thus, compressive 2D clouds are also disrupted over longer time-scales than solenoidal 2D clouds, despite being initialised with the same mass and average density.

\subsection{Cloud dynamics}
\label{subsec:Dynamics}
\begin{figure*}
\begin{center}
  \begin{tabular}{c c c}
  \hspace{-0.2cm}\resizebox{58mm}{!}{\includegraphics{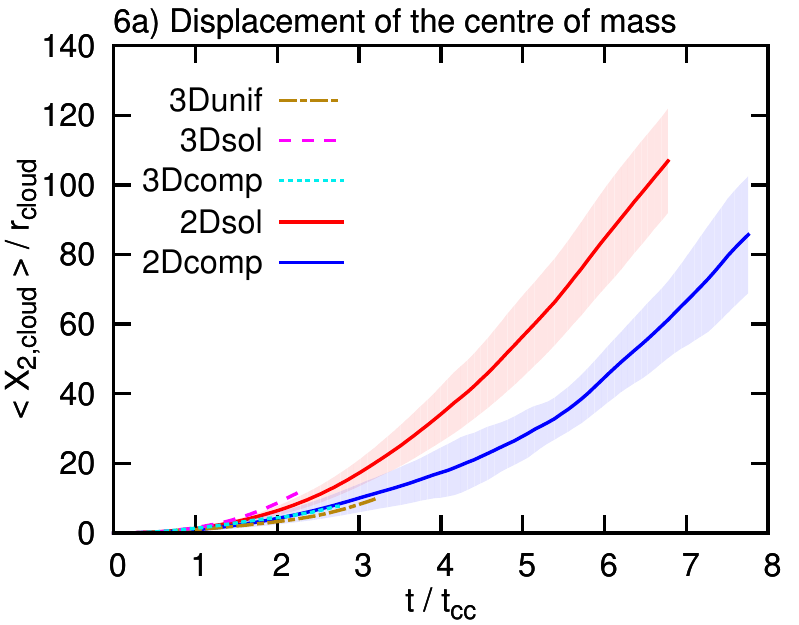}} & \hspace{-0.4cm}\resizebox{58mm}{!}{\includegraphics{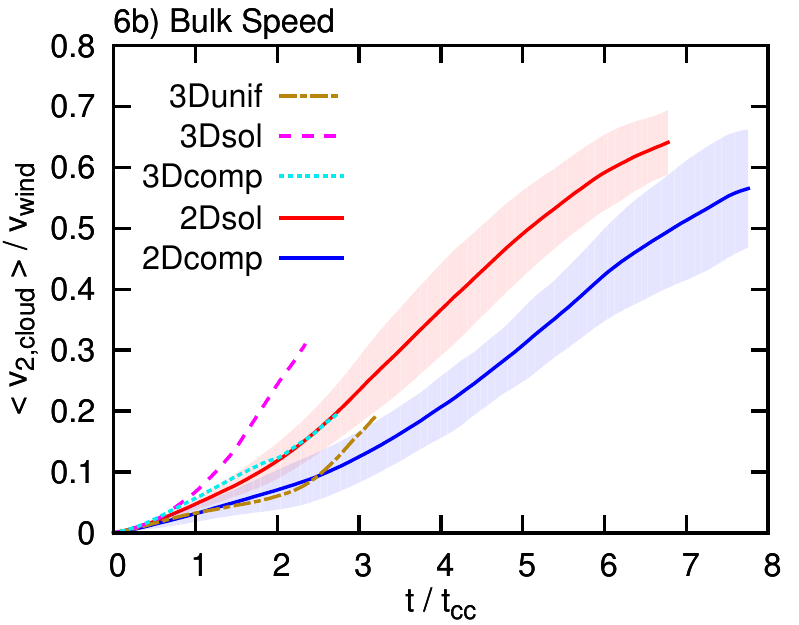}} & \hspace{-0.4cm}\resizebox{58mm}{!}{\includegraphics{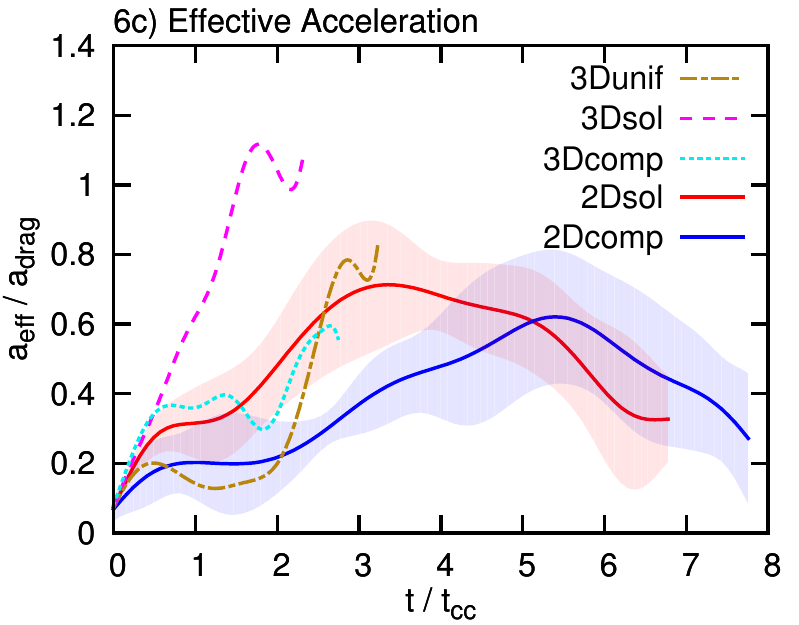}}\vspace{-0.1cm}\\
  \hspace{-0.2cm}\resizebox{58mm}{!}{\includegraphics{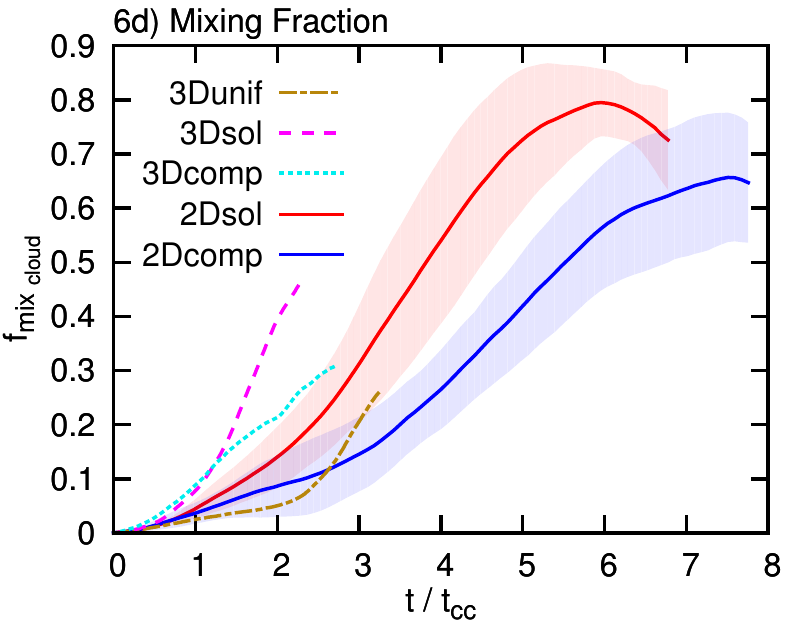}} & \hspace{-0.4cm}\resizebox{58mm}{!}{\includegraphics{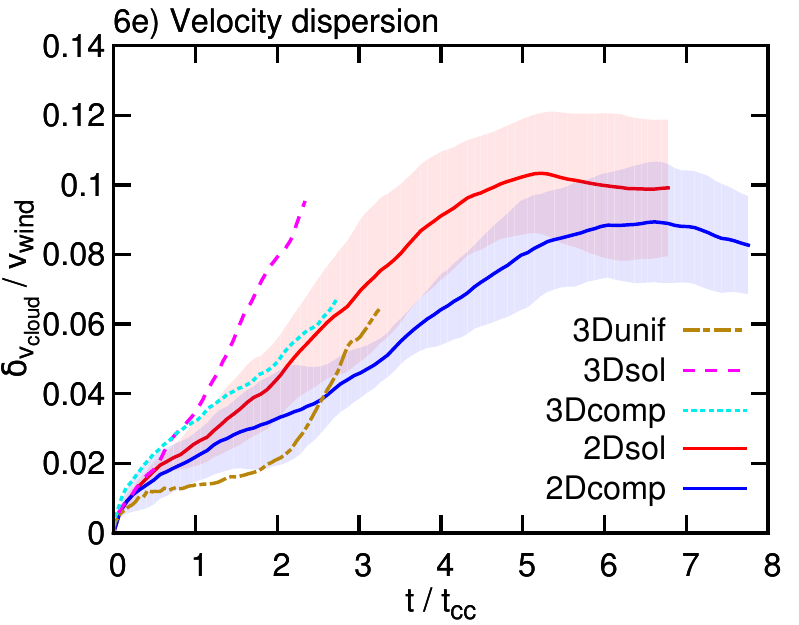}} & \hspace{-0.4cm}\resizebox{58mm}{!}{\includegraphics{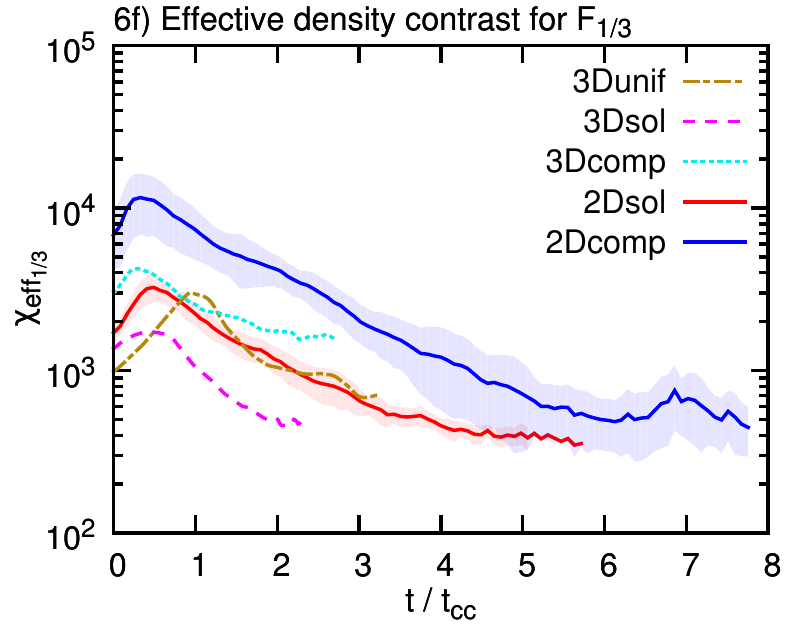}}\vspace{-0.2cm}\\
  \end{tabular}
  \caption{Time evolution of six diagnostics: the displacement of the centre of mass (panel 6a), the bulk speed (panel 6b), the acceleration (panel 6c), the mixing fraction (panel 6d), the transverse velocity dispersion (panel 6e), and the effective density contrast between cloud gas with densities above $\bar{\rho}_{\rm cloud}/3$ and the wind density $\rho_{\rm wind}$ (panel 6f) in all 3D and 2D uniform and fractal cloud models. The top panels show that solenoidal clouds reach larger distances, and attain higher speeds and accelerations than compressive clouds. The bottom panels show that the gas in solenoidal clouds mixes earlier, becomes more dispersed, and has a lower effective density contrast than in compressive cloud models. Kelvin-Helmholtz and Rayleigh-Taylor instabilities grow, therefore, slower in compressive clouds.}
  \label{Figure6}
\end{center}
\end{figure*}

In order to quantify the differences in the dynamical evolution of uniform, solenoidal, and compressive cloud models, we study several diagnostics in both 3D and 2D models. Figure \ref{Figure6} shows the evolution of the displacement of the centre of mass (panel 6a), the bulk speed (panel 6b), and the acceleration (panel 6c) of cloud material ($\rho C_{\rm cloud}$) in all models. The curves corresponding to the 3D models are displayed independently, while the curves corresponding to the 2D realisations are grouped together for each model, and represented by the average (with a solid line) of the diagnostic and the one-standard-deviation limits (with shaded areas around the average). These panels confirm the qualitative results discussed above and show that the dynamics of wind-swept clouds is coupled to the statistics of their initial density PDFs.\par

Panels 6a, 6b, and 6c of Figure \ref{Figure6} show that 3D fractal clouds reach larger distances, and acquire higher bulk speeds and accelerations than the 3D uniform cloud, over the same time-scales. Moreover, within the 3D and 2D samples of fractal clouds, these panels show that solenoidal clouds are more effectively accelerated than compressive clouds owing to their larger cross sectional areas. For instance, at $t/t_{\rm cc}$=2.0, solenoidal clouds are $\sim 1.7-1.9$ times faster and have reached distances $\sim 1.5-1.8$ times greater than their compressive counterparts (in both 3D and 2D geometries). The absolute differences between the two become more pronounced as time progresses and are statistically significant (i.e., outside one another's one-standard-deviation levels) in 2D models for times $t/t_{\rm cc}\gtrsim2.5$. At $t/t_{\rm cc}=6.0$, 2D solenoidal clouds have reached bulk speeds of $\sim 0.60\pm0.06\,v_{\rm wind}$ and distances of $\sim 88.1\pm14.6\,r_{\rm cloud}$, while compressive clouds have only reached bulk speeds of $\sim 0.44\pm0.07\,v_{\rm wind}$ and distances of $\sim 47.4\pm8.4\,r_{\rm cloud}$. These different cloud accelerations in solenoidal and compressive models imply that RT instabilities grow at different rates in each model (see Section \ref{subsec:CloudDestruction}).\par

\subsection{Gas mixing and dispersion}
\label{subsec:GasMixing}
Understanding how the cloud gas is mixed and dispersed into the ambient medium is another important aspect of wind-cloud interactions. Panels 6d, 6e, and 6f of Figure \ref{Figure6} present the evolution of the mixing fraction, the velocity dispersion, and the effective cloud-to-wind density contrast, respectively, in both 3D and 2D fractal cloud models. Panels 6d and 6e confirm that, on average, solenoidal clouds are more turbulent (i.e. have higher velocity dispersions) and mix faster with the ambient medium than compressive clouds. At $t/t_{\rm cc}$=2.0, solenoidal clouds are $\sim 1.6-1.8$ times more mixed and turbulent than their compressive counterparts (in both 3D and 2D geometries). By $t/t_{\rm cc}$=4.0, the absolute differences are more statistically significant in 2D models, with mixing fractions of $55.8\pm15.5\,\%$ and velocity dispersions of $0.092\pm0.017\,v_{\rm wind}$ in solenoidal models and of $27.7\pm7.5\,\%$ and $0.065\pm0.010\,v_{\rm wind}$ in compressive models.\par

The process of gas mixing and dispersion in wind-swept clouds is associated with the generation of vortical motions in it via KH instabilities. Since the effective cloud-to-wind density contrast in compressive cloud models is larger than in solenoidal cloud models (see panel 6f of Figure \ref{Figure6}), the growth of KH instabilities is delayed in such models. For comparison we use Equation (\ref{KHtime}) to compute the ratio of the KH instability growth time-scales for solenoidal and compressive clouds. We find that KH instabilities with long wavelengths ($k_{\rm KH}\,r_{\rm cloud}\sim 1$) grow $\sim 1.5-2$ times slower in compressive models ($t_{\rm KH}/t_{\rm cc}=0.88$ in 3D and $t_{\rm KH}/t_{\rm cc}=1.30\pm0.23$ in 2D) than in solenoidal models ($t_{\rm KH}/t_{\rm cc}=0.58$ in 3D and $t_{\rm KH}/t_{\rm cc}=0.65\pm0.05$ in 2D). This means that the mixing of gas in high-density nuclei in compressive cloud models is much harder to achieve than in solenoidal cloud models, so that the downstream tails in compressive models are less turbulent when compared to solenoidal models (over the same time-scales).\par

Note that the results from both 3D and 2D cloud models on the overlapping time-scales ($0\leq t/t_{\rm cc}\leq2.5$) are consistent with one another. All the diagnostics presented in Figure \ref{Figure6} show similar trends and solenoidal-to-compressive ratios. However, mixing processes are more effective in 3D models owing to their spherical geometry, to the extra degree of freedom intrinsic to this configuration (see also \citealt{1995ApJ...454..172X,2012MNRAS.425.2212A,2018MNRAS.tmp.2921S}), and to the softer standard deviations of the initial 3D density PDFs that we set up for these models. This is in agreement with previous studies, which show that 3D spherical clouds are more accelerated and mixed than their 2D and 3D cylindrical counterparts (e.g., see \citealt{2018MNRAS.tmp.2921S}), while 2.5D and 3D spherical clouds evolve similarly until non-azimuthal instabilities start to grow in 3D simulations (e.g., see \citealt{2016MNRAS.457.4470P}).

\subsection{Mass loss and cloud destruction}
\label{subsec:CloudDestruction}
The efficiency of gas mixing has a direct impact on cloud destruction, i.e., on how much of the original cloud mass survives pressure-gradient forces and instabilities after certain time. The process of cloud destruction in wind-cloud models is associated with the generation of vortical motions via RT instabilities growing at the leading edge of the cloud. Thus, for comparison, we use Equation (\ref{RTtime}) to compute the ratio of the RT instability growth time-scales for solenoidal and compressive clouds at $t/t_{\rm cc}=1$. We find that RT instabilities with long wavelengths ($k_{\rm RT}\,r_{\rm cloud}\sim 1$) grow $\sim 1.3$ times slower in compressive models ($t_{\rm RT}/t_{\rm cc}=0.48$ in 3D and $t_{\rm RT}/t_{\rm cc}=0.65\pm0.12$ in 2D) than in solenoidal models ($t_{\rm RT}/t_{\rm cc}=0.36$ in 3D and $t_{\rm RT}/t_{\rm cc}=0.51\pm0.07$ in 2D). The lower effective acceleration of the compressive cloud models delays the growth of long-wavelength RT perturbations, thus prolonging the cloud lifetimes in these models.\par

A common practice to define the cloud destruction time, $t_{\rm des}$, is to set a density threshold and designate all gas with densities at or above that threshold as pertaining to the cloud. Since the cloud mixes with its surroundings, the amount of cloud gas with densities higher than the threshold goes down until only a small percentage of it can still be qualified as cloud material. When the percentage of cloud mass goes below that threshold, the cloud is considered as destroyed.\par

In this context, \cite{1994ApJ...420..213K,2006ApJS..164..477N} defined the destruction time as the time when the mass of the cloud (or of its largest fragment) has dropped by a factor of $1/e$. \cite{2015ApJS..217...24S} characterised cloud destruction using a mixing time-scale defined as the time when the mass of the cloud above $2\,\rho_{\rm wind}$ has dropped by $50$ per cent. In addition, \cite{2015ApJ...805..158S,2016ApJ...822...31B} defined several time-scales with the function\footnote{Note that we have adjusted the original function to be consistent with our definition of cloud-crushing time in Equation (\ref{eq:CloudCrushing}). The same adjustment applies to Equations (\ref{eq:Distance}) and (\ref{eq:Speed}) below.}

\begin{equation}
t=0.5\,\alpha\,t_{\rm cc}\sqrt{1+M_{\rm wind}}
\label{eq:TimeScalingMach}
\end{equation}

\noindent to describe cloud destruction, namely $t_{90}$, $t_{75}$, $t_{50}$, and $t_{25}$, as the times when the mass of the cloud above $\bar{\rho}_{\rm cloud}/3$ has dropped to $90$, $75$, $50$, and $25$ per cent, respectively. They found that $\alpha=1.75$, $2.5$, $4$, and $6$ for $t_{90}$, $t_{75}$, $t_{50}$, and $t_{25}$, respectively. In hydrodynamical models the cloud destruction time depends on the Mach number of the shock/wind, the cloud density contrast, the level of environmental turbulence, and whether a shock or a wind is considered (e.g., see \citealt{2009MNRAS.394.1351P}; \citealt*{2010MNRAS.405..821P}; \citealt{2016MNRAS.457.4470P,2017MNRAS.470.2427G,2018MNRAS.476.2209G}). In adiabatic simulations clouds are typically destroyed on time-scales of the order of $t_{\rm des}/t_{\rm cc}\sim1.5-4$ (e.g, see \citealt{1994ApJ...420..213K}; \citealt{2002ApJ...576..832P}; \citealt{2006ApJS..164..477N}), while in models with radiative cooling clouds survive for longer time-scales, of the order of $t_{\rm des}/t_{\rm cc}\sim4-12$ (e.g., see \citealt{2005AA...443..495M,2015ApJ...805..158S,2017ApJ...834..144S}).\par

\begin{figure*}
\begin{center}
  \begin{tabular}{c c c}
          \hspace{-0.2cm}\resizebox{58mm}{!}{\includegraphics{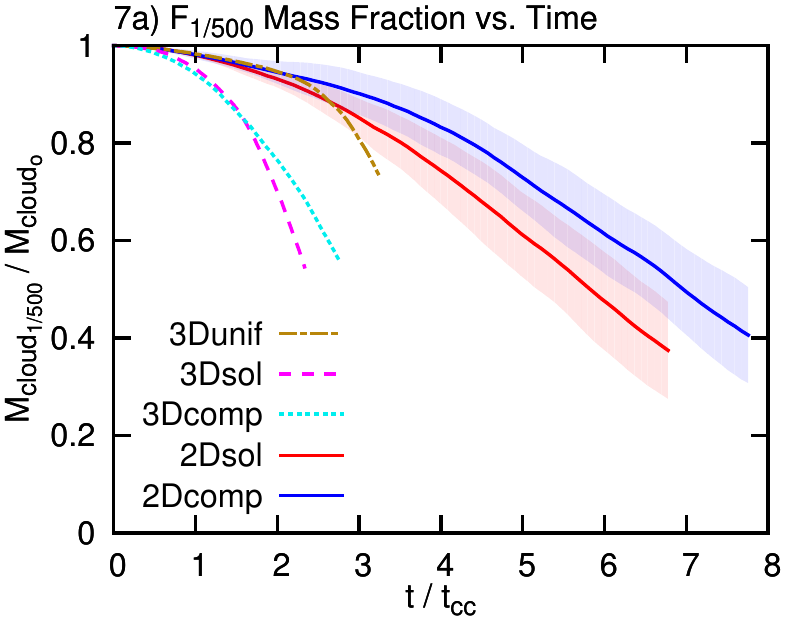}} & \hspace{-0.4cm}\resizebox{58mm}{!}{\includegraphics{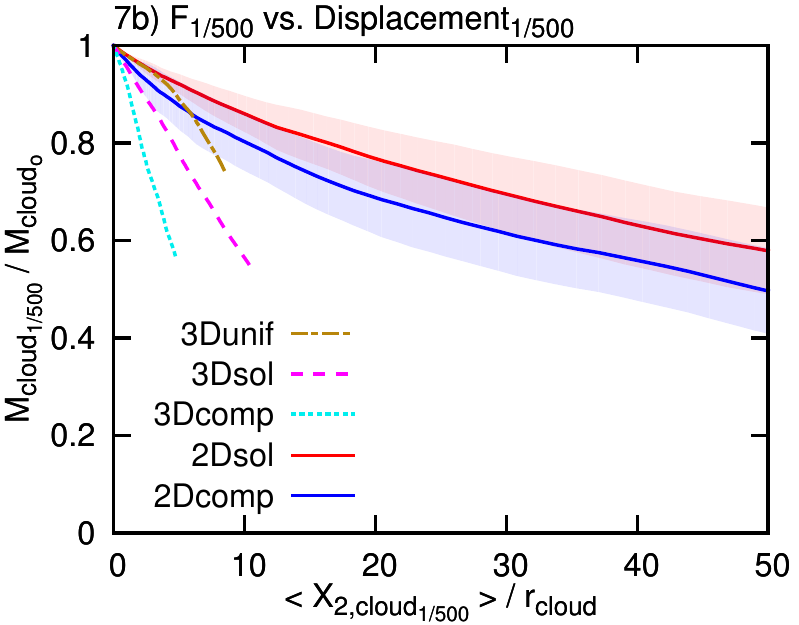}} & \hspace{-0.4cm}\resizebox{58mm}{!}{\includegraphics{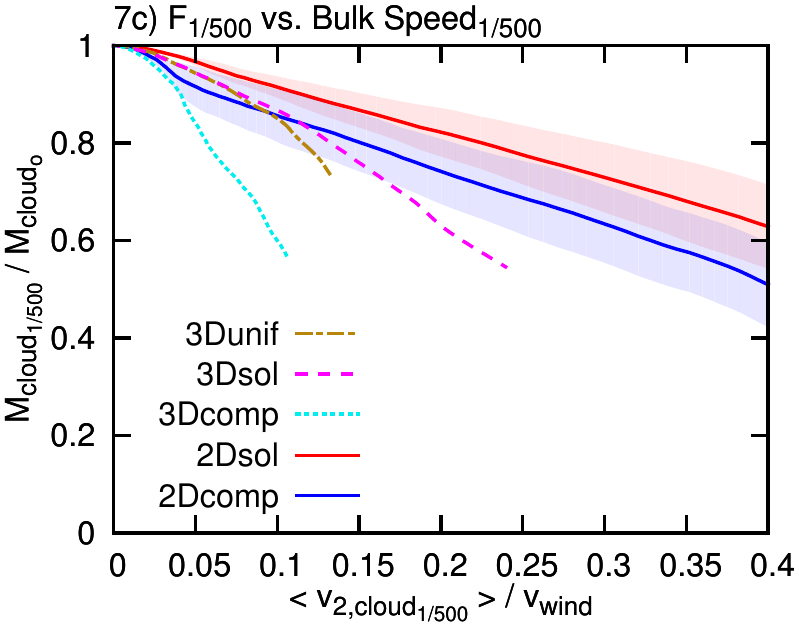}}\\
          \hspace{-0.2cm}\resizebox{58mm}{!}{\includegraphics{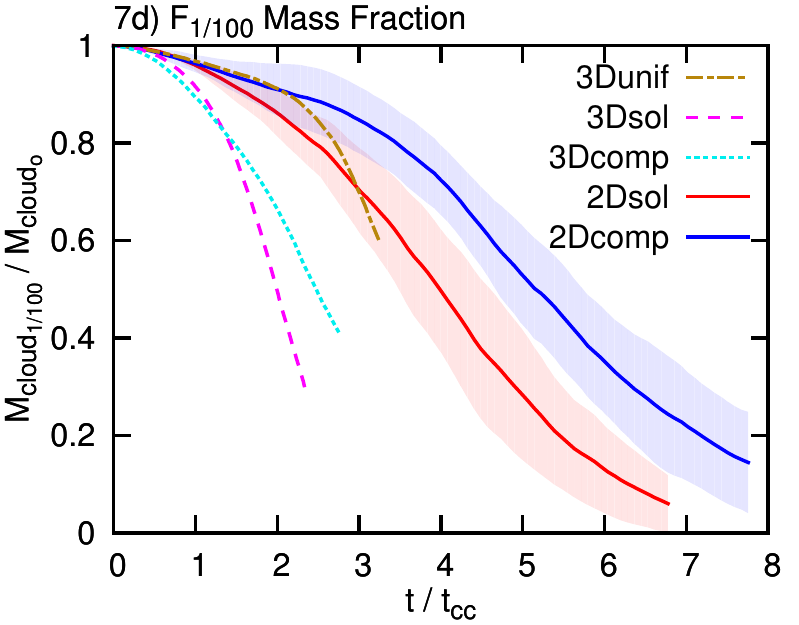}} & \hspace{-0.4cm}\resizebox{58mm}{!}{\includegraphics{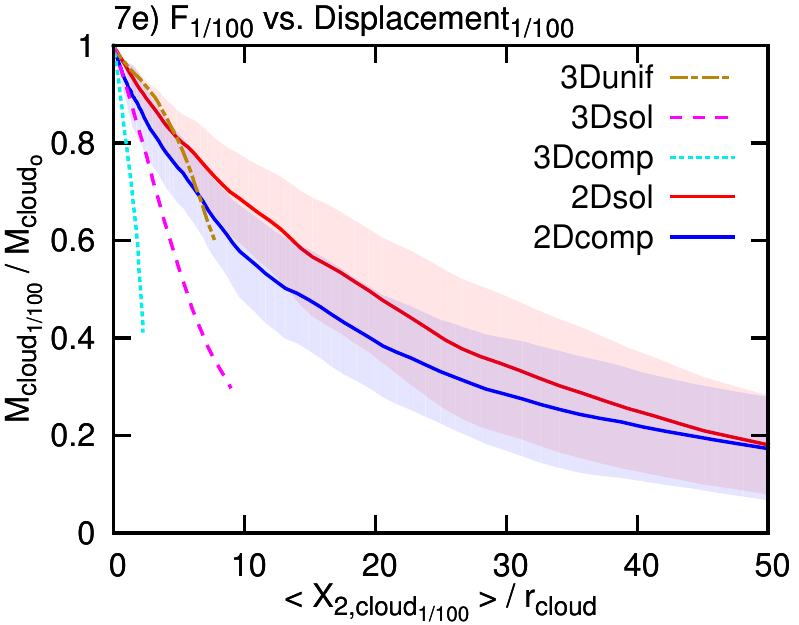}} & \hspace{-0.4cm}\resizebox{58mm}{!}{\includegraphics{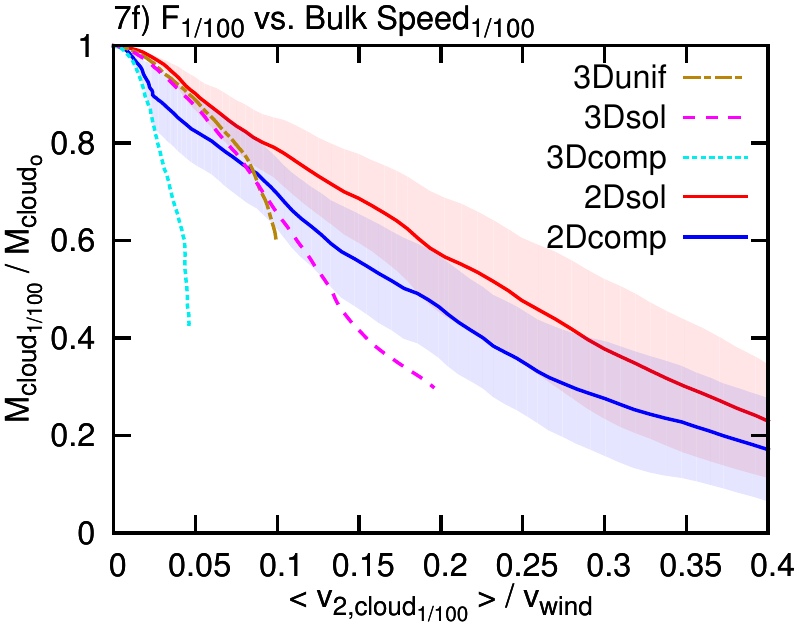}}\\
          \hspace{-0.2cm}\resizebox{58mm}{!}{\includegraphics{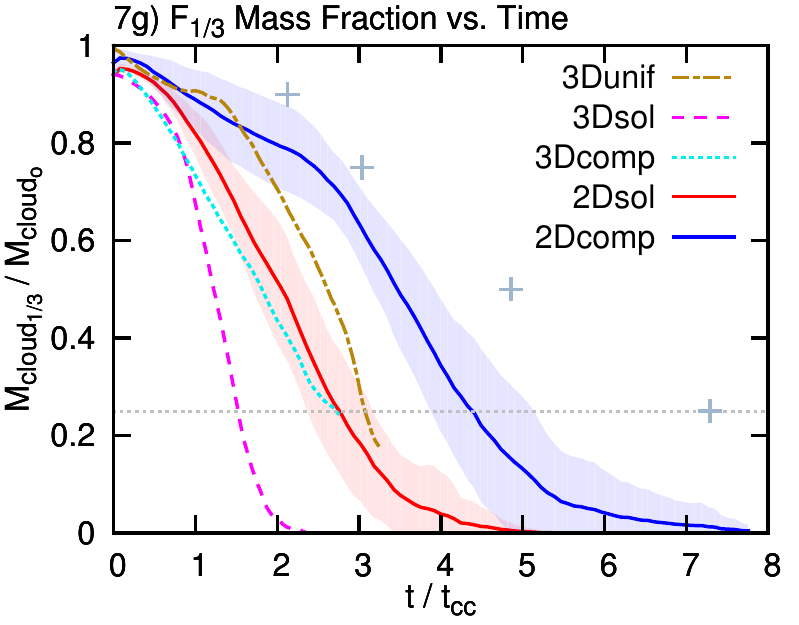}} & \hspace{-0.4cm}\resizebox{58mm}{!}{\includegraphics{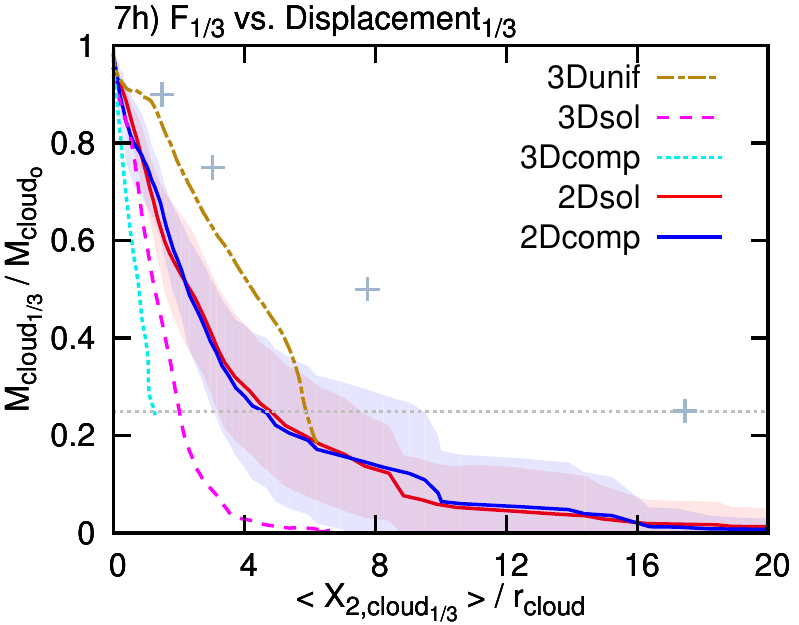}} & \hspace{-0.4cm}\resizebox{58mm}{!}{\includegraphics{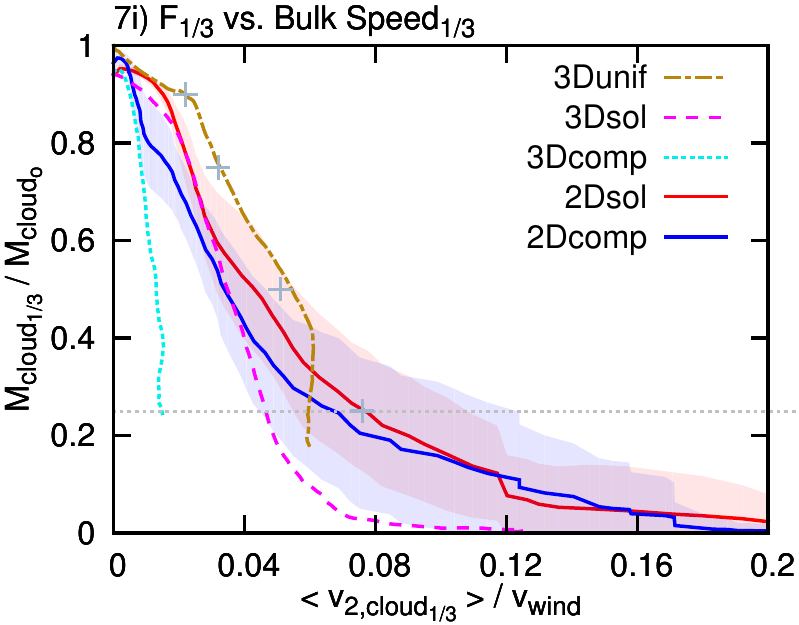}}\\
          \hspace{-0.2cm}\resizebox{58mm}{!}{\includegraphics{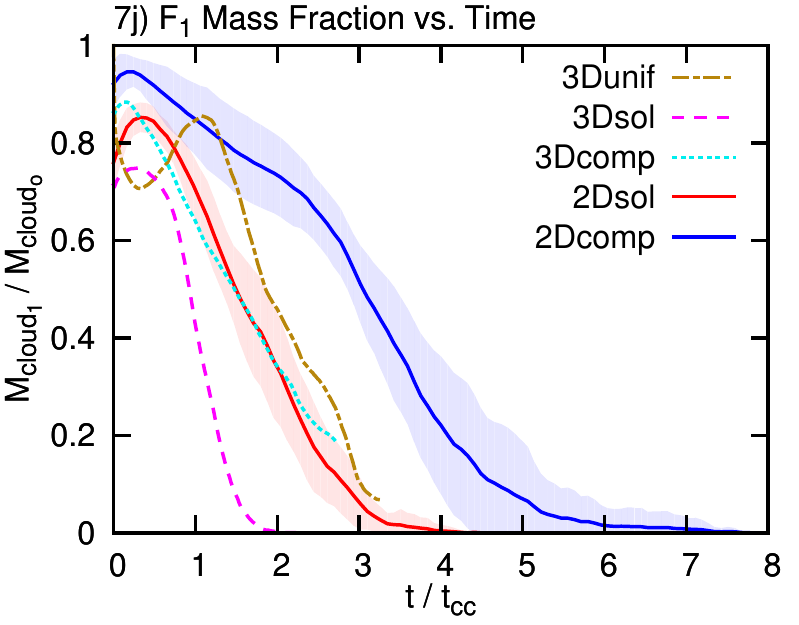}} & \hspace{-0.4cm}\resizebox{58mm}{!}{\includegraphics{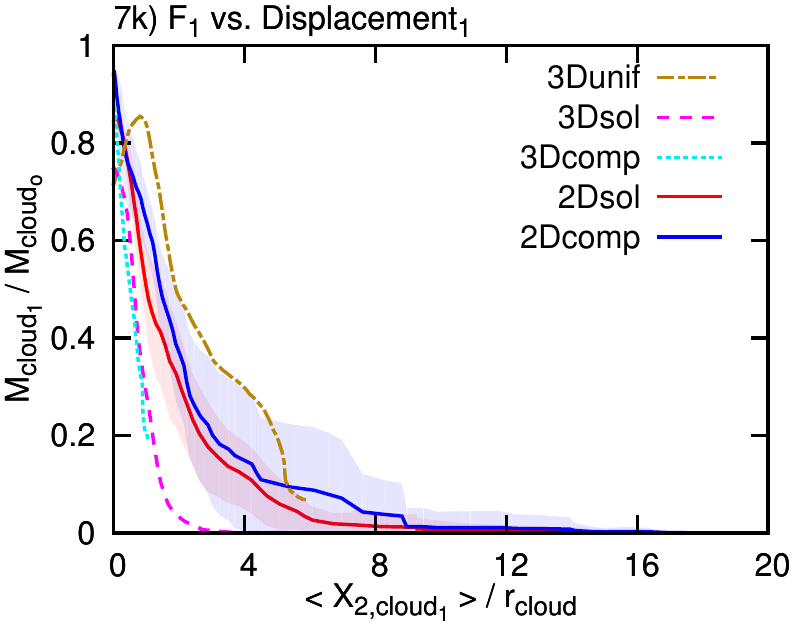}} & \hspace{-0.4cm}\resizebox{58mm}{!}{\includegraphics{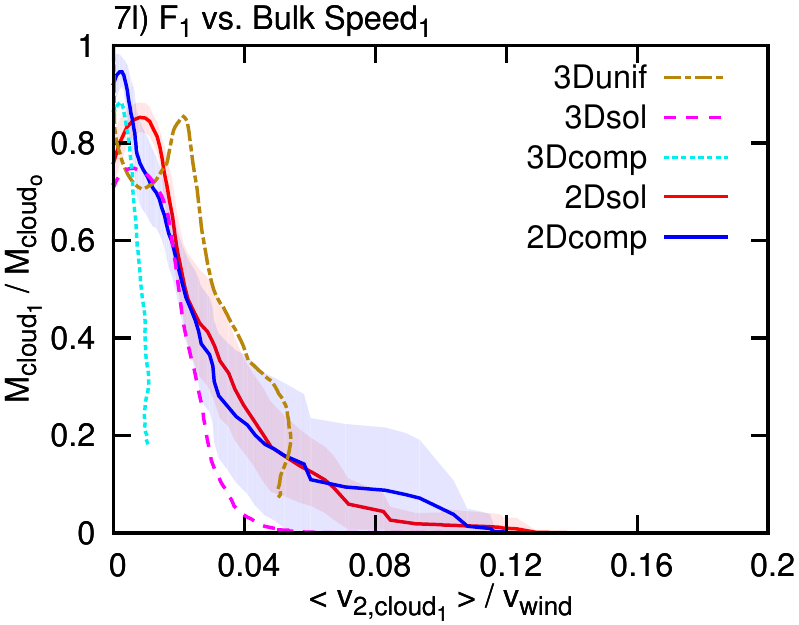}}\vspace{-0.25cm}\\
  \end{tabular}
  \caption{Evolution of cloud mass fractions at or above $\bar{\rho}_{\rm cloud}/500$, $\bar{\rho}_{\rm cloud}/100$, $\bar{\rho}_{\rm cloud}/3$, and $\bar{\rho}_{\rm cloud}$ as a function of time (left-hand side panels), travelled distance (middle panels), and bulk speed (right-hand side panels). Fractal clouds are destroyed earlier, and are less entrained than the uniform cloud at the destruction time. Within the fractal cloud sample, solenoidal clouds lose mass faster, are destroyed earlier, travel larger distances, and acquire higher bulk speeds than compressive clouds before destruction. While cloud survival mainly depends on the growth of dynamical instabilities, there is a correlation between the standard deviation of the initial density PDF in the clouds and entrainment efficiency: the wider the initial density distribution in the clouds, the harder it is for dense gas in them to become entrained in the wind. In the $F_{1/3}$ panels, the grey crosses show the predictions from the semi-analytical functions introduced by \citealt{2015ApJ...805..158S}, and the grey dotted lines indicate the mass-fraction threshold at which cloud destruction occurs.}
  \label{Figure7}
\end{center}
\end{figure*}

In order to compare our models with the results from previous works, we compute the fractions of cloud mass above different density thresholds. By doing this, we are also able to study the evolution and dynamics of cloud gas at different densities. Figure \ref{Figure7} presents the evolution of cloud mass fractions at or above the following density thresholds: $F_{1/500}$, $F_{1/100}$, $F_{1/3}$, and $F_{1}$, as a function of time (left-hand side panels), travelled distance (middle panels), and bulk speed (right-hand side panels). Following \cite{2015ApJ...805..158S} we define our clouds as destroyed when $F_{1/3}=0.25$, i.e., when only $25$ per cent of the initial cloud mass has densities above $1/3$ of the original average density in the cloud, $\bar{\rho}_{\rm cloud}$.\par

The left-hand side panels of Figure \ref{Figure7}, in both 3D and 2D, show that compressive clouds lose mass at a lower rate than solenoidal clouds regardless of the reference density threshold being considered. As the density threshold goes up, the absolute differences in mass between solenoidal and compressive models become more accentuated. The 3D solenoidal cloud is destroyed at $t/t_{\rm cc}=1.5$, while the 3D compressive cloud is destroyed at $t/t_{\rm cc}=2.7$. In addition, the 2D solenoidal clouds are destroyed, on average, at $t/t_{\rm cc}=2.8$, while the 2D compressive clouds are disrupted at $t/t_{\rm cc}=4.3$. Thus, on average, compressive clouds are destroyed $\sim 1.5-1.8$ times later than solenoidal clouds, and we also confirm earlier results by \cite{1995ApJ...454..172X,2012MNRAS.425.2212A,2018MNRAS.tmp.2921S} and find that 3D clouds are mixed and destroyed faster than 2D clouds, owing to dynamical instabilities being more effective in 3D.\par 

Figure \ref{Figure7} also reveals that 3D fractal clouds, in general, lose mass faster than the uniform cloud, which is only destroyed at $t/t_{\rm cc}=3.1$. Thus, 3D solenoidal and compressive fractal clouds are destroyed $\sim 2$ and $\sim 1.1$ times faster, respectively, than the uniform cloud. This result is in agreement with \cite{2009ApJ...703..330C,2015ApJS..217...24S,2017ApJ...834..144S}, who found that clouds with log-normal density distributions were disrupted and fragmented faster than uniform clouds. To reconcile the differences in disruption time-scales in uniform and fractal cloud models, \cite{2015ApJS..217...24S,2017ApJ...834..144S} (in the context of the shock-cloud formalism introduced by \citealt{1994ApJ...420..213K}) proposed that the initial median (instead of the average) cloud density be used when computing the cloud-crushing time-scale in Equation (\ref{eq:DensityContrast}).\par

We find that this redefinition works for fractal solenoidal clouds, but it fails to do so for compressive cloud models, as the latter have lower initial median densities than solenoidal clouds. We think that the classical definition of cloud-crushing time (originally envisaged for uniform cloud models) can still be used for fractal clouds and can be understood as an average value of a distribution of cloud-crushing times intrinsic to the gas density distribution. The differences seen in uniform versus fractal cloud models are physically motivated by distinctly-growing dynamical instabilities, which are associated to the evolution of the cloud-to-wind density contrast and the cloud acceleration.

\subsection{Dense gas entrainment}
\label{subsec:Entrainment}
The analyses in previous sections show that fractal clouds are more easily disrupted than uniform clouds, and, within the fractal sample, that compressive clouds live longer than solenoidal clouds. Does this different behaviour translate into fractal clouds being more effectively entrained in the wind than uniform clouds, or vice versa? Are the dynamical properties of uniform and fractal solenoidal and compressive clouds different at their individual destruction times? To answer these questions, we first study the distances and bulk speeds of cloud gas above different density thresholds as functions of mass loss rate (instead of time), and then we compare the density-velocity mass distribution of gas in the clouds at their respective destruction times.\par

In 3D, the middle and right-hand side panels of Figure \ref{Figure7} show that, at any given mass-loss value (for all mass fractions), the uniform cloud reaches larger distances and bulk speeds than its fractal counterparts, and that the solenoidal cloud attains higher displacements and velocities than the compressive cloud. In particular, panels 7h and 7i show that, at the destruction time, cloud gas with densities above $\bar{\rho}_{\rm cloud}/3$ has reached distances and bulks speeds of $\sim 5.9\,r_{\rm cloud}$ and $\sim 0.06\,v_{\rm wind}$ in the 3D uniform cloud, $\sim 2\,r_{\rm cloud}$ and $\sim 0.05\,v_{\rm wind}$ in the 3D solenoidal cloud, and $\sim 1.2\,r_{\rm cloud}$ and $\sim 0.02\,v_{\rm wind}$ in the 3D compressive cloud. Thus, we conclude that entrainment of dense gas in the wind is significantly more efficient in uniform clouds than in either of the fractal clouds, and is more efficient in solenoidal clouds than in compressive clouds. Even though it takes longer for the wind to accelerate uniform clouds, dense gas in such clouds is $\sim 1.3$ and $\sim 4$ times faster than in solenoidal and compressive clouds, respectively, at the destruction time.\par

In 2D, the $F_{1/500}$ mass fraction shows a behaviour similar to the 3D models, i.e., compressive clouds attain lower distances and bulks speeds than solenoidal clouds at a given mass-loss value. However, as the density threshold goes up the difference between solenoidal and compressive models becomes smaller. For instance, at the destruction time, cloud gas with densities above $\bar{\rho}_{\rm cloud}/3$ has reached distances and bulks speeds of $\sim 5.2\pm1.5\,r_{\rm cloud}$ and $\sim 0.08\pm0.03\,v_{\rm wind}$ in 2D solenoidal clouds, and $\sim 4.2\pm1.3\,r_{\rm cloud}$ and $\sim 0.06\pm0.03\,v_{\rm wind}$ in 2D compressive clouds. Thus, for the $F_{1/3}$ mass fraction, the average distances and bulks speeds attained by dense gas are, on average, also higher in 2D solenoidal clouds compared to 2D compressive clouds. However, the one-standard-deviation limits of both diagnostics partially overlap, suggesting that the dynamics of dense gas (particularly in the compressive cases) is different in 3D and 2D geometries. Dense gas can accelerate more in 2D than in 3D. In 2D, wind gas piles up in front of dense nuclei in the clouds, thus aiding momentum transfer, while, in 3D, wind gas can slip trough the sides of these nuclei, which become a footpoint with very low momentum.\par

\begin{figure*}
\begin{center}
  \begin{tabular}{c c c c}
\multicolumn{1}{l}{\hspace{-1.5mm}8a) 3Dunif \hspace{+25mm}$t_{\rm des}/t_{\rm cc}=3.1$} &\multicolumn{1}{l}{\hspace{-1.5mm}8b) 3Dsol \hspace{+21mm}$t_{\rm des}/t_{\rm cc}=1.5$} & \multicolumn{1}{l}{\hspace{-1.5mm}8c) 3Dcomp \hspace{+19mm}$t_{\rm des}/t_{\rm cc}=2.7$} & \hspace{-5.6mm}$\frac{M}{M_{\rm cloud,0}}$\\    
       \hspace{-0.26cm}\resizebox{58mm}{!}{\includegraphics{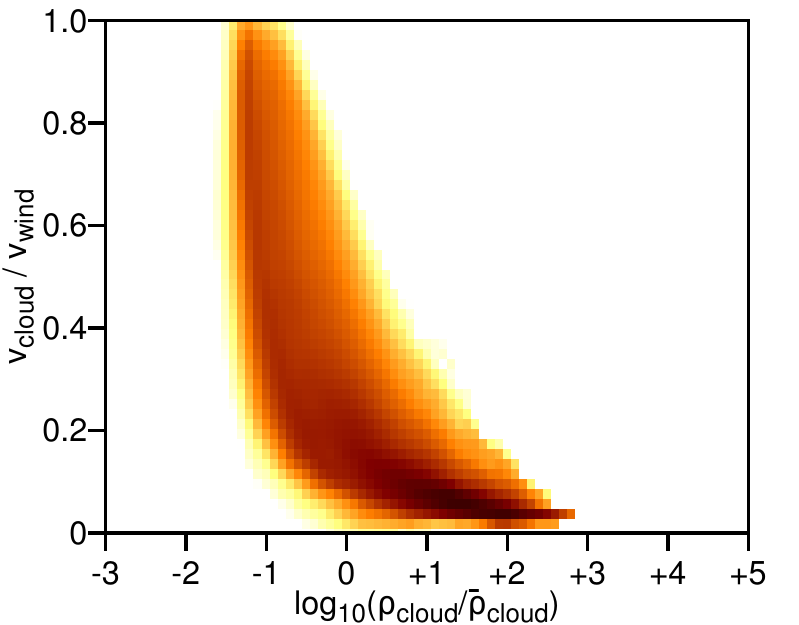}} & \hspace{-0.63cm}\resizebox{58mm}{!}{\includegraphics{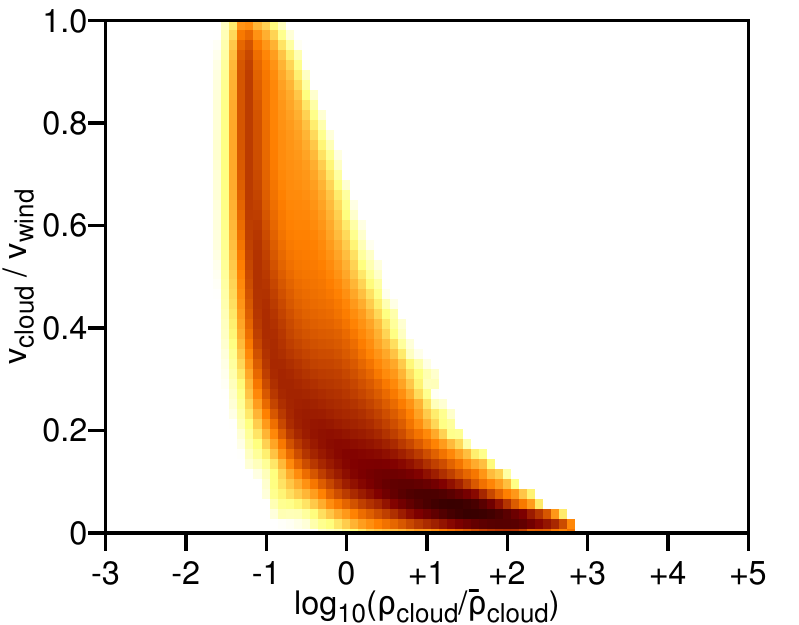}} & \hspace{-0.63cm}\resizebox{58mm}{!}{\includegraphics{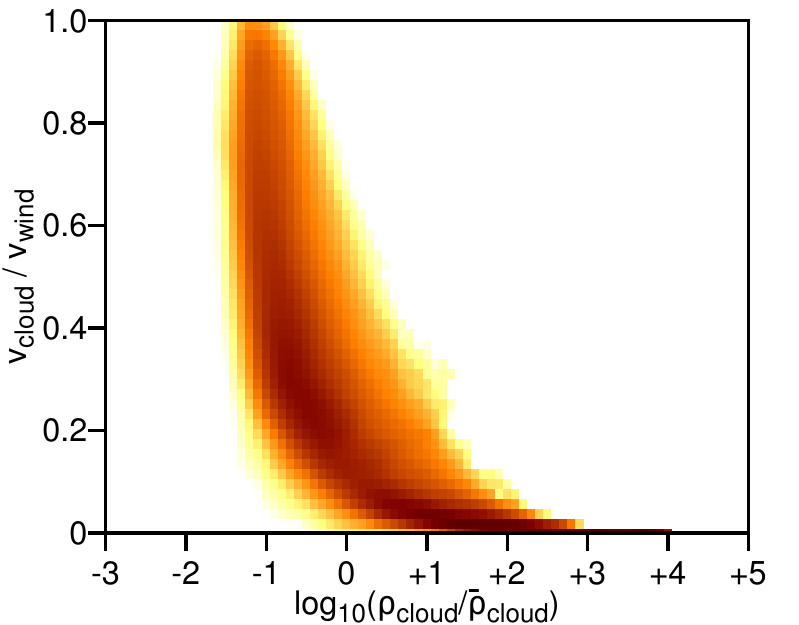}} & \\
 \multicolumn{1}{l}{\hspace{-1.5mm}8d) Bulk speed$_{1/3}$ vs. Standard Deviation} & \multicolumn{1}{l}{\hspace{-1.5mm}8e) 2Dsol \hspace{+21mm}$t_{\rm des}/t_{\rm cc}=2.8$}  & \multicolumn{1}{l}{\hspace{-1.5mm}8f) 2Dcomp \hspace{+19mm}$t_{\rm des}/t_{\rm cc}=4.3$} &\\     
    \hspace{-0.55cm}\resizebox{58mm}{!}{\includegraphics{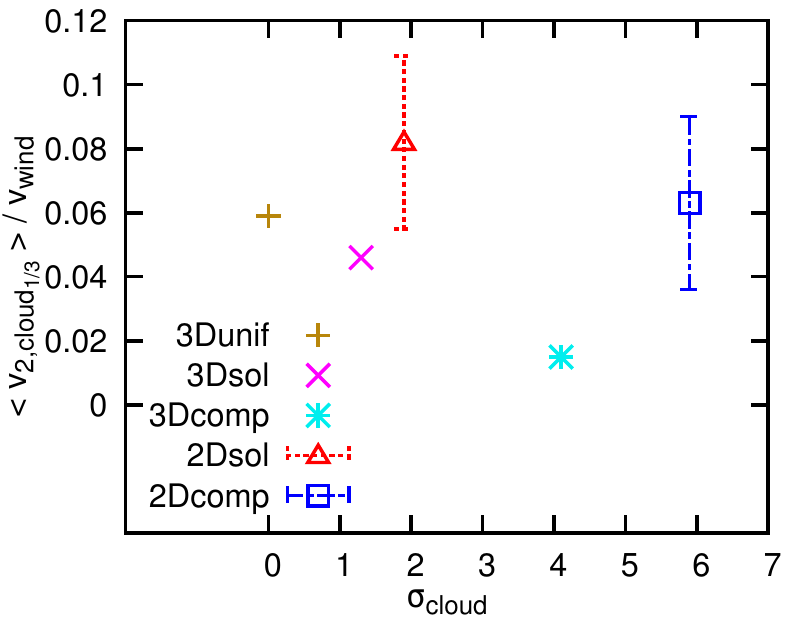}}   & \hspace{-0.63cm}\resizebox{58mm}{!}{\includegraphics{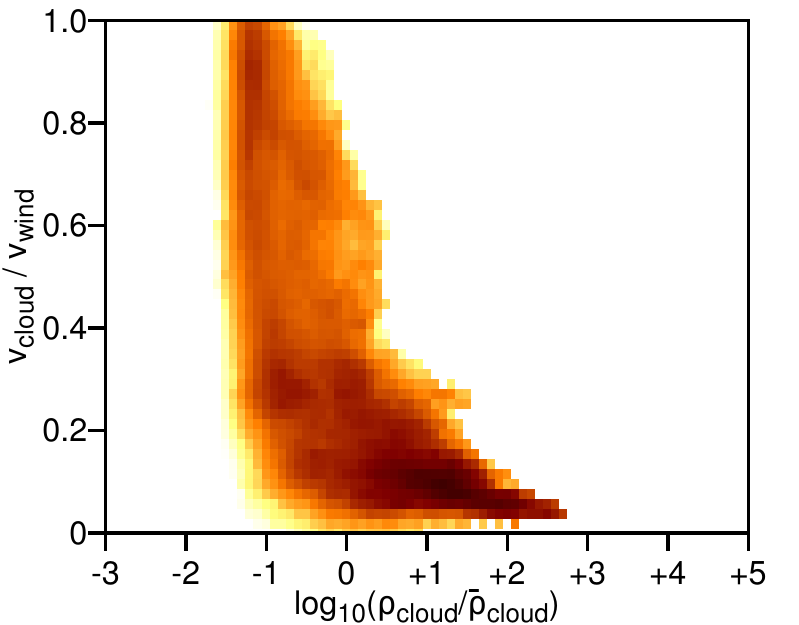}} & \hspace{-0.63cm}\resizebox{58mm}{!}{\includegraphics{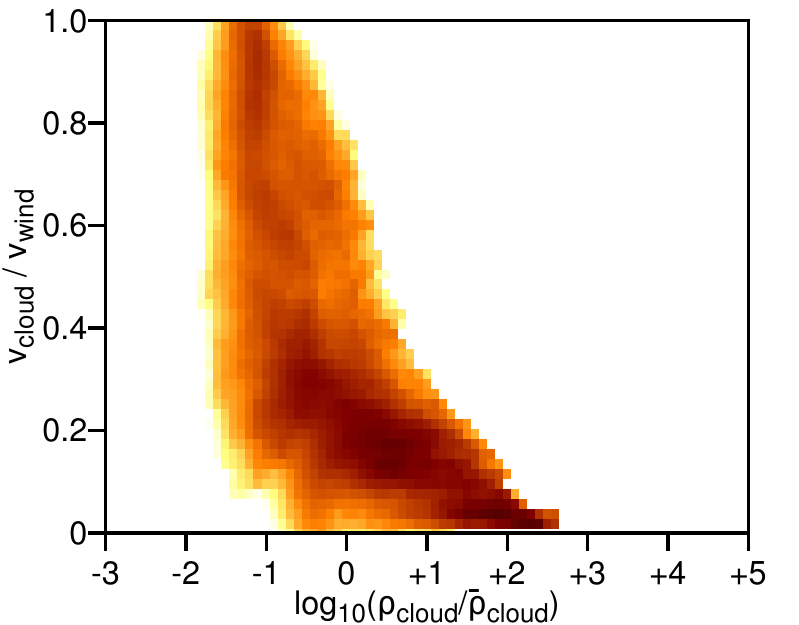}} &  \multirow[t]{3}{*}{\hspace{-0.51cm}{\includegraphics[width=8.8mm]{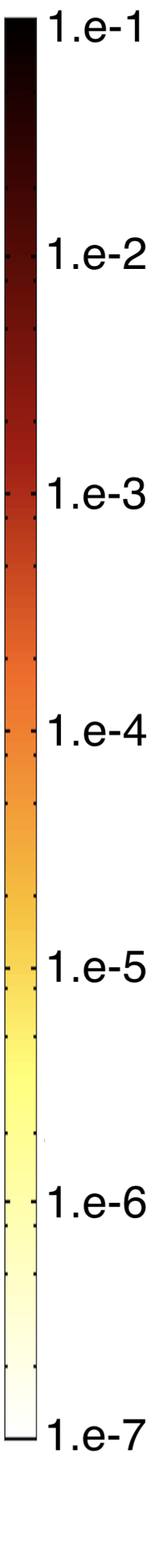}}}\\  
  \end{tabular}
  \caption{The top panels 8a, 8b, and 8c show the mass-weighted phase diagrams of velocity versus density in 3D uniform, solenoidal, and compressive cloud models at their respective destruction times. The bottom panels 8e and 8f show the density-velocity histograms for fiducial 2D solenoidal and compressive fractal cloud models, respectively, also at their destruction times. Overall, entrainment of dense gas is more effective in uniform clouds than in solenoidal or compressive clouds. Within the fractal sample, dense gas has accelerated more effectively in solenoidal cloud models than in compressive cloud models. Panel 8d shows that there is a correlation between the bulk speeds of gas denser than $\bar{\rho}_{\rm cloud}/3$ (measured at the destruction time) and the initial standard deviation of the density distributions in the clouds. Very dense gas with low momentum only survives in compressive cloud models.}
  \label{Figure8}
\end{center}
\end{figure*}

Figure \ref{Figure8} shows mass-weighted phase diagrams of the cloud gas velocity as a function of its density. These histograms allow us to further investigate entrainment efficiency. The top panels 8a, 8b, and 8c show the density-velocity histograms for the uniform, solenoidal, and compressive cloud models in 3D, at their respective destruction times (i.e., $t_{\rm des}/t_{\rm cc}=3.1$, $1.5$, and $2.7$). These diagrams confirm that entrainment of gas with $\rho\geq \bar{\rho}_{\rm cloud}/3$ is more efficient in the uniform cloud model than in either of the fractal cloud models. They also indicate that dense gas in the solenoidal cloud has higher velocities than in the compressive cloud, and that very dense gas ($\rho\geq 10^3\,\bar{\rho}_{\rm cloud}$) only survives in the compressive cloud model, but it has a very low momentum.\par

Panels 8e and 8f show the evolution of density-velocity histograms for fiducial solenoidal and compressive cloud models in 2D, at their respective destruction times (i.e., $t_{\rm des}/t_{\rm cc}=2.8$, and $4.3$). We find the same results as in the 3D scenarios, i.e. solenoidal cloud models favour entrainment of gas with densities $\rho\geq \bar{\rho}_{\rm cloud}/3$, while compressive cloud models retain gas with low momentum. Contrary to the results in 3D models, no gas with $\rho\geq 10^3\,\bar{\rho}_{\rm cloud}$ survives disruption in this 2D compressive cloud model. In general, however, both 3D and 2D models agree in that compressive clouds retain $\sim 20$ per cent of the original cloud mass in regions with densities $\geq \bar{\rho}_{\rm cloud}$ at the destruction time, while $<10$ per cent of gas with such densities survives in solenoidal and uniform models (see panels 7j, 7k, and 7l of Figure \ref{Figure7}).\par

Overall, the above results indicate that dense gas in both 3D and 2D fractal clouds can only travel modest distances and reach marginal fractions of the wind speed before destruction, thus implying that entrainment of high-density gas in galactic winds is overall not efficient in purely hydrodynamical wind-cloud set-ups (which do not include magnetic fields and turbulent velocity fields). In addition, entrainment is poorer in fractal cloud models than in the more widely studied uniform cloud models, and it becomes more difficult to attain as the standard deviation of the initial density PDF in the clouds distribution increases. Thus, while cloud survival mainly depends on the growth rates of KH and RT instabilities, the efficiency of dense gas entrainment in the wind is correlated with the initial density distribution of the cloud: the wider the initial density PDF, the less efficient entrainment becomes (see panel 8d of Figure \ref{Figure8}).

\subsection{Wind mass loading}
\label{subsec:MassLoading}
Studying how wind mass loading operates in different wind-cloud set-ups is also relevant to understanding the prevalence of dense gas phases in multi-phase galactic outflows. As mentioned in Section \ref{sec:Intro}, for thermal instabilities (see \citealt{2016MNRAS.455.1830T}) or condensation of warm gas (see \citealt{2018MNRAS.480L.111G}) to account for dense gas (re)formation, the wind has to be sufficiently mass loaded. If mass loading is efficient, i.e., if the wind gathers a large amount of mass from the destruction of dense gas clouds at the launching site, then thermal instabilities can operate at smaller radii from the launching site, and condensation of mixed, warm gas in the cloud tails can also be more effective, thus favouring the emergence of a new, high-velocity, dense gas phase in the wind. In this context, we compare, in our different models, the mass fractions of cloud material above different density thresholds that have been loaded into the wind at the destruction time.\par

The left-hand side panels of Figure \ref{Figure7} reveal that in 3D, at the destruction time, $\sim 43$ per cent of cloud material below $\bar{\rho}_{\rm cloud}/500$ has been loaded into the wind in the compressive model, while only $\sim 21$ per cent and $\sim 13$ per cent have been loaded in the uniform and solenoidal models, respectively. In 2D, we find similar patterns for wind mass loading: $\sim 20\pm6$ per cent of cloud material below $\bar{\rho}_{\rm cloud}/500$ has been loaded into the wind in compressive models, while only $\sim 13\pm4$ per cent has been loaded in solenoidal models at the destruction time. These results indicate that wind mass loading is much more efficient in compressive cloud models than in the other two cases.\par

Since the density distribution in compressive clouds is wider than in solenoidal models, low-density gas in compressive models is removed from the cloud and loaded into the wind very efficiently, while high-density gas is much more difficult to accelerate owing to its large column density. In solenoidal models, on the other hand, low and high densities are initially closer to the average cloud density, $\bar{\rho}_{\rm cloud}$, so momentum is transferred more uniformly to gas at different densities. This reduces mass loading of low-density gas into the wind in solenoidal models, but it facilitates entrainment of high-density gas (as discussed in Section \ref{subsec:Entrainment}), compared to compressive models. Thus, considering the turbulent nature of interstellar clouds in hydrodynamical wind-cloud models facilitates wind mass loading and can explain the survival of very dense gas (either advected or reformed) in galactic winds.

\subsection{Quasi-isothermal vs. radiative clouds}
\label{subsec:Quasi-radiative}
In this section we compare quasi-isothermal and radiative cloud models. How good are quasi-isothermal models at mimicking the dynamics of radiative wind-swept clouds? The data points shown in the middle and right-hand side $F_{1/3}$ panels of Figure \ref{Figure7} correspond to the predicted distances and cloud velocities using the formalism derived by \cite{2015ApJ...805..158S,2016ApJ...822...31B} for radiative, uniform clouds. They found the following functional fits for the distance and bulk speed of the clouds:

\begin{equation}
\frac{d_{\rm cloud}}{r_{\rm cloud}}=0.34\,\alpha^2\,(1+{\cal M}_{\rm wind})^{1-\beta}
\label{eq:Distance}
\end{equation}

\begin{equation}
\frac{v_{\rm cloud}}{v_{\rm wind}}=0.68\,\alpha\,\chi^{-0.5}(1+{\cal M}_{\rm wind})^{0.5-\beta}
\label{eq:Speed}
\end{equation}

\noindent where $\alpha$ is defined as in Equation (\ref{eq:TimeScalingMach}) for each evolutionary stage, and $\beta=0.8$ for $t_{90}$ and $t_{75}$, and $\beta=0.9$ for $t_{50}$ and $t_{25}$. For cloud distances, we find a good agreement between quasi-isothermal and radiative clouds at early times, $t_{90}$ and $t_{75}$, but at late times, $t_{50}$ and $t_{25}$, our 3D quasi-isothermal, uniform cloud reaches lower distances than what the above expressions predict for ${\cal M}_{\rm wind}=4.9$. For cloud bulk speeds, the overall trend of the curve is well captured by our quasi-isothermal model up to $t_{50}$. The discrepancies at late times are due to radiative cooling producing a stronger cloud compression, fragmentation, and dense gas reformation in the radiative models by \cite{2015ApJ...805..158S} than what we can capture with our quasi-isothermal model. Thus, our self-similar, quasi-isothermal approximation mimics well the dynamics of radiative clouds at early stages of the evolution, but it does not capture the complex kinematics of radiative warm, mixed gas downstream.

\section{Caveats and future work}
\label{sec:Future}
The primary goal of this paper is to show how changing the initial PDF of the cloud density influences the dynamics and survival of wind-swept clouds. In order to isolate these effects, we concentrated our analysis on fractal cloud models that excluded ingredients that are known to be dynamically important in wind-cloud systems, such as radiative cooling, thermal conduction, self-gravity, and magnetic fields. In particular, we showed in Section \ref{subsec:Quasi-radiative} that our quasi-isothermal models allow us to mimic the shear-layer stability imparted to the cloud by radiative cooling (\citealt{2009ApJ...703..330C}), but we do not capture the physics of cooling-induced shattering (\citealt*{2002AA...395L..13M}; \citealt{2004ApJ...604...74F,2018MNRAS.473.5407M,2018MNRAS.tmp.2921S}).\par

Similarly, \cite{2016ApJ...822...31B,2017MNRAS.470..114A}; \cite*{2018ApJ...864...96C} showed that isotropic, electron thermal conduction is efficient in high-column density clouds and compresses them into compact filaments that survive for long dynamical time-scales. Given the morphological differences between uniform and fractal clouds seen in our hydrodynamic cloud models, we expect both the shattering process induced by cooling and the protective effects of thermal conduction to also act differently on solenoidal and compressive cloud populations. In addition, including self-gravity in wind-cloud models is also needed to study shock-triggered star formation in these systems (\citealt{2004ApJ...604...74F}; \citealt*{2013ApJ...774..133L}). We shall study systems of winds and fractal clouds including radiative cooling, self-gravity, and sink particles (\citealt{2010ApJ...713..269F}) in future work.\par

Magnetic fields have also been shown to delay cloud destruction by stabilising wind-cloud shear layers (\citealt{1994ApJ...433..757M,2008ApJ...680..336S,2017ApJ...845...69G,2016MNRAS.455.1309B,2018MNRAS.473.3454B,2018ApJ...865...64G}) and favour RT-induced sub-filamentation (\citealt{1999ApJ...527L.113G,2000ApJ...543..775G,2016MNRAS.455.1309B}). Since the density PDF is correlated with the magnetic field distribution in turbulent clouds (\citealt{2012ApJ...761..156F}), we also expect the evolution of wind-swept solenoidal and compressive fractal cloud models to be different. Thus, future magnetohydrodynamic studies of such clouds are also warranted.

\section{Conclusions}
\label{sec:Conclusions}
Wind-cloud models have been widely studied in the context of galactic winds, but clouds are usually assumed to have uniform or smooth density distributions. In this paper we have compared uniform and turbulent fractal clouds immersed in supersonic winds. This is the first study to look into the density distribution of clouds by taking into account the statistical properties of two regimes of interstellar supersonic turbulence, namely solenoidal (divergence-free) and compressive (curl-free). We have shown that the evolution of wind-cloud models in purely hydrodynamical cases depends on the initial distribution of the cloud density field, which is quantified by the initial standard deviation of the log-normal density PDF. We summarise the conclusions drawn from our study below:

\begin{enumerate}
	\item \textit{Uniform versus fractal clouds:} In agreement with \cite{2009ApJ...703..330C,2017ApJ...834..144S,2018MNRAS.473.3454B}, we found that log-normal, fractal clouds accelerate, mix, and are disrupted earlier than uniform clouds. Aided by their intrinsic porosity, internal refracted shocks can more easily propagate through fractal clouds, thus expanding their cross sectional areas and accelerating them earlier than their uniform counterparts.
	\item \textit{Solenoidal versus compressive fractal clouds:} Within the fractal cloud population we find that compressive wind-swept clouds are more confined, less accelerated, and have lower velocity dispersions than their solenoidal counterparts. While solenoidal clouds are rapidly and steadily disrupted, compressive clouds are supported for longer time-scales by high-density nuclei, which act as footpoints for long-lived, downstream filaments. 
	\item \textit{Evolution of the density PDF:} The density PDFs of wind-swept fractal clouds evolve with time. For diffuse gas the evolution is similar in both solenoidal and compressive regimes as mixing processes rapidly flatten the low-density tail of the PDF, thus dismantling the initial log-normality. For dense gas, on the other hand, the high-density tail of the PDFs in solenoidal models moves towards lower densities much faster than in compressive models (which retain dense gas in their nuclei). At the destruction time, very dense gas accounts for $\sim 20$ per cent of the original cloud mass in compressive clouds, while $<10$ per cent survives in the other two models.
	\item \textit{Cloud survival:} Cloud survival depends on how pressure-gradient forces and dynamical KH and RT instabilities stretch and strip mass from the cloud. Uniform clouds are destroyed $\sim 2$ and $\sim 1.1$ times later than solenoidal and compressive clouds, respectively. Within the fractal cloud population, compressive clouds are less prone to both KH and RT instabilities than their solenoidal counterparts (in both 2D and 3D), owing to their higher density nuclei and lower accelerations, respectively. Thus, compressive clouds develop less turbulence and mix with the ambient medium later than their solenoidal counterparts. This translates into compressive clouds surviving $\sim 1.5-1.8$ times longer than solenoidal clouds.
	\item \textit{Dense gas entrainment:} The efficiency of gas entrainment depends on the initial standard deviation of the cloud density distribution: the wider the initial density PDF, the harder it is for dense gas in the clouds to become entrained in the wind. At the destruction time, in 3D, dense gas reaches distances of $\sim 5.9\,r_{\rm cloud}$ and $\sim 0.06\,v_{\rm wind}$ in uniform clouds ($\sigma_{\rm cloud}=0$), $\sim 2.0\,r_{\rm cloud}$ and $\sim 0.05\,v_{\rm wind}$ in solenoidal clouds ($\sigma_{\rm cloud}=1.3$), and $\sim 1.2\,r_{\rm cloud}$ and $\sim 0.02\,v_{\rm wind}$ in compressive clouds ($\sigma_{\rm cloud}=4.1$). Thus, entrainment of dense gas in supersonic winds is: a) more effective in uniform cloud models than in either of the fractal cloud models, and b) more effective in solenoidal models than in compressive models.
	\item \textit{Wind mass loading:} The efficiency of wind mass loading depends on both the growth rate of dynamical instabilities and the initial standard deviation of the cloud density distribution. At the destruction time, in 3D, $\sim 43$ per cent of low-density gas has been loaded into the wind in compressive models, while only $\sim 21$ per cent and $\sim 13$ per cent have been loaded in uniform and solenoidal models, respectively. Thus, mass loading is more effective in compressive clouds than in the other two models.
	\item \textit{Galactic winds:} Taking into account the turbulent nature of interstellar clouds can explain the survival of very dense gas in hot flows, but it makes entrainment even more difficult than in uniform clouds. Moreover, wind mass loading, which can seed thermal instabilities and allow dense gas reformation downstream, is more effective in fractal models. We have linked the efficiency of entrainment and mass loading to the initial density distributions in the clouds, and we have shown that compressive clouds, which are expected to inhabit star-forming regions and shocked gas in galactic winds, retain high-density gas even after destruction.
	\item \textit{Numerical resolution and statistics:} Our choice of numerical resolutions of $R_{64}$ and $R_{128}$ for the 3D and 2D sets, respectively, adequately capture the evolution of fractal clouds for the parameter space explored in this paper (see Appendix \ref{sec:AppendixB}). In addition, the results presented hold for 3D and 2D geometries, and for differently-seeded and distinctly-oriented 2D fractal clouds.
\end{enumerate}

\section*{Acknowledgements}
We thank the anonymous referee for a timely and constructive report. This numerical work was supported by the Deutsche Forschungsgemeinschaft (DFG) via grant BR2026125 and by the Australian National Computational Infrastructure and the Pawsey Supercomputing Centre via grant ek9, with funding from the Australian Government and the Government of Western Australia, in the framework of the National Computational Merit Allocation Scheme and the ANU Allocation Scheme. Some of the 2D simulations and data post-processing were carried out on the computing cluster made available to us by the Ecuadorian Network for Research and Education RED CEDIA, and on the Hummel supercomputer at Universit\"at Hamburg in Germany. W.~B-B.~thanks H. D\'enes for helpful discussions on the simulations and the National Secretariat of Higher Education, Science, Technology, and Innovation of Ecuador, SENESCYT. C.~F.~acknowledges funding provided by the Australian Research Council (Discovery Projects DP170100603 and Future Fellowship FT180100495), the Australia-Germany Joint Research Cooperation Scheme (UA-DAAD), the Leibniz Rechenzentrum and the Gauss Centre for Supercomputing (grants~pr32lo, pr48pi and GCS Large-scale project~10391), and the Partnership for Advanced Computing in Europe (PRACE grant pr89mu). This work has made use of the Pluto code (\url{http://plutocode.ph.unito.it}), the pyFC package (\url{https://bitbucket.org/pandante/pyfc}), the Functional Data Analysis R package \citep{RprojectFDA}, the VisIt visualisation software \citep{HPV:VisIt}, and the gnuplot program (\url{http://www.gnuplot.info}).

%%%%%%%%%%%%%%%%%%%%%%%%%%%%%%%%%%%%%%%%%%%%%%%%%%



\bibliographystyle{mnras}
\bibliography{wlady} % if your bibtex file is called example.bib

%\begin{thebibliography}{99}
%\end{thebibliography}

%%%%%%%%%%%%%%%%%%%%%%%%%%%%%%%%%%%%%%%%%%%%%%%%%%

%%%%%%%%%%%%%%%%% APPENDICES %%%%%%%%%%%%%%%%%%%%%

\appendix

\section{The wind-cloud and shock-cloud problems in the literature}
\label{sec:AppendixA}
In this Appendix, we show Table \ref{TableA1}, which contains a summary of the parameter space explored by previous numerical studies on wind-cloud and shock-cloud interactions. The reader is referred to \cite{2016PhDT.......154B,2016MNRAS.455.1309B,2018MNRAS.473.3454B} and \cite{2016MNRAS.457.4470P} for recent reviews of the literature on wind-cloud and shock-cloud interactions.

\begin{table*}\centering
\caption{Parameter space explored in previous studies and in this paper. Columns 1 and 2 indicate the references, dimensions and simulation types: hydrodynamic (HD), magnetohydrodynamic (MHD), or both (M/HD). Column 3 indicates the cloud geometry: spherical (Sph), cylindrical (Cyl), elliptical (Ell), elongated (Elo), non-uniform (Nun), fractal (Fra), or turbulent (Tur). Column 4 indicates the resolutions (number of cells per cloud radius, $R_{\rm x}$). Columns 5, 6, 7, and 8 indicate the polytropic indices ($\gamma$), density contrasts ($\chi$), Mach numbers (\textbf{$\cal M_{\rm wind}$}), and plasma betas ($\beta$). Column 9 indicates the magnetic field configuration: tangled (Ta), turbulent (Tu), poloidal (Po), toroidal (To), or aligned (Al), transverse (Tr), and oblique (Ob) with respect to the wind.}
\begin{tabular}{*9c}
\hline
\textbf{(1)} & \textbf{(2)} & \textbf{(3)} & \textbf{(4)} & \textbf{(5)} & \textbf{(6)} & \textbf{(7)} & \textbf{(8)} & \textbf{(9)}\\
\textbf{Reference} & \textbf{Type} & \textbf{Cloud} & \textbf{Resolution} & \textbf{$\gamma$} & \textbf{$\chi$} & \textbf{$\cal M_{\rm wind}$} & \textbf{$\beta$} & \textbf{Topology}\\ \hline
\cite{1992ApJ...390L..17S}  & 3D HD & Sph & $R_{60}$ & $1.67$ & $10$ & $10$ & $\infty$ & --\Tstrut \\
\cite{1993ApJ...407..588M}  & 2D HD & Cyl & $R_{25}$ & $1.67$ & $500$, $10^3$ & $0.25$ - $1$ & $\infty$ & --\\ 
\cite{1994ApJ...420..213K}  & 2.5D HD & Sph & $R_{60}$ - $R_{240}$ & $1.67$, $1.1$ & $3$ - $400$ & $10$ - $10^3$ & $\infty$ & --\\ 
\cite{1994ApJ...433..757M}  & 2.5D MHD & Cyl, Sph & $R_{25}$ - $R_{240}$ & $1.67$ & $10$ & $10$ - $100$ & $0.01$, $1$ & Al, Tr\\ 
\cite{1994ApJ...432..194J}  & 2D HD  & Cyl & $R_{43}$ & $1.67$ & $30$, $100$ & $3$, $10$ & $\infty$ & --\\
\cite{1994ApJ...436..776D}  & 2D MHD & Cyl & $R_{46}$ & $1.67$ & $2$ - $40$ & $2$ - $50$ & $\geq0.5$ & Ob\\ 
\cite{1995ApJ...439..237S}  & 2/2.5D HD  & Cyl, Sph & $R_{128}$ - $R_{270}$ & $1.67$ & $10$ - $2000$ & $10$ & $\infty$ & --\\
\cite{1995ApJ...454..172X}  & 3D HD & Sph & $R_{11}$ - $R_{64}$ & $1.67$ & $10$ & $10$ & $\infty$ & --\\ 
\cite{1996ApJ...473..365J}  & 2D M/HD & Cyl & $R_{50}$, $R_{100}$ & $1.67$ & $10$, $40$, $100$ & $10$ & $1$ - $256$, $\infty$ & Al, Tr\\ 
\cite{1999ApJ...517..242M} & 2D MHD & Cyl & $R_{26}$ & $1.67$ & $10$, $100$ & $1.5$,$10$ & $4$ & Ob  \\ 
\cite{1999ApJ...527L.113G} & 3D M/HD & Sph & $R_{26}$ & $1.67$ & $100$ & $1.5$ & $4$, $100$, $\infty$ & Tr  \\ 
\cite{2000ApJ...543..775G}  & 3D MHD & Sph & $R_{26}$ & $1.67$ & $100$ & $1.5$ & $4$, $100$ & Tr  \\
\cite{2002AA...395L..13M}  & 2/2.5D HD & Ell, Sph & $R_{200}$ & $1.67$ & $10^3$ & $10$ & $\infty$ & --\\ 
\cite{2002ApJ...576..832P}  & 2D HD & Cyl & $R_{32}$ & $1.67$ & $500$ & $10$ & $\infty$ & --\\ 
\cite{2004ApJ...604...74F}  & 2D HD & Cyl & $R_{200}$ & $1.67$ & $10^3$ & $5$ - $40$ & $\infty$ & --\\ 
\cite{2004ApJ...613..387P}  & 2.5D HD & Sph & $R_{128}$ & $1.67$ & $100$ & $10$ - $200$ & $\infty$ & --\\
\cite{2005ApJ...619..327F}  & 2D MHD & Cyl & $R_{100}$, $R_{200}$ & $1.67$ & $10^3$ & $10$ & $1$ - $100$ & Al, Tr\\ 
\cite{2005ApJ...633..240P}  & 2D HD & Cyl, Nun & $R_{300-500}$ & $1.67$ & $3-15$ & $10$, $20$ & $\infty$ & --\\ 
\cite{2005AA...443..495M}  & 3D HD & Sph & $R_{32}$ & $1.67$ & $100$, $500$ & $7$ & $\infty$ & --\\ 
\cite{2005RMxAA..41...45R}  & 3D HD & Sph & $R_{25}$ & $1.0$ & $50$ & $2.6$ & $\infty$ & --  \\
\cite{2005MNRAS.361.1077P} & 2D HD & Cyl & $R_{<32}$ & $1.0$ & $\leq 350$ & $1$, $20$ & $\infty$ &  --  \\
\cite{2005MNRAS.362..626M}  & 2.5D HD & Sph & $R_{75}$, $R_{150}$ & $1.67$ & $100$, $500$ & $3$, $6.7$ & $\infty$ & --\\ 
\cite{2006ApJS..164..477N}  & 2.5/3D HD & Sph & $R_{30}$ - $R_{960}$ & $1.67$, $1.1$ & $10$, $100$ & $1.5$ - $10^3$ & $\infty$ & --\\ 
\cite{2006AA...457..545O}  & 2/3D HD & Sph & $R_{105}$, $R_{132}$ & $1.67$ & $10$ & $30$, $50$ & $\infty$ & --\\ 
\cite{2007AA...471..213V}  & 2.5D MHD & Sph & $R_{640}$ & $1.67$ & $45$ & $1.5$ - $5$ & $1$ & Al\\ 
\cite{2007ApJ...668..310R}  & 3D HD & Sph & $R_{76}$ & $1.67$ & $10$ & $242$ & $\infty$ & --\\ 
\cite{2007AA...472..141V}  & 2.5D HD & Sph & $R_{28}$ -$R_{33}$ & $1.67$ & $1$ - $10^4$ & $0.3$ & $\infty$ & --\\ 
\cite{2008ApJ...678..274O}  & 2.5D M/HD & Sph & $R_{132}$ - $R_{528}$ & $1.67$ & $10$ & $50$ & $1$ - $100$ & Al, Tr\\ 
\cite{2008ApJ...680..336S}  & 3D MHD & Sph & $R_{120}$ & $1.67$ & $10$ & $10$ & $0.5$ - $10$ & Al, Tr, Ob\\ 
\cite{2009MNRAS.394.1351P}  & 2.5D HD & Sph & $R_{16}$ -  $R_{256}$ & $1.67$ & $10$ - $10^3$ & $10$ & $\infty$ & --\\ 
\cite{2009ApJ...703..330C}& 3D HD & Fra & $R_{6}$ - $R_{38}$ & $1.67$ & $630$ - $1260$ & $4.6$ & $\infty$ &  -- \\
\cite{2010MNRAS.405..821P}  & 2.5D HD & Sph & $R_{128}$ & $1.67$ & $10$-$10^3$ & $1.5$ - $10$ & $\infty$ & --\\ 
\cite{2010ApJ...722..412Y}  & 2.5D HD & Sph & $R_{12}$ - $R_{1536}$ & $1.67$ & $100$ & $50$ & $\infty$ & --\\ 
\cite{2011ApJ...739...30K} &  2D HD & Sph & $R_{<64}$ & $1.67$ & $10^3$ & $0.6$ - $2$ & $\infty$ &  -- \\
\cite{2011ApSS.336..239P}  & 2.5D HD & Sph & $R_{128}$ & $1.67$ & $10^3$ & $1.5$, $3$ & $\infty$ & --\\ 
\cite{2012MNRAS.425.2212A}  & 2/3D HD & Cyl, Sph & $R_{8}$ - $R_{256}$ & $1.67$ & $10$ - $10^3$ & $1.5$ - $10$ & $\infty$ & --\\ 
\cite{2013ApJ...766...45J}  & 3D MHD & Sph & $R_{100}$ & $1.67$ & $100$ & $30$ & $1$ - $10^3$ & Al, Tr\\ 
\cite{2013ApJ...774..133L}  & 3D MHD & Sph & $R_{54}$ & $1.67$ & $100$ & $10$ & $0.25$, $1$ & Po, To\\ 
\cite{2014MNRAS.444..971A}  & 2D MHD & Cyl & $R_{32}$, $R_{128}$ & $1.67$ & $100$ & $3$ & $0.5$ - $5$ & Al, Tr, Ob\\
\cite{2015MNRAS.449....2M} & 3D MHD & Sph & $R_{32}$ & $1.67$ & $50$ & $1.5$ & $0.1$ - $10$ &  Ta \\
\cite{2015ApJ...805..158S} & 3D HD & Sph & $R_{32}$ - $R_{128}$ & $1.67$ & $300$ - $10^4$ & $0.5$ - $11.4$ & $\infty$ &  -- \\
\cite{2016ApJ...822...31B} & 3D HD & Sph & $R_{32}$ - $R_{96}$ & $1.67$ & $300$ - $10^4$ & $0.99$ - $11.4$ & $\infty$ &  -- \\
\cite{2016MNRAS.455.1309B} & 3D M/HD& Sph & $R_{128}$ & $1.67$, $1.1$ & $10^3$ & $4$, $4.9$ & $10$, $100$, $\infty$ & Al, Tr, Ob \\
\cite{2016MNRAS.457.4470P}  & 2.5/3D HD & Sph & $R_{8}$ - $R_{128}$ & $1.67$ & $10$ - $10^3$ & $1.5$ - $10$ & $\infty$ & --\\
\cite{2016MNRAS.458.1139P}  & 3D HD & Sph, Elo & $R_{4}$ - $R_{64}$ & $1.67$ & $10$ - $10^3$ & $1.5$ - $10$ & $\infty$ & --\\
\cite{2016MNRAS.461..578G}  & 3D MHD & Sph, Elo & $R_{4}$ - $R_{64}$ & $1.67$ & $10$ - $10^3$ & $1.5$ - $10$ & $0.5$ - $10$ & Al, Tr, Ob\\
\cite{2017MNRAS.468.3184G}  & 3D HD & Sph & $R_{6}$ - $R_{200}$ & $1.67$ & $10$ & $10$ & $\infty$ & -- \\
\cite{2017MNRAS.470..114A} & 2D HD& Cyl & $R_{25}$ - $R_{250}$ & $1.67$ & $200$ & $0.5$ - $1.5$ & $\infty$ & -- \\
\cite{2017MNRAS.470.2427G} & 2.5D HD& Sph & $R_{128}$ & $1.67$ & $10$ & $1.36$ - $43$ & $\infty$ & -- \\
\cite{2017ApJ...834..144S}  & 3D HD & Fra & $R_{32}$ - $R_{128}$ & $1.67$ & $19$ - $190$ & $5$ & $\infty$ & -- \\
\cite{2017ApJ...839..103D}  & 3D HD & Sph & $R_{115}$ & $1.67$ & $30$ - $600$ & $10^3$ - $10^4$ & $\infty$ & -- \\
\cite{2017ApJ...845...69G} & 3D MHD& Sph & $R_{16}$ - $R_{32}$ & $1.67$ & $500$, $2500$ & $1.5$, $3.5$ & $2\times 10^3$ & Tr \\
\cite{2018MNRAS.476.2209G} & 2.5D HD& Sph & $R_{128}$ & $1.67$ & $10^3$ & $1.36$ - $43$ & $\infty$ & -- \\
\cite{2018MNRAS.473.3454B} & 3D MHD & Sph, Tur & $R_{128}$ & $1.67$, $1.1$ & $10^3$ & $4$, $4.9$ & $0.04$ - $100$ & Ob, Tu\\
\cite{2018ApJ...865...64G} & 3D MHD & Sph & $R_{32}$ - $R_{128}$ & $1.67$ & $200$, $400$ & $0.45$ & $7$, $700$ & Al, Tr, Ob\\
\cite{2018ApJ...864...96C} & 3D HD & Sph & $R_{64}$ & $1.67$ & $300$ - $10^4$ & $0.5$ - $11.4$ & $\infty$ &  -- \\
\cite{2018MNRAS.480L.111G} & 3D HD & Sph & $R_{8}$ - $R_{64}$ & $1.67$ & $100$ - $10^3$ & $1.5$ & $\infty$ &  -- \\
\cite{2018MNRAS.tmp.2921S} & 2/3D HD & Cyl, Sph & $R_{76}$ - $R_{607}$ & $1.67$ & $10^3$ & $1.5$ & $\infty$ &  -- \\
\cite{2018arXiv181012925F} & 3D HD & Sph & $R_{16}$ & $1.67$ & $100$ & $0.31$ - $1$ & $\infty$ &  -- \\
This paper & 2/3D HD & Sph, Fra & $R_{32}$ - $R_{256}$ & $1.1$ & $10^3$ & $4.9$ & $\infty$ &  -- \\
\hline
\end{tabular}
\label{TableA1}
\end{table*}

\section{On the effects of the numerical resolution}
\label{sec:AppendixB}

\begin{figure*}
\begin{center}
  \begin{tabular}{c c c}
  \hspace{-0.2cm}\resizebox{58mm}{!}{\includegraphics{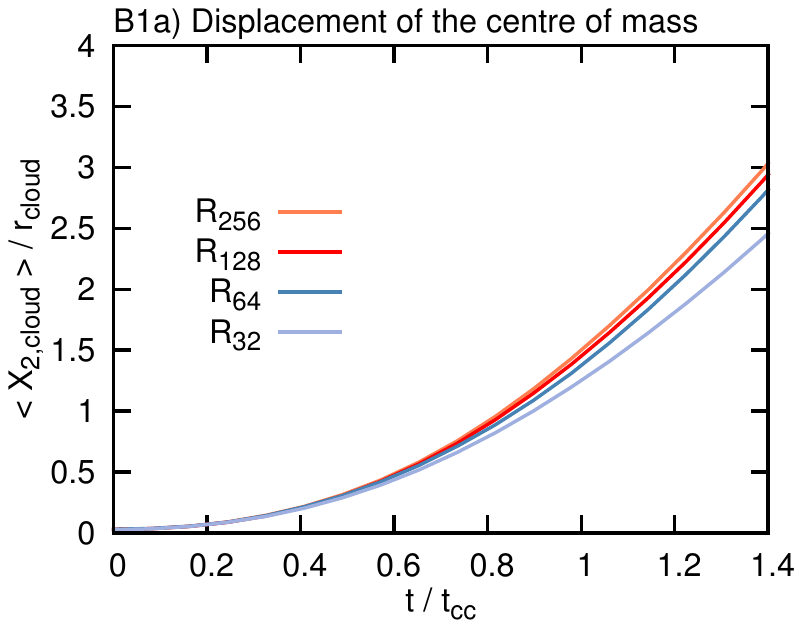}} & \hspace{-0.4cm}\resizebox{58mm}{!}{\includegraphics{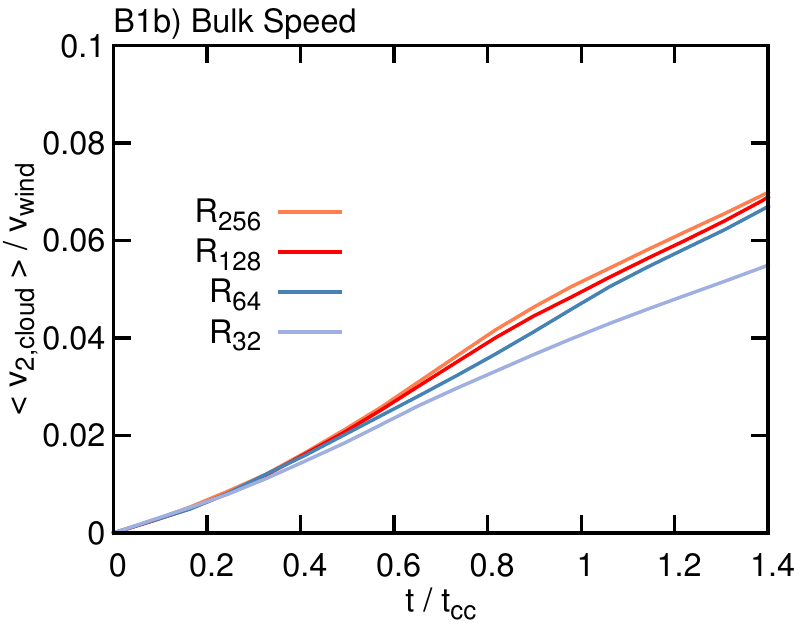}} & \hspace{-0.4cm}\resizebox{58mm}{!}{\includegraphics{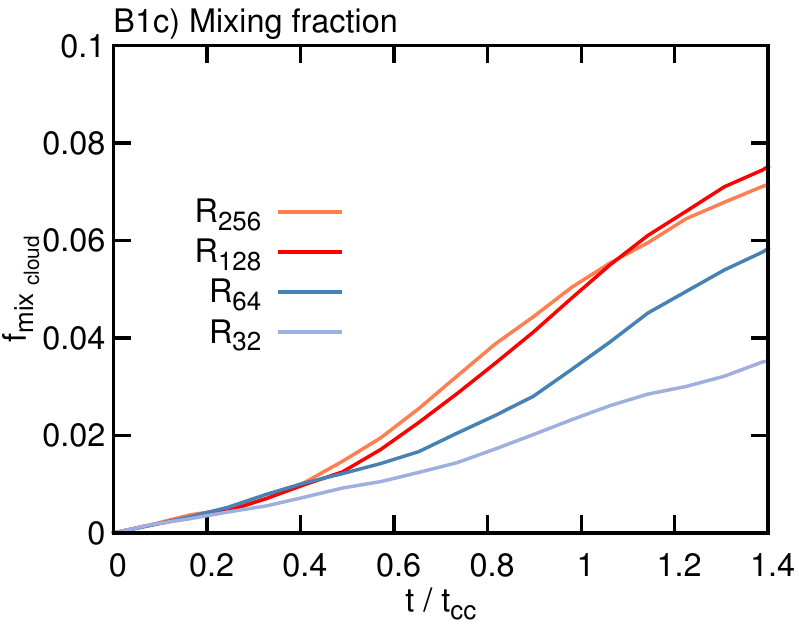}}\\
    \hspace{-0.2cm}\resizebox{58mm}{!}{\includegraphics{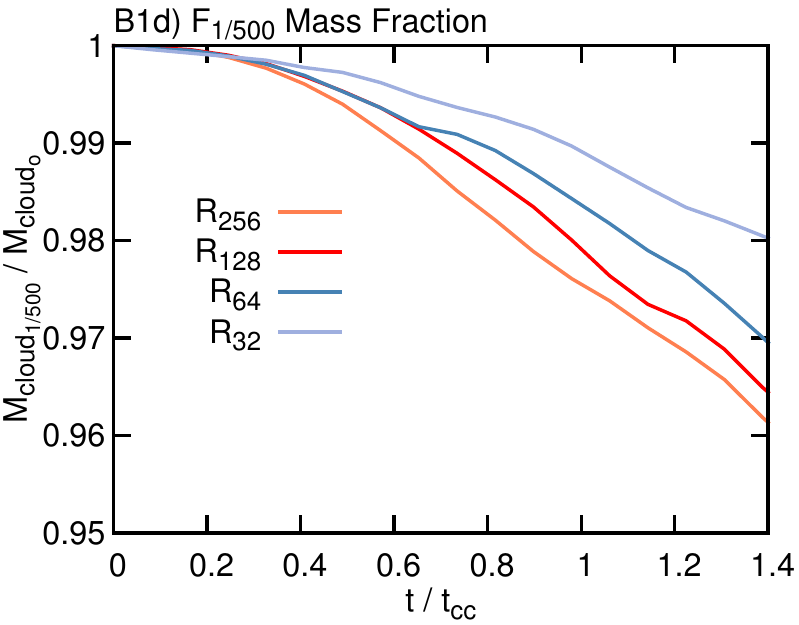}} & \hspace{-0.4cm}\resizebox{58mm}{!}{\includegraphics{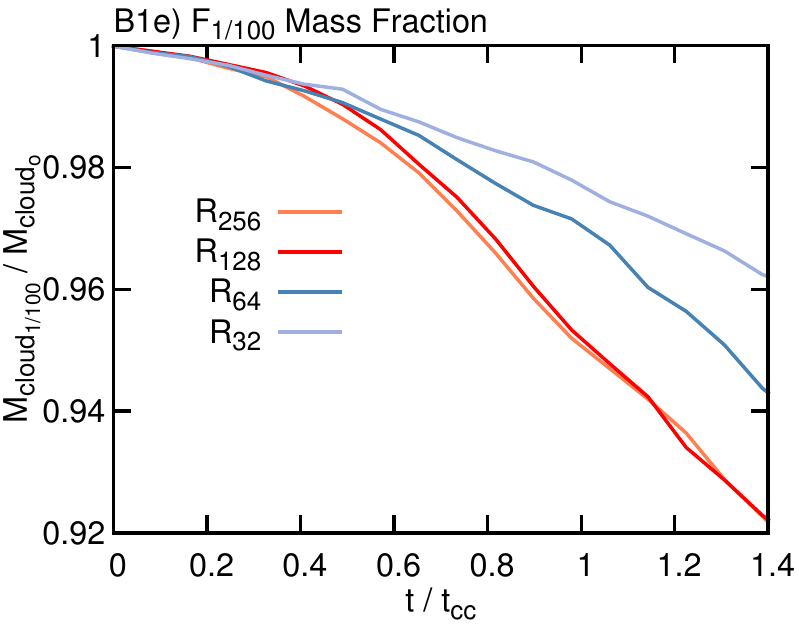}} & \hspace{-0.4cm}\resizebox{58mm}{!}{\includegraphics{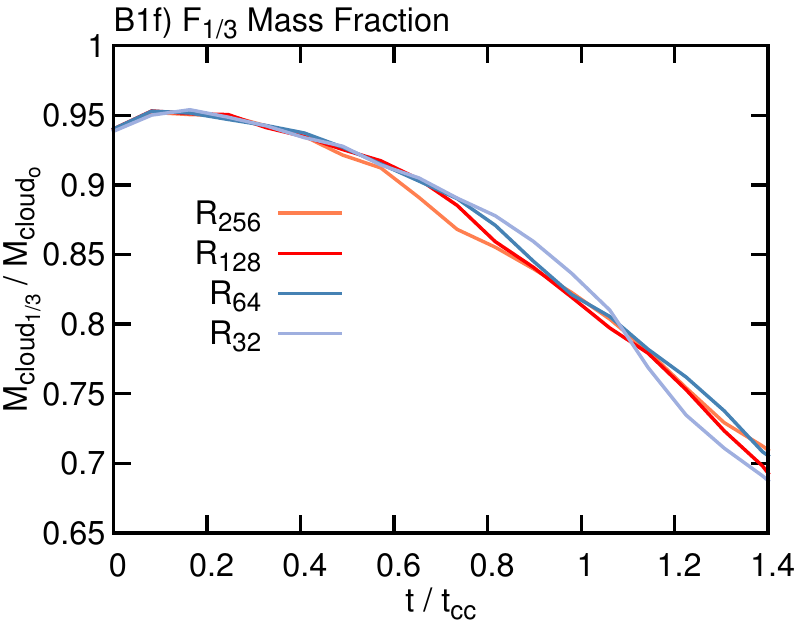}}\vspace{-0.3cm}\\
  \end{tabular}
  \caption{Time evolution of six diagnostics at four different resolutions: $R_{32}$-$R_{256}$. The top panels show the displacement of the centre of mass, the bulk speed, and the mixing fraction, while the bottom panels show the $F_{1/500}$, $F_{1/100}$, $F_{1/3}$ mass fractions. Overall the curves show convergence at a resolution of $R_{128}$, corresponding to $128$ cells per cloud radius. This is for 2D clouds, while \citealt{2018MNRAS.473.3454B} showed that a resolution of at least $64$ cells per cloud radius ($R_{64}$) is required in 3D cloud models.}
  \label{FigureB1}
\end{center}
\end{figure*}

In order to assess how our choice of numerical resolution influences the evolution of fractal clouds, we carried out an extra set of simulations of a 2D solenoidal fractal cloud model at varying resolutions ($R_{\rm 32}$-$R_{\rm 256}$). In order to do a clean comparison we set up all these models in a smaller domain, which consists of a rectangular area with a spatial range $-20\,r_{\rm cloud}\leq X_1\leq20\,r_{\rm cloud}$, $-2\,r_{\rm cloud}\leq X_2\leq78\,r_{\rm cloud}$, and we run the simulations for times $t/t_{\rm cc}\leq1.4$ to ensure that no cloud material leaves the computational domain.\par

Figure \ref{FigureB1} presents six panels: the top panels show the displacement of the centre of mass, the bulk speed, and the mixing fraction between wind and cloud material; while the bottom panels show the evolution of the cloud mass fractions $F_{1/500}$, $F_{1/100}$, and $F_{1/3}$. We find that the standard resolution of $R_{128}$, employed in the 2D models, is adequate to describe the morphological and dynamical evolution of fractal clouds. The mixing fraction and the $F_{1/500}$ mass fraction are the least converged as they greatly depend on the growth of small-scale perturbations at fluid interfaces. As the resolution increases, smaller wavelengths of such perturbations are resolved. Thus, as a general trend, higher resolutions allow low-density gas to mix more effectively, thus reducing the mass of the cloud at such densities.  Despite this, the $F_{1/100}$ and $F_{1/3}$ mass fractions show convergence at $R_{128}$, indicating that our standard resolution is adequate to describe both the dynamics and mixing processes concerning high-density gas in the cloud.\par

The reader is referred to Appendix A in \cite{2018MNRAS.473.3454B} for a resolution study of 3D uniform and fractal cloud models with initial conditions similar to the ones we employed in the models presented in this paper. There we showed that a resolution of $R_{64}$ is adequate to describe the morphology and dynamics of dense gas in 3D wind-swept clouds. This is in agreement with \cite{2016MNRAS.457.4470P} who conducted a shock-cloud resolution study in 3D. They concluded that resolutions of $R_{32}$-$R_{64}$ are the minimum required to achieve convergence, and also showed that poorly resolved clouds can accelerate and mix up to $\sim 5$ times faster than their high-resolution counterparts.

\section{Reflected and refracted shocks}
\label{sec:AppendixC}
\begin{figure}
\begin{center}
  \begin{tabular}{c c c c}
    C1a) 3Dunif  & C1b) 3Dsol & C1c) 3Dcomp & \hspace{-0.8cm} $\frac{P}{P_{\rm wind}}$\\
       \hspace{-0.35cm}\resizebox{26mm}{!}{\includegraphics{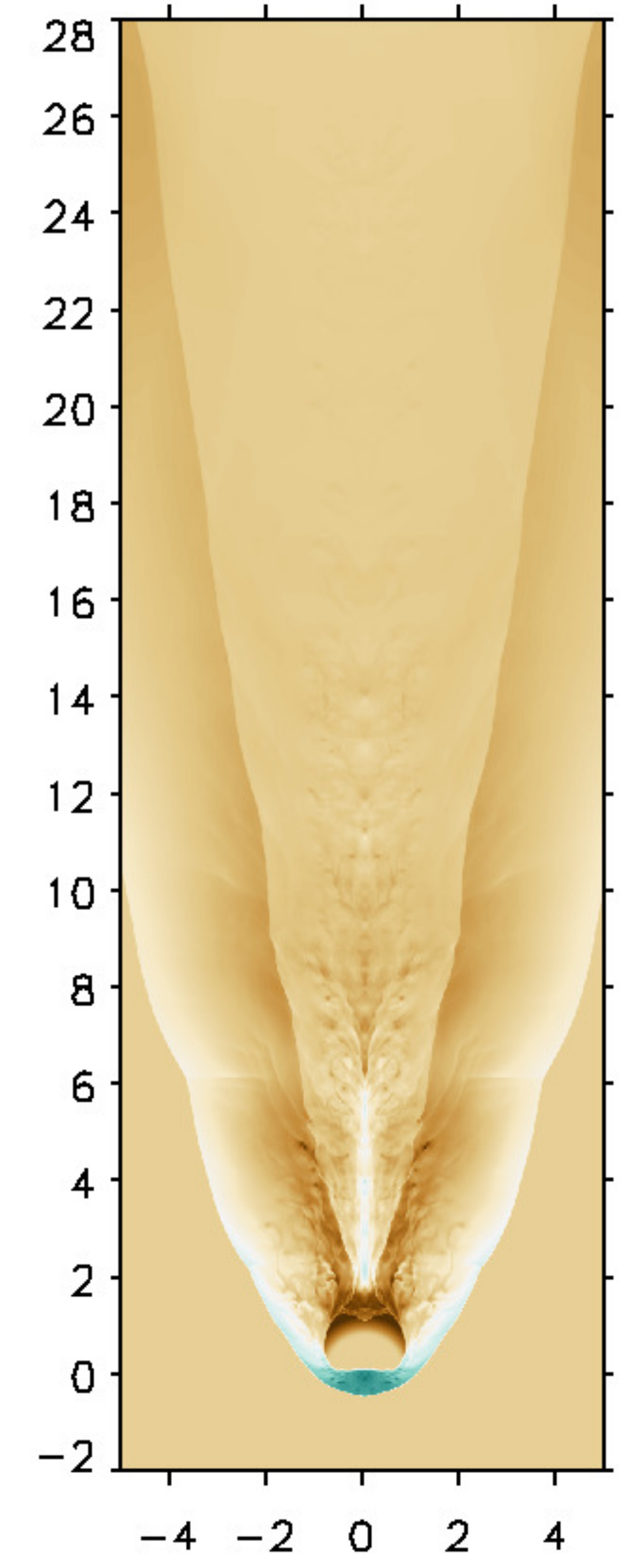}} & \hspace{-0.52cm}\resizebox{26mm}{!}{\includegraphics{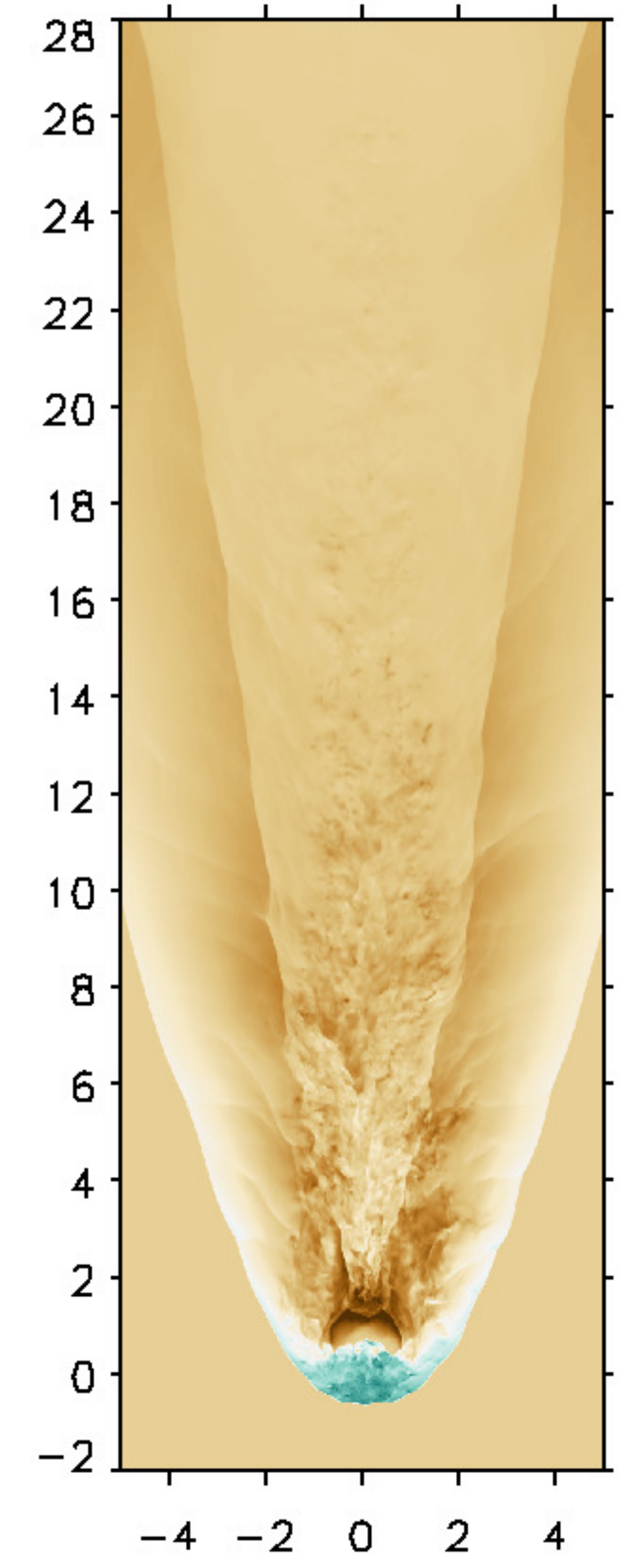}} & \hspace{-0.52cm}\resizebox{26mm}{!}{\includegraphics{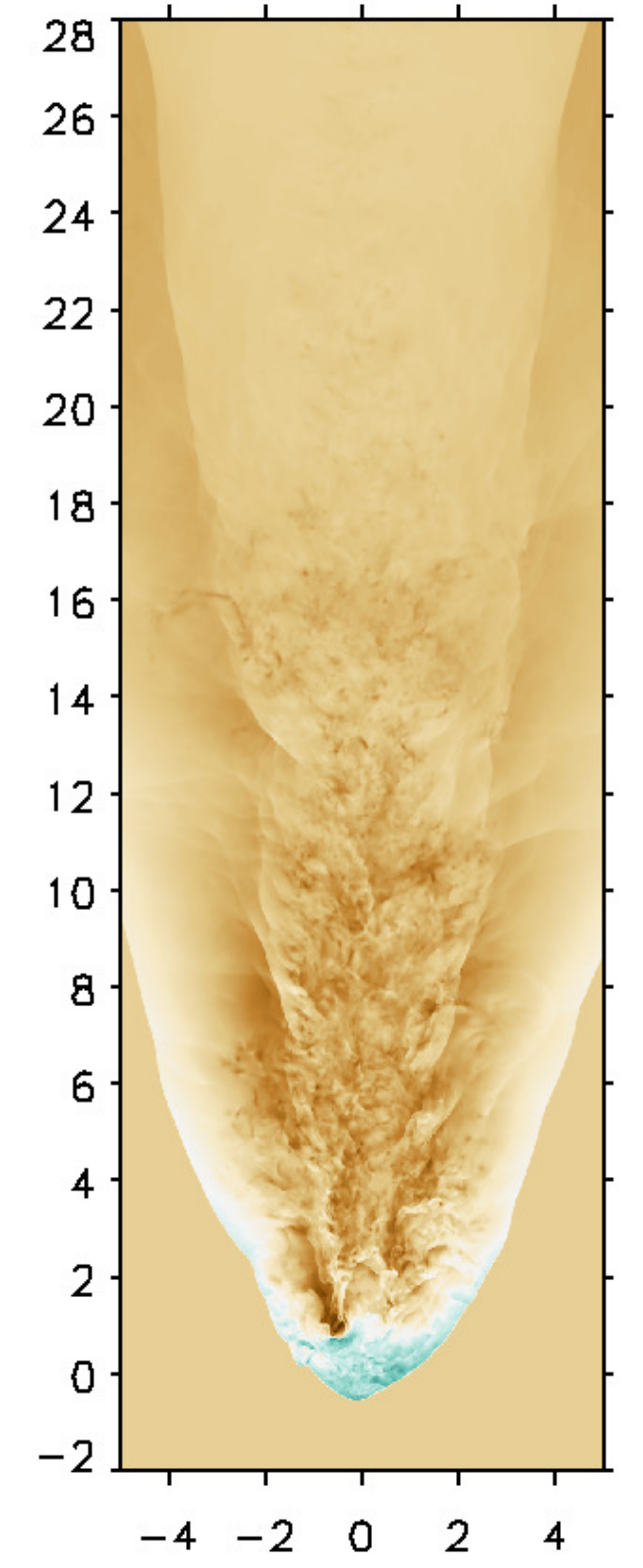}} & \hspace{-0.4cm}{\includegraphics[width=8.1mm]{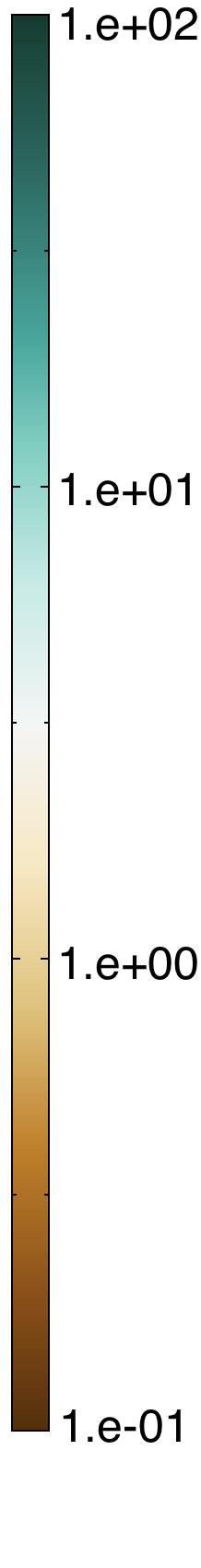}}\\
  \end{tabular}
  \caption{2D slices at $X_3=0$ of the 3D uniform (panel C1a), and 3D fractal solenoidal (panel C1b) and compressive (panel C1c) cloud models showing the thermal gas pressure normalised with respect to the wind pressure at $t/t_{\rm cc}=0.5$. These plots show that fractal clouds favour the formation of anisotropic bow shocks and trigger internal shock splitting.} 
  \label{FigureC1}
\end{center}
\end{figure}

In this Appendix, we show the shock structure in the interior of 3D uniform and fractal cloud models. Figure \ref{FigureC1} shows that in fractal cloud models an anisotropic bow shock forms at the leading edge of the cloud and several refracted shocks travel through the cloud, in contrast to the symmetric bow shock and the single refracted shock characteristic of the uniform cloud model.

%%%%%%%%%%%%%%%%%%%%%%%%%%%%%%%%%%%%%%%%%%%%%%%%%%

% Don't change these lines
\bsp	% typesetting comment
\label{lastpage}
\end{document}